\documentclass[sigconf,10pt]{acmart}
\pdfoutput=1

\renewcommand\footnotetextcopyrightpermission[1]{}

\usepackage{ifpdf}

\usepackage{balance} 

\usepackage{url}
\usepackage{booktabs} 
\usepackage{epstopdf}
\usepackage{graphicx}
\usepackage[subrefformat=parens, labelformat=parens]{subfig}
\usepackage[font=footnotesize, labelfont=footnotesize]{caption}
\usepackage{caption}
\usepackage{amsmath,amssymb,amsfonts}
\usepackage{multirow,makecell}

\usepackage{siunitx} 
\DeclareSIUnit \MHz{ MHz}
\DeclareSIUnit \pF{ pF}
\DeclareSIUnit \nH{ nH}
\DeclareSIUnit \dB{ dB}
\DeclareSIUnit \dBm{ dBm}
\DeclareSIUnit \deg{ deg}

\usepackage{algorithm, algorithmicx, algpseudocode}

\usepackage{tabularx}
\usepackage{enumitem}
\usepackage{flushend}

\newif\ifshowedit
\showedittrue

\usepackage[normalem]{ulem}
\ifshowedit
\newcommand{\remove}[1]{\textcolor{red}{{\sout{#1}}}}

\else
\newcommand{\remove}[1]{}

\fi

\begin{document}

\sloppy

\title{Wideband Full-Duplex Wireless via Frequency- Domain Equalization: Design and Experimentation}






\author{
Tingjun Chen$^\dag$, Mahmood Baraani Dastjerdi$^\dag$, Jin Zhou$^\ddag$, Harish Krishnaswamy$^\dag$, Gil Zussman$^\dag$ \vspace{-0.8\baselineskip}}
\affiliation{
  \institution{
$^\dag$Electrical Engineering, Columbia University \\
$^\ddag$Electrical and Computer Engineering, University of Illinois at Urbana-Champaign \\
\{tingjun@ee, b.mahmood, harish@ee, gil@ee\}.columbia.edu, jinzhou@illinois.edu
}
}



\renewcommand{\shorttitle}{}
\renewcommand{\shortauthors}{}

\begin{abstract}
Full-duplex (FD) wireless can significantly enhance spectrum efficiency but requires tremendous amount of self-interference (SI) cancellation. Recent advances in the RFIC community enabled wideband RF SI cancellation (SIC) in \emph{integrated circuits (ICs)} via frequency-domain equalization (FDE), where RF filters channelize the SI signal path. Unlike other FD implementations, that mostly rely on delay lines, FDE-based cancellers \emph{can be realized in small-form-factor devices}. However, the fundamental limits and higher layer challenges associated with these cancellers were not explored yet. Therefore, and in order to support the integration with a software-defined radio (SDR) and to facilitate experimentation in a testbed with several nodes, we design and implement an FDE-based RF canceller on a printed circuit board (PCB). We derive and experimentally validate the PCB canceller model and present a canceller configuration scheme based on an optimization problem. We then extensively evaluate the performance of the FDE-based FD radio in the SDR testbed. Experiments show that it achieves $\SI{95}{dB}$ overall SIC ($\SI{52}{dB}$ from RF SIC) across $\SI{20}{MHz}$ bandwidth, and an average link-level FD gain of $1.87\times$. We also conduct experiments in: (i) uplink-downlink networks with inter-user interference, and (ii) heterogeneous networks with half-duplex and FD users. The experimental FD gains in the two types of networks confirm previous analytical results. They depend on the users' SNR values and the number of FD users, and are 1.14$\times$--1.25$\times$ and 1.25$\times$--1.73$\times$, respectively. Finally, we numerically evaluate and compare the RFIC and PCB implementations and study various design tradeoffs.
\end{abstract}

%


\keywords{Full-duplex wireless, frequency-domain equalization, wideband self-interference cancellation, software-defined radios}



\setlength{\abovedisplayskip}{3pt}
\setlength{\belowdisplayskip}{3pt}

\maketitle

\thispagestyle{fancy}

\newcommand{\littlesum}{\mathop{\textstyle\sum}}
\newcommand{\littleint}{\mathop{\textstyle\int}}

\newcommand{\SIChnlTF}{H_{\textrm{SI}}}
\newcommand{\SIChnlTFvec}{\mathbf{H}_{\textrm{SI}}}
\newcommand{\AntTF}{H_{\textrm{SI}}}

\newcommand{\NumTap}{M} 
\newcommand{\NumChnl}{K}
\newcommand{\BW}{B}
\newcommand{\RFCancTF}{H}
\newcommand{\RFCancTapTF}[1]{H_{#1}}
\newcommand{\ResTF}{H_{\textrm{res}}}
\newcommand{\ResTapTF}[1]{H_{\textrm{res}, #1}}
\newcommand{\OptProblem}{\textsf{(P1)}}
\newcommand{\OptProblemIC}{\textsf{(P3)}}
\newcommand{\OptProblemPCB}{\textsf{(P2)}}
\newcommand{\OptProblemICPCB}{\textsf{(P4)}}

\newcommand{\FDECancTF}{H^{\textrm{FDE}}}
\newcommand{\FDETapTF}[1]{H_{#1}^{\textrm{FDE}}}

\newcommand{\TF}{H}
\newcommand{\TFApprox}{\widetilde{H}}
\newcommand{\ICTF}{H^{\textrm{I}}}
\newcommand{\ICResTF}{H_{\textrm{res}}^{\textrm{I}}}
\newcommand{\ICTapTF}[1]{H_{#1}^{\textrm{I}}}
\newcommand{\ICTapTFApprox}[1]{\widetilde{H}_{#1}^{\textrm{I}}}
\newcommand{\ICTapAmp}[1]{A_{#1}^{\textrm{I}}}
\newcommand{\ICTapAmpMin}{A_{\textrm{min}}^{\textrm{I}}}
\newcommand{\ICTapAmpMax}{A_{\textrm{max}}^{\textrm{I}}}
\newcommand{\ICTapPhase}[1]{\phi_{#1}^{\textrm{I}}}
\newcommand{\ICTapCF}[1]{f_{\textrm{c},#1}}
\newcommand{\ICTapCFMin}{f_{\textrm{c,min}}}
\newcommand{\ICTapCFMax}{f_{\textrm{c,max}}}
\newcommand{\ICTapCFNoidx}{f_{\textrm{c}}}
\newcommand{\ICTapQF}[1]{Q_{#1}}
\newcommand{\ICTapQFMin}{Q_{\textrm{min}}}
\newcommand{\ICTapQFMax}{Q_{\textrm{max}}}
\newcommand{\PCBTF}{H^{\textrm{P}}}
\newcommand{\PCBResTF}{H_{\textrm{res}}^{\textrm{P}}}
\newcommand{\PCBTapTF}[1]{H_{#1}^{\textrm{P}}}
\newcommand{\BPFTapTF}[1]{H_{#1}^{\textrm{B}}}
\newcommand{\PCBTapTFApprox}[1]{\widetilde{H}_{#1}^{\textrm{P}}}
\newcommand{\PCBTapAmp}[1]{A_{#1}^{\textrm{P}}}
\newcommand{\PCBTapAmpMin}{A_{\textrm{min}}^{\textrm{P}}}
\newcommand{\PCBTapAmpMax}{A_{\textrm{max}}^{\textrm{P}}}
\newcommand{\PCBTapPhase}[1]{\phi_{#1}^{\textrm{P}}}
\newcommand{\PCBTapCF}[1]{f_{c,#1}^{\textrm{P}}}
\newcommand{\PCBTapCFCapNoidx}{C_{\textrm{F}}}
\newcommand{\PCBTapCFCap}[1]{C_{\textrm{F},#1}}
\newcommand{\PCBTapCFCapMin}{C_{\textrm{F,min}}}
\newcommand{\PCBTapCFCapMax}{C_{\textrm{F,max}}}
\newcommand{\PCBTapQFCap}[1]{C_{\textrm{Q},#1}}
\newcommand{\PCBTapQFCapNoidx}{C_{\textrm{Q}}}
\newcommand{\PCBTapQFCapMin}{C_{\textrm{Q,min}}}
\newcommand{\PCBTapQFCapMax}{C_{\textrm{Q,max}}}
\newcommand{\ICPCBTF}{H^{\textrm{I/P}}}
\newcommand{\ICPCBTapTF}[1]{H_{#1}^{\textrm{I/P}}}

\newcommand{\CenterFreq}[1]{f_{c,#1}}
\newcommand{\CenterFreqSingle}{f_c}
\newcommand{\CenterFreqVec}{\mathbf{f}_{c}}
\newcommand{\QFactor}[1]{Q_{#1}}
\newcommand{\QFactorVec}{\mathbf{Q}}
\newcommand{\Amp}[1]{A_{#1}}
\newcommand{\AmpVec}{\mathbf{A}}
\newcommand{\Phase}[1]{\phi_{#1}}
\newcommand{\PhaseVec}{\bm{\upphi}}
\newcommand{\IterSICFreq}[1]{f_{#1}^{\textrm{SIC}}}
\newcommand{\IterSICFreqVec}{\mathbf{f}^{\textrm{SIC}}}
\newcommand{\CF}[1]{f_{c,#1}}
\newcommand{\CFNoidx}{f_c}
\newcommand{\QF}[1]{Q}

\newcommand{\CapCF}[1]{C_F}
\newcommand{\CapCFVec}{\mathbf{C}_F}
\newcommand{\CapQF}[1]{C_Q}
\newcommand{\CapQFVec}{\mathbf{C}_Q}
\newcommand{\Att}[1]{A_{#1}}
\newcommand{\AttVec}{\mathbf{A}}
\newcommand{\PS}[1]{\phi_{#1}}
\newcommand{\PSVec}{\bm{\upphi}}

\newcommand{\PtxNoIdx}{P_{\textrm{tx}}}
\newcommand{\Ptx}[1]{P_{\textrm{tx},{#1}}}
\newcommand{\Prx}[1]{P_{\textrm{rx},{#1}}}
\newcommand{\Dist}[1]{d_{#1}}
\newcommand{\PathGain}{G}
\newcommand{\Nrx}{N_{\textrm{rx}}}
\newcommand{\SNR}[1]{\gamma_{#1}}
\newcommand{\SNRUL}{\gamma_{\textrm{UL}}}
\newcommand{\SNRDL}{\gamma_{\textrm{DL}}}
\newcommand{\XINR}{\gamma_{\textrm{Self}}}
\newcommand{\IUI}{\gamma_{\textrm{IUI}}}
\newcommand{\DataRate}[1]{r_{#1}}

\newcommand{\MyNormTwo}[1]{\left| #1 \right|_2}
\newcommand{\NormTwo}[1]{\left| #1 \right|}


\section{Introduction}
\label{sec:intro}
\begin{figure}[!t]
\centering
\subfloat[]{
\label{fig:intro-pcb}
\includegraphics[height=1in]{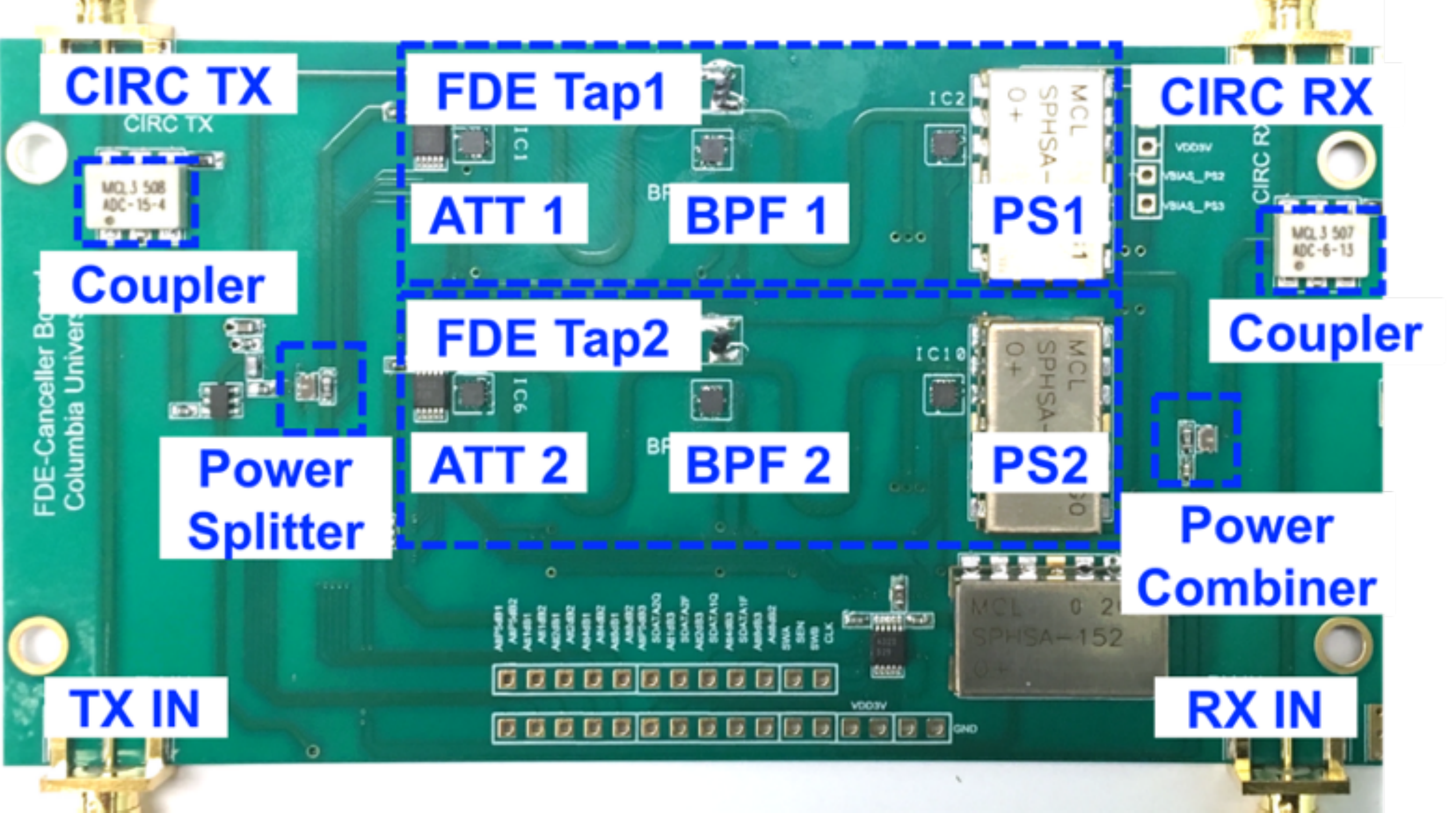}}
\hspace{-6pt} \hfill
\subfloat[]{
\label{fig:intro-fd-radio}
\includegraphics[height=1in]{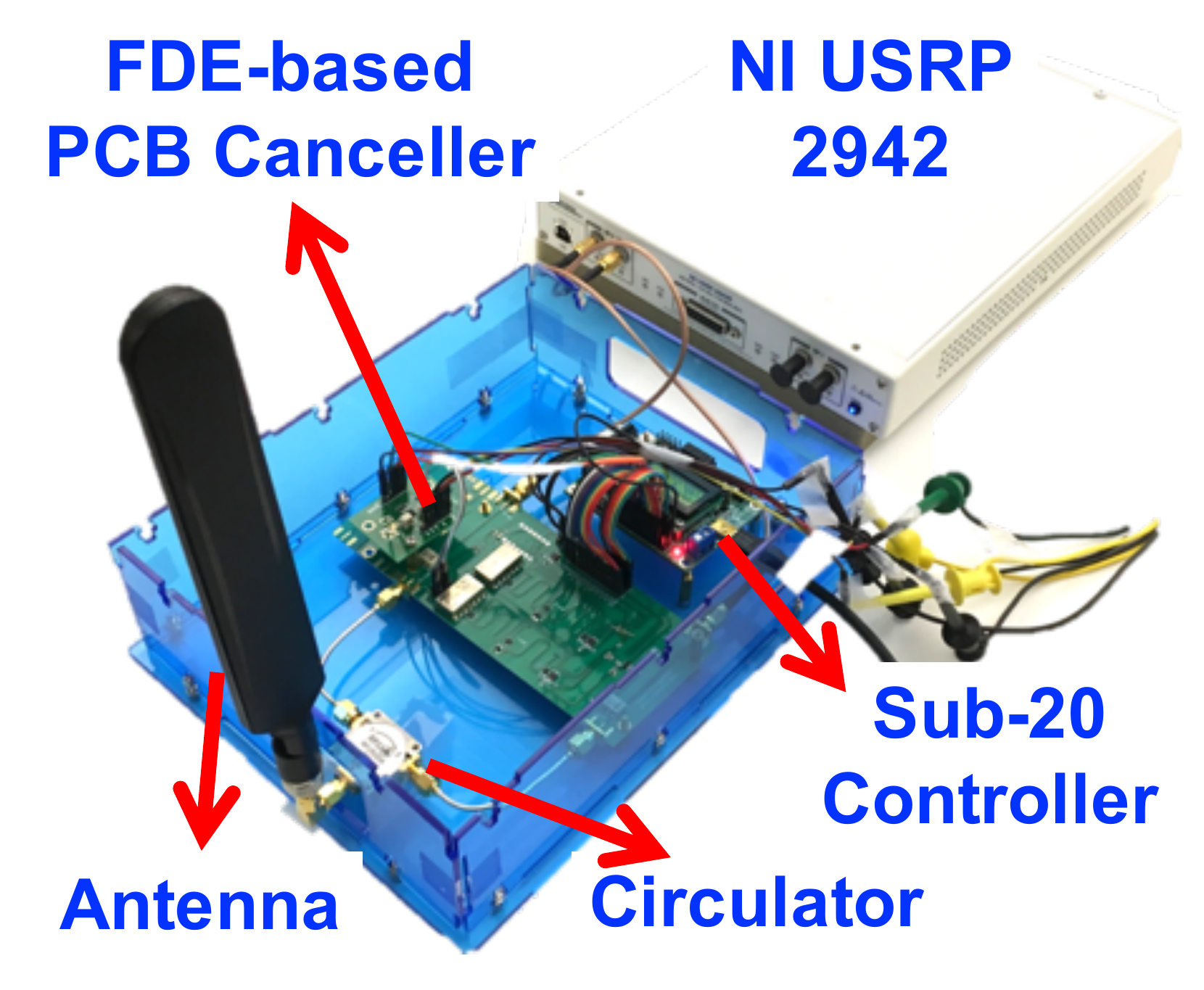}}
\\ \vspace{-.75\baselineskip}
\subfloat[]{
\label{fig:intro-fd-net}
\includegraphics[width=0.9\columnwidth]{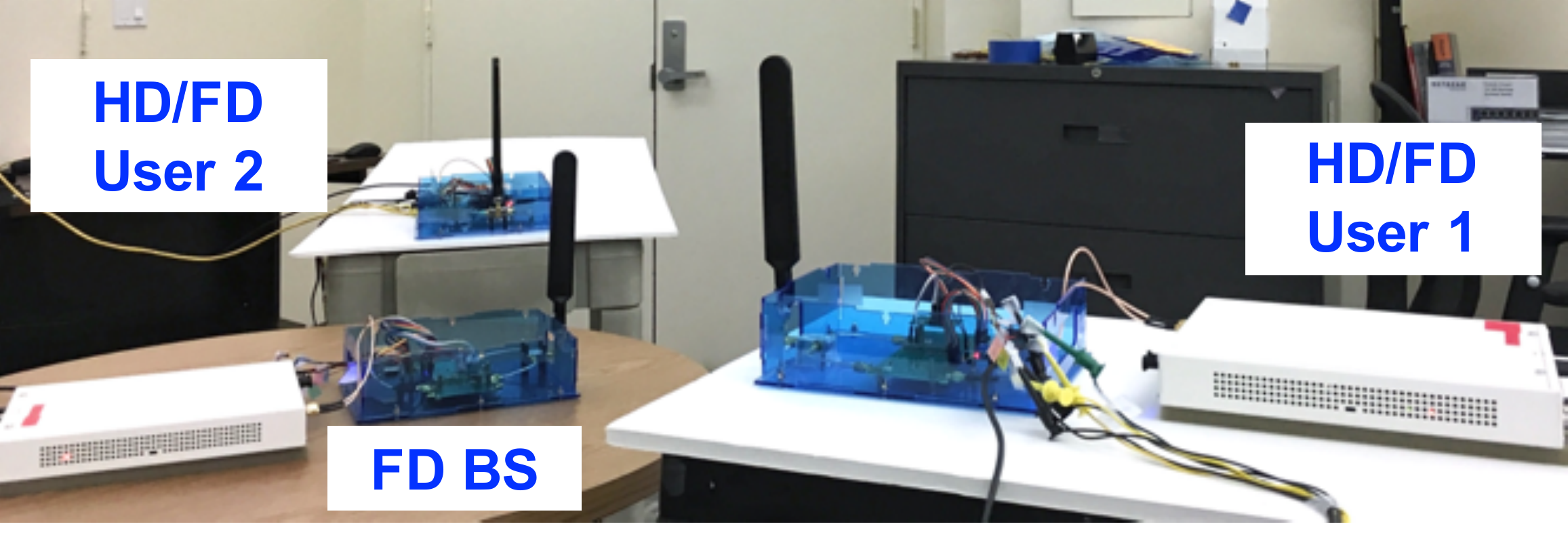}}
\vspace{-0.5\baselineskip}
\caption{(a) The frequency-domain equalization- (FDE-) based wideband RF canceller implemented using discrete components on a PCB, (b) the implemented FDE-based FD radio, and (c) the experimental testbed consisting of an FD base station (BS) and 2 users that can operate in either half-duplex (HD) or FD mode.}
\label{fig:intro}
\vspace{-\baselineskip}
\end{figure}

Full-duplex (FD) wireless -- simultaneous transmission and reception on the same frequency channel -- can significantly improve spectrum efficiency at the physical (PHY) layer and provide many other benefits at the higher layers~\cite{sabharwal2014band,krishnaswamy2016spectrum}. The main challenge associated with FD is the extremely strong self-interference (SI) signal that needs to be suppressed, requiring $90$--$\SI{110}{dB}$ of SI cancellation (SIC). 

Recent work leveraging off-the-shelf components and software-defined radios (SDRs) has established the feasibility of FD wireless through SI suppression at the antenna interface, and SIC in analog/RF and digital domains~\cite{choi2010achieving,duarte2010full,khojastepour2011case,bharadia2013full,korpi2016full}. However, RF cancellers achieving wideband SIC (e.g.,~\cite{bharadia2013full, korpi2016full}) rely on transmission-line delays, which cannot be realized in small-form-factor nodes and/or integrated circuits (ICs) due to the required length for generating nanosecond-scale time delays and the lossy nature of the silicon substrate.\footnote{For instance, obtaining a nanosecond delay in silicon typically requires a $\SI{15}{cm}$-long delay line.}

A \emph{compact IC-based} design is necessary for supporting FD in hand-held devices (e.g., handsets and tablets)~\cite{yang2015wideband,Zhou_WBSIC_JSSC15,krishnaswamy2016spectrum,zhou2017integrated}. Specifically, recent advances in the RFIC community allowed achieving wideband RF SIC in IC implementations based on the technique of frequency-domain equalization (FDE)~\cite{Zhou_WBSIC_JSSC15}. In contrast to the delay line-based approaches (which are essentially performing time-domain equalization), the FDE-based RF canceller utilizes tunable, reconfigurable, high quality factor $2^{\textrm{nd}}$-order bandpass filters (BPFs) with amplitude and phase controls to emulate the frequency-selective antenna interface. In general, tunable, high quality factor BPFs are perhaps as hard to implement on an IC as nanosecond-scale delay lines. However, $N$-path filters represent an exciting advance that has enabled their implementation in nanoscale CMOS over the past few years~\cite{ghaffari2011tunable,reiskarimian2016analysis}.

While major advances have been made at the IC level, existing work (e.g.,~\cite{Zhou_WBSIC_JSSC15}) has several limitations: (i) the fundamental limits of the achievable RF SIC based on the technique of FDE have not been fully understood, (ii) configuration schemes for this new type of RF canceller need to be developed in order to achieve optimized and adaptive RF SIC in real-world environments, and (iii) the system-level performance of such IC-based FD radios has not been evaluated in different network settings. Therefore, in this paper we focus on FDE-based RF cancellers.
 
Since interfacing an RFIC canceller to an SDR presents numerous technical challenges, we design and implement an FDE-based RF canceller using discrete components on a printed circuit board (PCB). This canceller appears in Fig.~\ref{fig:intro}\subref{fig:intro-pcb} (we refer to it as the \emph{PCB canceller}) and it emulates its RFIC counterpart.\footnote{The PCB canceller design is available at~\cite{flexicon_research_gen2}.} This FDE-based PCB canceller facilitates the evaluation of the canceller configuration scheme and the experimentation using SDRs in a network with multiple FD nodes. Moreover, the PCB canceller is more robust and stable than its IC counterpart and as such can be integrated in the future in the open-access ORBIT~\cite{orbit} and COSMOS~\cite{COSMOS} testbeds to allow the community to experiment with wideband compact FD wireless. For example, our previous \emph{narrowband} RF canceller emulating its RFIC counterpart~\cite{Zhou_NCSIC_JSSC14} is implemented on a PCB and is integrated in the ORBIT testbed~\cite{flexicon_orbit_gen1,flexicon_orbit_arxiv}.

We present a realistic model of the PCB canceller. We then present its configuration scheme based on an optimization problem, which allows efficient adaption of the canceller to environmental changes. The PCB canceller model is experimentally validated and is shown to have high accuracy. We implement an FDE-based FD radio by integrating the PCB canceller with an NI USRP SDR, as depicted in Fig.~\ref{fig:intro}\subref{fig:intro-fd-radio}.\footnote{A preliminary version of the system was demonstrated in~\cite{fd_demo_infocom17}.} This FD radio achieves $\SI{95}{dB}$ overall SIC across $\SI{20}{MHz}$ real-time bandwidth, enabling an FD link budget of $\SI{10}{dBm}$ average TX power level and $-\SI{85}{dBm}$ RX noise floor. In particular, $\SI{52}{dB}$ RF SIC is achieved, from which $\SI{20}{dB}$ is obtained from the antenna interface isolation.

We also evaluate the performance and robustness of the FDE-based FD radio at the link-level in terms of packet reception ratio (PRR) and FD throughput gain, in both line-of-sight (LOS) and non-line-of-sight (NLOS) settings. The results show that the FDE-based FD radio achieves an average FD link throughput gain of 1.85$\times$--1.91$\times$. Moreover, the link SNR difference when the radio operates in half-duplex (HD) and FD modes is less than $\SI{1}{dB}$.

Using our testbed (see Fig.~\ref{fig:intro}\subref{fig:intro-fd-net}), we extensively evaluate the network-level FD gain and confirm analytical results in two types of networks: (i) \emph{UL-DL networks} consisting of one FD base station (BS) and two half-duplex (HD) users with inter-user interference (IUI), and (ii) \emph{heterogeneous HD-FD networks} consisting of one FD BS and co-existing HD and FD users. For UL-DL networks, we show experimentally that the throughput gain is between  1.14$\times$--1.25$\times$ compared to 1.22$\times$--1.3$\times$ predicted by analysis. We discuss the relationship between the FD gain and UL and DL SNR values, as well as the IUI levels. For heterogeneous HD-FD networks, we demonstrate via experiments the impact of different user SNR values and the number of FD users on the FD gain. For example, in a 4-node network consisting of an FD BS and 3 users with various user locations and SNR values, median experimental FD gains of 1.25$\times$ and 1.52$\times$ can be achieved when one and two users become FD-capable, respectively.

To the best of our knowledge, this is the first experimental study of FD gains in such networks using a testbed composed of both HD and FD radios. The results demonstrate the practicality and performance of FDE-based FD radios, which are suitable for small-form-factor devices. The results can also serve as building blocks for developing higher layer (e.g., MAC) protocols.

Finally, we numerically evaluate the FDE-based cancellers based on measurements and validated canceller models. We compute achievable RF SIC under practical constraints and discuss various canceller design tradeoffs. We also compare the performance of the RFIC and PCB cancellers. We show that our optimized canceller configuration scheme achives an order of magnitude higher RF SIC than the heuristic scheme used in the RFIC canceller~\cite{Zhou_WBSIC_JSSC15}.

To summarize, the main contributions of the paper are:
\begin{enumerate}[leftmargin=*,topsep=3pt]
\item[1.]
We present the design, implementation, modeling, and validation of the FDE-based PCB canceller, as well as an optimized canceller configuration scheme;
\item[2.]
We experimentally evaluate the performance of our FDE-based FD radio with the PCB canceller and the optimized canceller configuration, including the achieved overall SIC and link-level FD gain; 
\item[3.]
We experimentally evaluate the FD throughput gain in various network settings with different user capabilities (i.e., HD or FD) and user SNR values.
\end{enumerate}

The rest of the paper is organized as follows. Section~\ref{sec:related} reviews related work. In Section~\ref{sec:background}, we present the problem formulation and RF canceller designs. We present the design, implementation, and model of the FDE-based PCB canceller, as well as the optimized canceller configuration scheme in Section~\ref{sec:impl}. The canceller model is experimentally validated in Section~\ref{sec:pcb-validation}. The performance of the FDE-based FD radio is experimentally evaluated in Sections~\ref{sec:exp}. In Section~\ref{sec:sensitivity}, we numerically evaluate the FDE-based cancellers, and compare the RFIC and PCB implementations. We conclude and discuss future directions in Section~\ref{sec:conclusion}.


\begin{table}[!t]
\caption{Nomenclature}
\label{table:notation}
\vspace{-0.5\baselineskip}
\scriptsize
\begin{center}
\begin{tabular}{| p{0.15\columnwidth} | p{0.75\columnwidth} |}
\hline
$|z|$, $\angle z$ & Amplitude and phase of a complex value $z=x+jy$ ($x,y\in\mathbb{R}$), where $|z| = \sqrt{x^2+y^2}$ and $\angle z = \tan^{-1} \left(\frac{y}{x}\right)$ \\
$\BW$ & Total wireless bandwidth/desired RF SIC bandwidth \\
$\NumChnl$, $k$ & Total number of frequency channels and channel index \\
$f_k$ & Center frequency of the $k^{\textrm{th}}$ frequency channel \\
$\NumTap$ & Number of FDE taps in an FDE-based RF canceller \\
$\AntTF(f_k)$ & Frequency response of the antenna interface \\
$\PCBTF(f_k)$ & Frequency response of the FDE-based PCB canceller \\
$\PCBTapTF{i}(f_k)$  & Frequency response of the $i^{\textrm{th}}$ FDE tap in the PCB canceller \\
$\PCBTapAmp{i}$, $\PCBTapPhase{i}$ & Amplitude and phase controls of the $i^{\textrm{th}}$ FDE tap in the PCB canceller \\
$\PCBTapCFCap{i}$, $\PCBTapQFCap{i}$ & Digitally tunable capacitors that control the center frequency and quality factor of the $i^{\textrm{th}}$ FDE tap in the PCB canceller \\
\hline
\end{tabular}
\end{center}
\vspace{-\baselineskip}
\end{table}


\section{Related Work}
\label{sec:related}
Extensive research related to FD wireless is summarized in~\cite{sabharwal2014band}, including implementations of FD radios and systems, analysis of rate gains, and resource allocation at the higher layers. Below, we briefly review the related work.

\subsubsection*{RF Canceller and FD Radio Designs}
RF SIC typically involves two stages: (i) isolation at the antenna interface, and (ii) SIC in the RF domain using cancellation circuitry. While a separate TX/RX antenna pair can provide good isolation and can be used to achieve cancellation~\cite{radunovic2010rethinking,choi2010achieving,khojastepour2011case,jain2011practical,aryafar2012midu,aryafar2015fd}, a shared antenna interface such as a circulator is more appropriate for single-antenna implementations~\cite{bharadia2013full,chung2015prototyping} and is compatible with FD MIMO systems. Existing designs of analog/RF SIC circuitry are mostly based on a time-domain interpolation approach~\cite{bharadia2013full,korpi2016full}. In particular, real delay lines with different lengths and amplitude weighting~\cite{bharadia2013full} and phase controls~\cite{korpi2016full} are used and their configurations are optimized to best emulate the SI channel. This essentially represents an RF implementation of a finite impulse response (FIR) filter. Based on the same RF SIC approach, several FD MIMO radio designs are presented~\cite{aryafar2012midu,bharadia2014full,chen2015flexradio,chung2017compact}. FD relays have also been successfully demonstrated in~\cite{hsu2016full,bharadia2015fastforward,chen2015airexpress,chen2017bipass}. Moreover, SIC can be achieved via digital/analog beamforming in FD massive-antenna systems~\cite{everett2016softnull,aryafar2018pafd}.
The techniques utilized in these works are incompatible with IC implementations, which are required for small-form-factor devices.
In this paper, we focus on an FDE-based canceller, which builds on our previous work towards the design of such an RFIC canceller~\cite{Zhou_WBSIC_JSSC15}. However, existing IC-based FD radios (e.g.,~\cite{Zhou_WBSIC_JSSC15}) have not been evaluated at the system-level in different network settings.

\subsubsection*{FD Gain at the Link- and Network-level}
At the higher layers, recent work focuses on characterizing the capacity region and rate gains, as well as developing resource allocation algorithms under both perfect~\cite{ahmed2013rate,Sabharwal_DistributedSideChannel_TWC13} and imperfect SIC~\cite{goyal2015full,marasevic2017resource,diakonikolas2017rate}. Similar problems are considered in FD multi-antenna/MIMO systems~\cite{zheng2015joint,everett2016softnull,qian2017concurrent}. Medium access control (MAC) algorithms are studied in networks with all HD users~\cite{choi2015power,chen2017probabilistic} or with heterogeneous HD and FD users~\cite{chen2018hybrid}. Moreover, network-level FD gain is analyzed in~\cite{radunovic2010rethinking,yang2014characterizing,xie2014does,wang2017fundamental} and experimentally evaluated in~\cite{jain2011practical,kim2013janus} where all the users are HD or FD. Finally,~\cite{hsu2017inter} proposes a scheme to suppress IUI using an emulated FD radio.

To the best of our knowledge, this is \emph{the first thorough study of wideband RF SIC achieved via a frequency-domain-based approach (which is suitable for compact implementations) that is grounded in real-world implementation and includes extensive system- and network-level experimentation}.


\section{Background and Formulation}
\label{sec:background}
In this section, we review concepts related to FD wireless and RF canceller configuration optimization. We also present different RF canceller designs and specificaully the design of the FDE-based RF canceller. Summary of the main notation is provided in Table~\ref{table:notation}.

\subsection{FD Background and Notation}
\begin{figure}[!t]
\centering
\includegraphics[width=0.85\columnwidth]{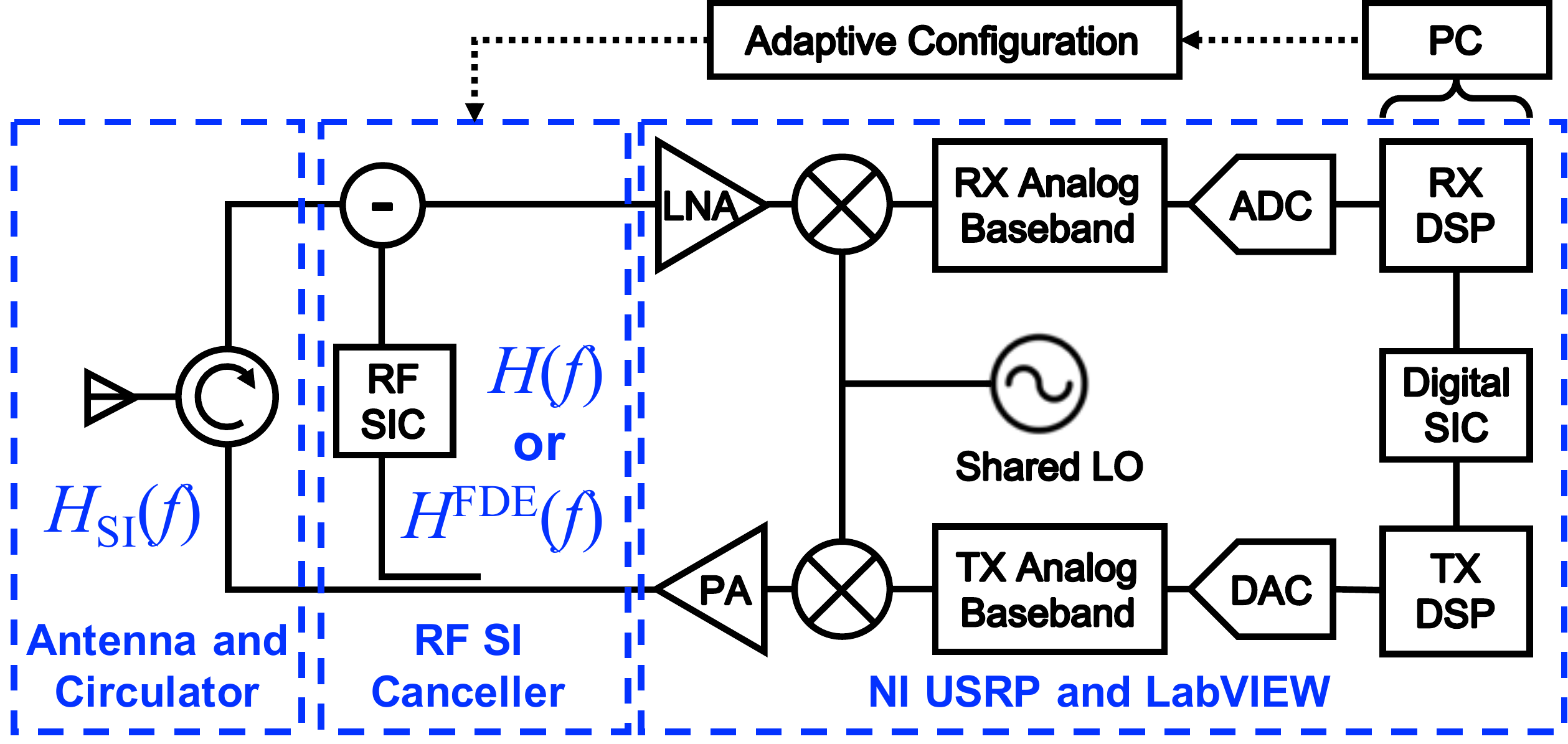}
\vspace{-0.5\baselineskip}
\caption{Block diagram of an FD radio.}
\label{fig:diagram}
\vspace{-\baselineskip}
\end{figure}

Fig.~\ref{fig:diagram} shows the block diagram of a single-antenna FD radio using a circulator at the antenna interface. Due to the extremely strong SI power level and the limited dynamic range of the analog-to-digital converter (ADC) at the RX, a total amount of $90$--$\SI{110}{dB}$ SIC must be achieved across the antenna, RF, and digital domains. Specifically, (i) SI suppression is first performed at the antenna interface, (ii) an RF SI canceller then taps a reference signal at the output of the TX power amplifier (PA) and performs SIC at the input of the low-noise amplifier (LNA) at the RX, and (iii) a digital SI canceller further suppresses the residual SI.

Consider a wireless bandwidth of $\BW$ that is divided into $\NumChnl$ orthogonal frequency channels. The channels are indexed by $k \in \{1,\dots,\NumChnl\}$ and denote the center frequency of the $k^{\textrm{th}}$ channel by $f_k$.\footnote{We use discrete frequency values $\{f_k\}$ since in practical systems, the antenna interface response is measured at discrete points (e.g., per OFDM subcarrier). However, the presented model can also be applied to cases with continuous frequency values.}
We denote the antenna interface response by $\AntTF(f_k)$ with amplitude $|\AntTF(f_k)|$ and phase $\angle\AntTF(f_k)$. Note that the actual SI channel includes the TX-RX leakage from the antenna interface as well as the TX and RX transfer functions at the baseband from the perspective of the digital canceller. Since the paper focuses on achieving wideband RF SIC, we use $\AntTF(f_k)$ to denote the antenna interface response and also refer to it as the \emph{SI channel}. We refer to \emph{TX/RX isolation} as the ratio (in $\SI{}{dB}$, usually a negative value) between the residual SI power at the RX input and the TX output power, which includes the amount of TX/RX isolation achieved by both the antenna interface and the RF canceller/circuitry. We then refer to \emph{RF SIC} as the absolute value of the TX/RX isolation. We also refer to \emph{overall SIC} as the total amount of SIC achieved in both the RF and digital domains. The antenna interface used in our experiments typically provides a TX/RX isolation of around $-\SI{20}{dB}$.

\subsection{Problem Formulation}
Ideally, an RF canceller is designed to best emulate the antenna interface, $\AntTF(f_k)$, across a desired bandwidth, $\BW=[f_1, f_\NumChnl]$. We denote by $\RFCancTF(f_k)$ the frequency response of an RF canceller and by $\ResTF(f_k) := \AntTF(f_k) - \RFCancTF(f_k)$ the \emph{residual SI channel response}. The optimized RF canceller configuration is obtained by solving {\OptProblem}:
\begin{align}
\textsf{(P1)}\
& \textrm{min:}\
\littlesum\limits_{k=1}^{\NumChnl} \NormTwo{ \ResTF(f_k) }^2 = \littlesum\limits_{k=1}^{\NumChnl} \NormTwo{ \AntTF(f_k) - \RFCancTF(f_k) }^2 \nonumber \\
\textrm{s.t.:}\ & \textrm{constraints on configuration parameters of}\ \RFCancTF(f_k),\ \forall k. \nonumber
\end{align}

The RF canceller configuration obtained by solving $\OptProblem$ minimizes the residual SI power referred to the TX output. As described in Section~\ref{sec:intro}, one main challenge associated with the design of the RF canceller with response $\RFCancTF(f_k)$ to achieve wideband SIC is due to the highly frequency-selective antenna interface, $\AntTF(f_k)$. Moreover, an efficient RF canceller configuration scheme needs to be designed so that the canceller can adapt to time-varying $\AntTF(f_k)$.

\subsection{RF Canceller Designs}
\label{ssec:background-previous-work}

\subsubsection*{Delay Line-based RF Cancellers}
An RF canceller design introduced in~\cite{bharadia2013full} involves using $\NumTap$ delay line taps. Specifically, the $i^{\textrm{th}}$ tap is associated with a time delay of $\tau_i$, which is \emph{pre-selected} and \emph{fixed} depending on the selected circulator and antenna, and an amplitude control of $\Amp{i}$. Since the Fourier transform of a delay of $\tau_i$ is $e^{-2\pi f\tau_i}$, an $\NumTap$-tap delay line-based RF canceller has a frequency response of $\RFCancTF^{\textrm{DL}}(f_k) = \sum_{i=1}^{\NumTap} \Amp{i} e^{-j2\pi f_k \tau_i}$. The configurations of the amplitude controls, $\{\Amp{i}\}$, are obtained by solving $\OptProblem$ with $\RFCancTF(f_k) = \RFCancTF^{\textrm{DL}}(f_k)$. In~\cite{bharadia2013full}, an RF canceller of $\NumTap=16$ delay line taps is implemented. In~\cite{korpi2016full}, a similar approach is considered with $\NumTap=3$ and an additional phase control, $\Phase{i}$, on each tap, resulting in an RF canceller model of $\RFCancTF^{\textrm{DL}}(f_k) = \sum_{i=1}^{3} \Amp{i} e^{-j(2\pi f_k \tau_i + \Phase{i})}$. As mentioned in Section~\ref{sec:intro}, although such cancellers can achieve wideband SIC, this approach is more suitable for large-form-factor nodes than for compact/small-form-factor implementations.

\subsubsection*{Amplitude- and Phase-based RF Cancellers}
A compact design that is based on an amplitude- and phase-based RF canceller realized in an RFIC implementation is presented in~\cite{Zhou_NCSIC_JSSC14}. This canceller has a single-tap with one amplitude and frequency control, $(\Amp{0}, \Phase{0})$, which can emulate the antenna interface, $\AntTF(f_k)$, at \emph{only one} given cancellation frequency $f_{1}$ by setting $\Amp{0} = |\AntTF(f_{1})|$ and $\Phase{0} = \angle \AntTF(f_{1})$. The same design is also realized using discrete components on a PCB (without using any length delay lines), and is integrated in the ORBIT testbed for open-access FD research~\cite{flexicon_orbit_arxiv}. However, this type of RF cancellers has limited RF SIC perfromacne and bandwidth, since it can only emulate the antenna interface at a single frequency.

\begin{figure}[!t]
\centering
\vspace{-\baselineskip}
\subfloat[]{
\label{fig:fde-concept-diagram}
\includegraphics[height=1.15in]{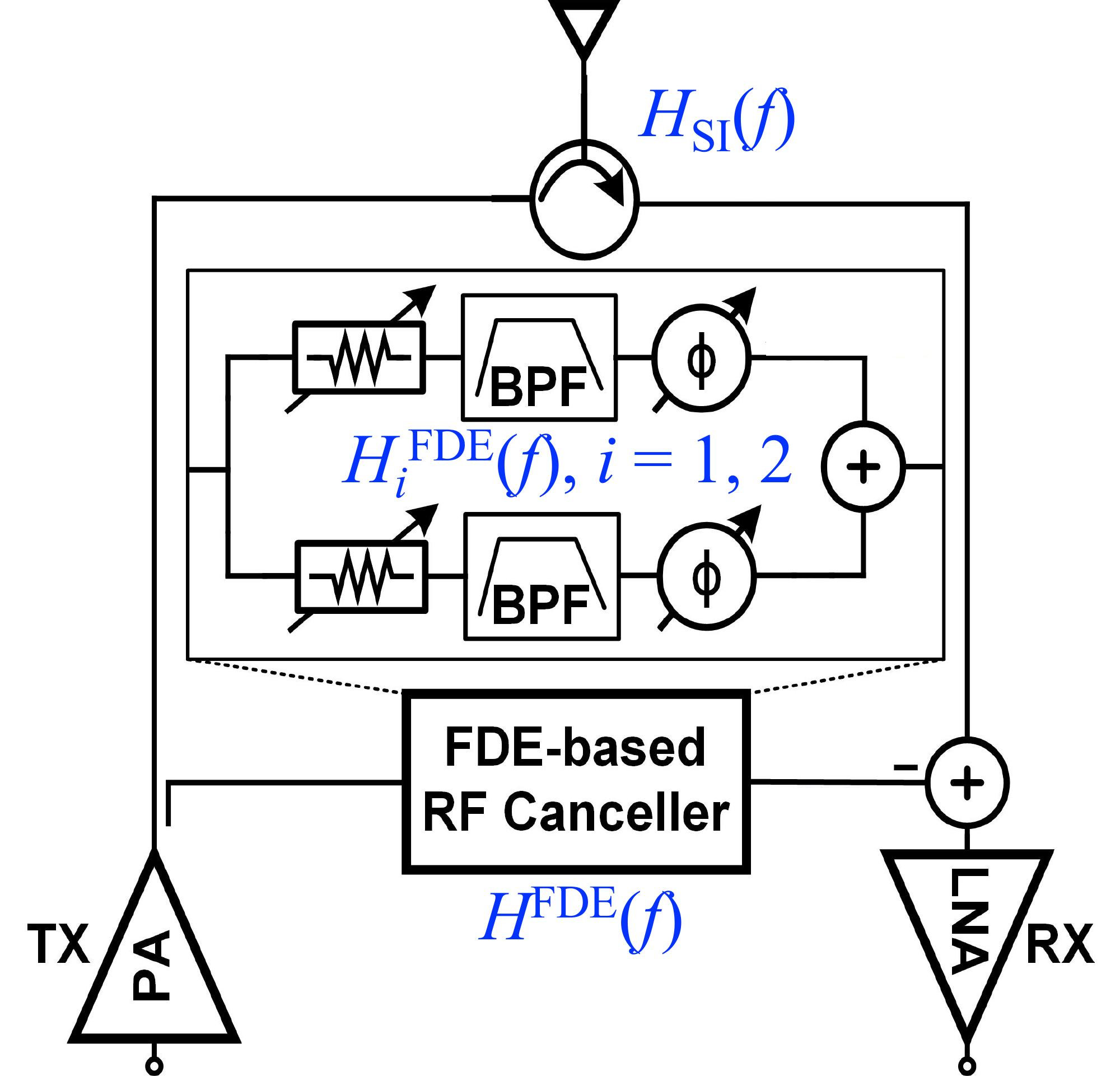}}
\hspace{-12pt} \hfill
\subfloat[]{
\label{fig:fde-concept-bpf}
\includegraphics[height=1.15in]{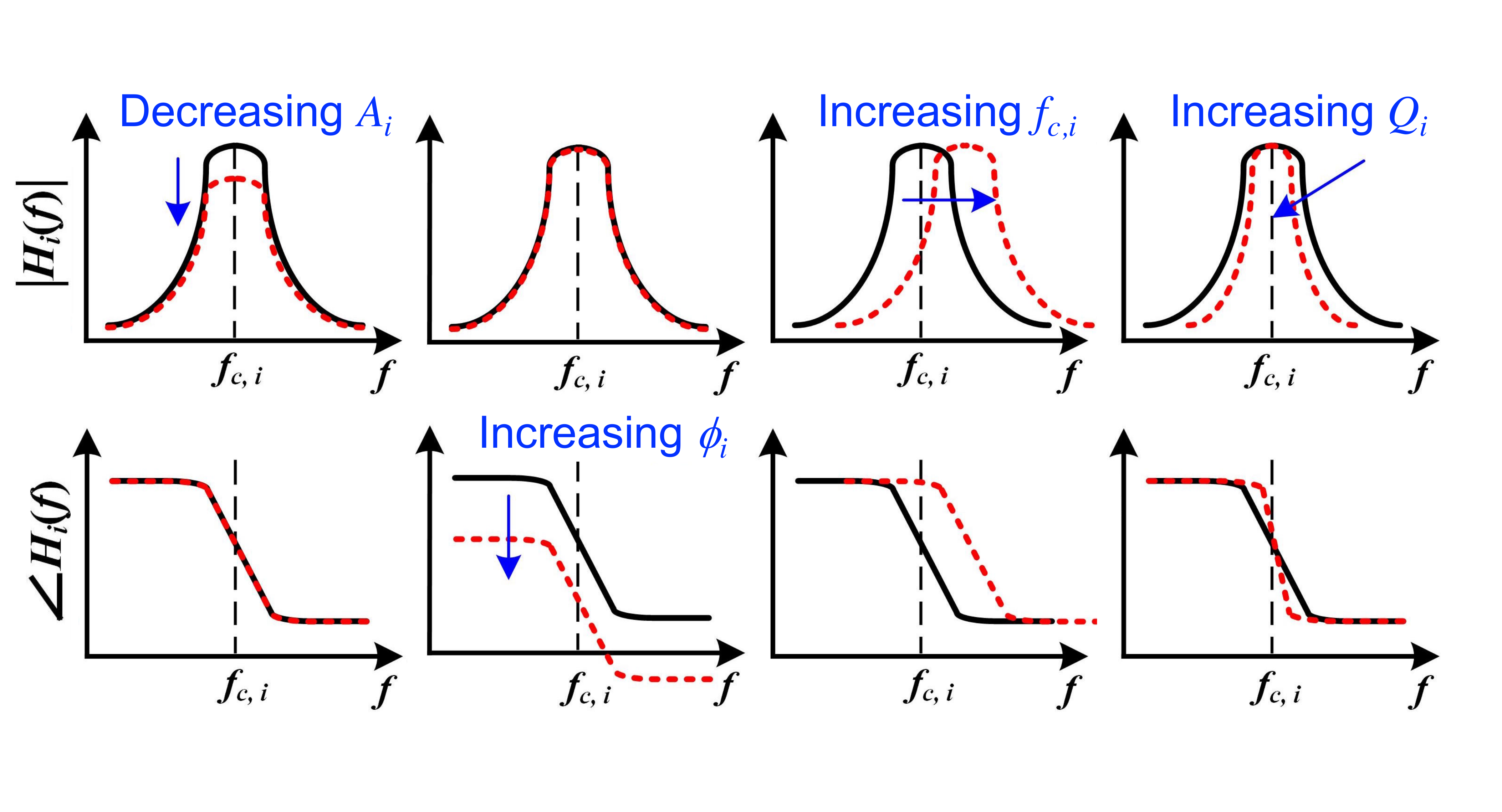}}
\vspace{-0.5\baselineskip}
\caption{(a) Block diagram of an FDE-based RF canceller with $\NumTap=2$ FDE taps, and (b) illustration of amplitude and phase responses of an ideal $2^{\textrm{nd}}$-order BPF with amplitude, phase, center frequency, and quality factor (i.e., group delay) controls.}
\label{fig:fde-concept}
\vspace{-\baselineskip}
\end{figure}

\subsubsection*{An FDE-based RF Canceller}
One compact design to achieve significantly enhanced performance and bandwidth of RF SIC is based on the technique of frequency-domain equalization (FDE) and was implemented in an RFIC~\cite{Zhou_WBSIC_JSSC15}. Fig.~\ref{fig:fde-concept}\subref{fig:fde-concept-diagram} shows the diagram of an FDE-based canceller, where parallel reconfigurable bandpass filters (BPFs) are used to emulate the antenna interface response across wide bandwidth. We denote the frequency response of a general FDE-based RF canceller consisting of $\NumTap$ FDE taps by
\begin{align}
\label{eq:fde-canc-tf}
\FDECancTF(f_k) & = \littlesum\limits_{i=1}^{\NumTap} \FDETapTF{i}(f_k),
\end{align}
where $\FDETapTF{i}(f_k)$ is the frequency response of the $i^{\textrm{th}}$ FDE tap containing a reconfigurable BPF with amplitude and phase controls. Theoretically, any $m^{\textrm{th}}$-order RF BPF ($m=1,2,\dots$) can be used. Fig.~\ref{fig:fde-concept}\subref{fig:fde-concept-bpf} illustrates the amplitude and phase of a $2^{\textrm{nd}}$-order BPF with different control parameters. For example, increased BPF quality factors result in ``sharper'' BPF amplitudes and increased group delay.
Since it is shown~\cite{ghaffari2011tunable,Zhou_WBSIC_JSSC15} that a $2^{\textrm{nd}}$-order BPF can accurately model the FDE $N$-path filter, the frequency response of an FDE-based RF\underline{I}C canceller with $\NumTap$ FDE taps is given by
\begin{align}
\label{eq:rfic-tf}
\ICTF(f_k) = \littlesum_{i=1}^{\NumTap} \ICTapTF{i}(f_k)
= \littlesum_{i=1}^{\NumTap} \frac{\ICTapAmp{i} \cdot e^{-j\ICTapPhase{i}}}{1 - j\ICTapQF{i} \cdot \left( \ICTapCF{i}/f_k-f_k/\ICTapCF{i}\right)}.
\end{align}
Within the $i^{\textrm{th}}$ FDE tap, $\ICTapTF{i}(f_k)$, $\ICTapAmp{i}$ and $\ICTapPhase{i}$ are the amplitude and phase controls, and $\ICTapCF{i}$ and $\ICTapQF{i}$ are the center frequency and quality factor of the $2^{\textrm{nd}}$-order BPF (see Fig.~\ref{fig:fde-concept}\subref{fig:fde-concept-bpf}). In the RFIC canceller, $\ICTapCF{i}$ and $\ICTapQF{i}$ are adjusted through a reconfigurable baseband capacitor and transconductors, respectively.

As Fig.~\ref{fig:fde-concept}\subref{fig:fde-concept-bpf} and {\eqref{eq:rfic-tf}} suggest, one FDE tap features four degrees of freedom (DoF) so that the antenna interface, $\AntTF(f_k)$, can be emulated \emph{not only in amplitude and phase, but also in the slope of amplitude and the slope of phase (i.e., group delay)}. Therefore, the RF SIC bandwidth can be significantly enhanced through FDE when compared with the amplitude- and phased-based RF cancellers.


\section{Design and Optimization}
\label{sec:impl}
In this section, we present our design and implementation of an FDE-based canceller using discrete components on a PCB (referred to as the \emph{PCB canceller}). Recall that the motivation is to facilitate integration with an SDR platform, the experimentation of FD at the link/network level, and integration with open-access wireless testbeds. We then present a realistic PCB canceller model, which is later validated (Section~\ref{sec:pcb-validation}) and used in the experimental and numerical evaluations (Sections~\ref{sec:exp} and~\ref{sec:sensitivity}).

\begin{figure}[!t]
\centering
\includegraphics[width=\columnwidth]{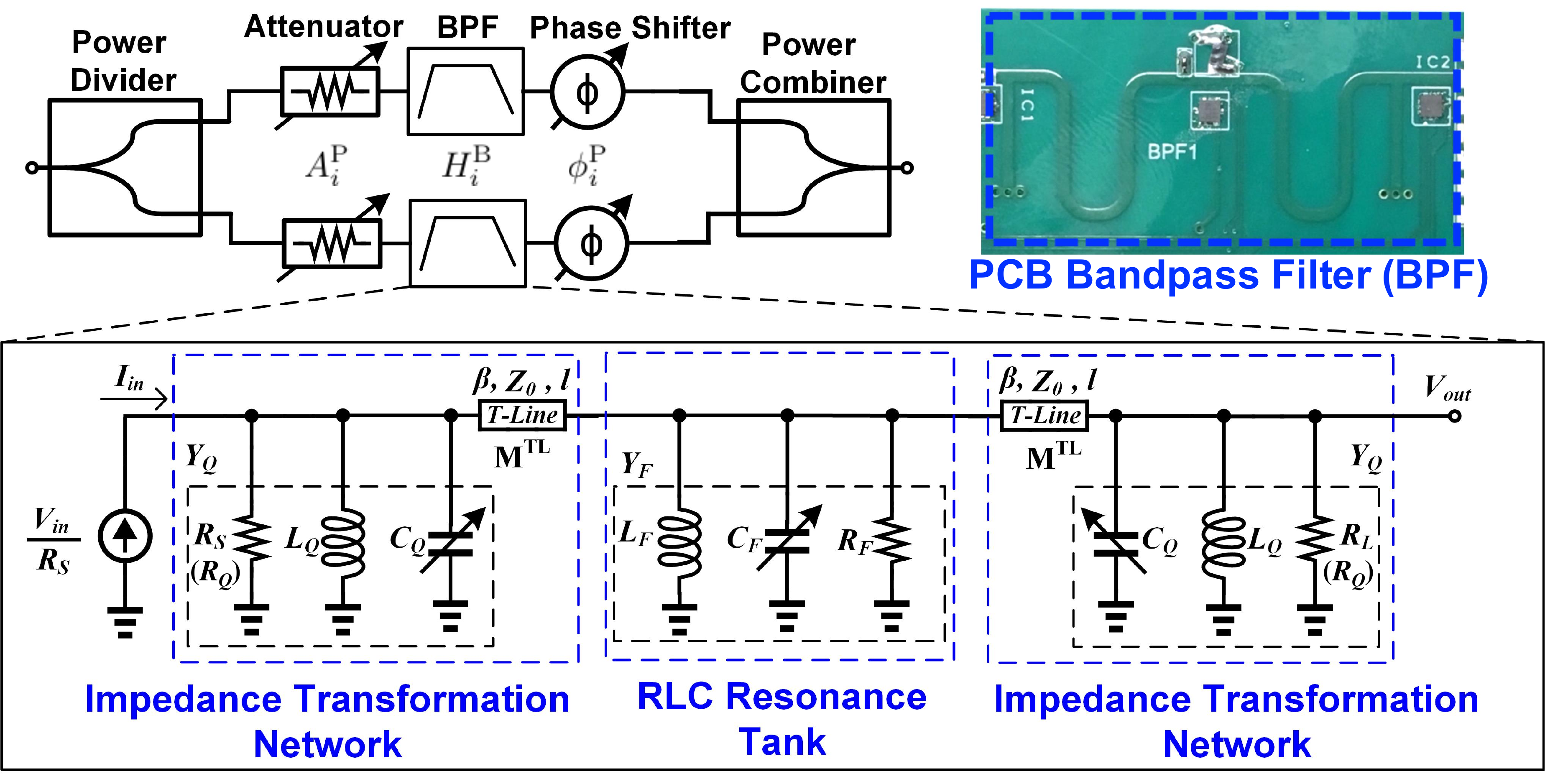}
%
\vspace{-\baselineskip}
\caption{Block diagram of the implemented $\NumTap=2$ FDE taps in the PCB canceller (see Fig.~\ref{fig:fde-concept}(a)), each of which consists of an RLC bandpass filter (BPF), an attenuator for amplitude control, and a phase shifter for phase control.}
\label{fig:diagram-pcb}
\vspace{-\baselineskip}
\end{figure}
\subsection{FDE PCB Canceller Implementation}
\label{ssec:impl-pcb}

Fig.~\ref{fig:intro}\subref{fig:intro-pcb} and Fig.~\ref{fig:fde-concept}\subref{fig:fde-concept-diagram} show the implementation and block diagram of the PCB canceller with 2 FDE taps. In particular, a reference signal is tapped from the TX input using a coupler and is split into two FDE taps through a power divider. Then, the signals after each FDE tap are combined and RF SIC is performed at the RX input. Each FDE tap consists of a reconfigurable $2^{\textrm{nd}}$-order BPF, as well as an attenuator and phase shifter for amplitude and phase controls. We refer to the BPF here as the \emph{PCB BPF} to distinguish from the one in the RFIC canceller {\eqref{eq:rfic-tf}}. The PCB BPF (with size of $\SI{1.5}{\cm}\times\SI{4}{\cm}$, see Fig.~\ref{fig:diagram-pcb}) is implemented as an RLC filter with impedance transformation networks and is optimized around $\SI{900}{MHz}$ operating frequency.\footnote{We select $\SI{900}{MHz}$ around the Region 2 $902$--$\SI{928}{MHz}$ ISM band as the operating frequency but the approach can be easily extended to other bands (e.g., $\SI{2.4}{GHz}$) with slight modification of the hardware design and proper choice of the frequency-dependent components.} When compared to the $N$-path filter used in the RFIC canceller~\cite{Zhou_WBSIC_JSSC15} that consumes certain amount of DC power, this discrete component-based passive RLC BPF has zero DC power consumption and can support higher TX power levels. Moreover, it has a lower noise level than the RFIC implementation.

The PCB BPF center frequency in the $i^{\textrm{th}}$ FDE tap can be adjusted through the capacitor, $\PCBTapCFCap{i}$, in the RLC resonance tank. In order to achieve a high and adjustable BPF quality factor, impedance transformation networks including transmission-lines (T-Lines) and digitally tunable capacitors, $\PCBTapQFCap{i}$, are introduced. In our implementation, $\PCBTapCFCap{i}$ consists of two parallel capacitors: a fixed $\SI{8.2}{pF}$ capacitor and a Peregrine Semiconductor PE64909 digitally tunable capacitor ($4$-bit) with a resolution of $\SI{0.12}{pF}$. For $\PCBTapQFCap{i}$, we use the Peregrine Semiconductor PE64102 digitally tunable capacitor ($5$-bit) with a resolution of $\SI{0.39}{pF}$. In addition, the programmable attenuator has a tuning range of $0$--$\SI{15.5}{dB}$ with a $\SI{0.5}{dB}$ resolution, and the passive phase shifter is controlled by a $8$-bit digital-to-analog converter (DAC) and covers full $\SI{360}{\degree}$ range.


\subsection{FDE PCB Canceller Model}
\label{ssec:impl-model}
Ideally, the PCB BPF has a $2^{\textrm{nd}}$-order frequency response from the RLC resonance tank. However, in practical implementation, its response deviates from that used in the FDE-based RFIC canceller {\eqref{eq:rfic-tf}}. Therefore, there is a need for a valid model tailored for evaluating the performance and optimized configuration of the PCB canceller. Based on the circuit diagram in Fig.~\ref{fig:diagram-pcb}, we derive a realistic model for the frequency response of the PCB BPF, $\BPFTapTF{i}(f_k)$, given by\footnote{The details can be found in Appendix~\ref{append:pcb-model}.}
\begin{align}
\label{eq:pcb-bpf-tf}
\BPFTapTF{i}(f_k) = & R_s^{-1} \Big[ j\sin(2\beta l) Z_0Y_{\textrm{F},i}(f_k)Y_{\textrm{Q},i}(f_k) \nonumber \\
& + \cos^2(\beta l)Y_{\textrm{F},i}(f_k) + 2\cos(2\beta l)Y_{\textrm{Q},i}(f_k) \nonumber \\
& + j\sin(2\beta l)/Z_0 + 0.5j\sin(2\beta l)Z_0(Y_{\textrm{Q},i}(f_k))^2 \nonumber \\
& - \sin^2(\beta l)Z_0^2Y_{\textrm{F},i}(f_k)(Y_{\textrm{Q},i}(f_k))^2  \Big]^{-1},
\end{align}
where $Y_{\textrm{F},i}(f_k)$ and $Y_{\textrm{Q},i}(f_k)$ are the admittance of the RLC resonance tank and impedance transformation networks, i.e., 
\begin{equation}
\left\{
\begin{aligned}
\label{eq:pcb-admittance}
Y_{\textrm{F},i}(f_k) & = 1/R_{\textrm{F}} + j2\pi \PCBTapCFCap{i} f_k + 1/(j2\pi L_{\textrm{F}} f_k), \\
Y_{\textrm{Q},i}(f_k) & = 1/R_{\textrm{Q}} + j2\pi \PCBTapQFCap{i} f_k + 1/(j2\pi L_{\textrm{Q}} f_k).
\end{aligned}
\right.
\end{equation}
In particular, to have perfect matching with the source and load impedance of the RLC resonance tank, $R_{\textrm{S}}$ and $R_{\textrm{L}}$ are set to be the same value of $R_\textrm{Q} = 50\Omega$ (see Fig.~\ref{fig:diagram-pcb}). $\beta$ and $Z_0$ are the wavenumber and characteristic impedance of the T-Line with length $l$ (see Fig.~\ref{fig:diagram-pcb}). In our implementation, $L_{\textrm{F}} = \SI{1.65}{nH}$, $L_{\textrm{Q}} = \SI{2.85}{nH}$, $\beta l \approx \SI{1.37}{rad}$, and $Z_0 = 50\Omega$.

In addition, other components in the PCB canceller (e.g., couplers and power divider/combiner) can introduce extra attenuation and group delay, due to implementation losses. Based on the S-Parameters of the components and measurements, we observed that the attenuation and group delay introduced, denoted by $\Amp{0}^{\textrm{P}}$ and $\tau_0^{\textrm{P}}$, are constant across frequency in the desired bandwidth. Hence, we empirically set $\Amp{0} = \SI{-4.1}{dB}$ and $\tau_0 = \SI{4.2}{ns}$. Recall that each FDE tap is also associated with amplitude and phase controls, $\PCBTapAmp{i}$ and $\PCBTapPhase{i}$, the PCB canceller with two FDE taps is modeled by
\begin{align}
\label{eq:pcb-tf-calibrated}
\PCBTF(f_k) & = \Amp{0}^{\textrm{P}} e^{-j2\pi f_k\tau_0^{\textrm{P}}} \left[ \littlesum_{i=1}^{2} \PCBTapAmp{i} e^{-j\PCBTapPhase{i}} \BPFTapTF{i}(f_k) \right],
\end{align}
where $\BPFTapTF{i}(f_k)$ is the PCB BPF model given by {\eqref{eq:pcb-bpf-tf}}. As a result, the $i^{\textrm{th}}$ FDE tap in the PCB canceller {\eqref{eq:pcb-tf-calibrated}} has configuration parameters $\{\PCBTapAmp{i}, \PCBTapPhase{i}, \PCBTapCFCap{i}, \PCBTapQFCap{i}\}$, featuring 4 DoF.

\subsection{Optimization of Canceller Configuration}
\label{ssec:impl-opt}
Based on {\OptProblem}, we now present a general FDE-based canceller configuration scheme that jointly optimizes all the FDE taps in the canceller.\footnote{The RFIC canceller presented in~\cite{Zhou_WBSIC_JSSC15} is configured based on heuristics. In Section~\ref{sec:sensitivity}, we show that the optimized configuration scheme can significantly improve the RFIC canceller performance.} Although our implemented PCB canceller has only 2 FDE taps, both its model and the configuration scheme can be easily extended to the case with a larger number of FDE taps, as described in Section~\ref{sec:sensitivity}.

The inputs to the FDE-based canceller configuration scheme are: (i) the PCB canceller model {\eqref{eq:pcb-tf-calibrated}} with given number of FDE taps, $\NumTap$, (ii) the antenna interface response, $\AntTF(f_k)$, and (iii) the desired RF SIC bandwidth, $f_k \in [f_1,f_\NumChnl]$. 
Then, the optimized canceller configuration is obtained by solving {\OptProblemPCB}.
\begin{align*}
\OptProblemPCB\ & \min: \littlesum_{k=1}^{\NumChnl} \NormTwo{\PCBResTF(f_k)} = \littlesum_{k=1}^{\NumChnl} \NormTwo{ \AntTF(f_k) - \PCBTF(f_k) }^2 \\
\textrm{s.t.:}\ & \PCBTapAmp{i} \in [\PCBTapAmpMin, \PCBTapAmpMax],\ \PCBTapPhase{i} \in [-\pi, \pi], \\
& \PCBTapCFCap{i} \in [\PCBTapCFCapMin, \PCBTapCFCapMax],\ \PCBTapQFCap{i} \in [\PCBTapQFCapMin, \PCBTapQFCapMax],\ \forall i.
\end{align*}

Note that {\OptProblemPCB} is challenging to solve due to its non-convexity and non-linearity, as opposed to the linear program used in the delay line-based RF canceller~\cite{bharadia2013full}. This is due to the specific forms of the configuration parameters in {\eqref{eq:pcb-tf-calibrated}} such as (i) the higher-order terms introduced by $f_k$, and (ii) the trigonometric term introduced by the phase control, $\PCBTapPhase{i}$. In addition, the antenna interface response, $\AntTF(f_k)$, is also frequency-selective and time-varying.

In general, it is difficult to maintain analytical tractability of {\OptProblemPCB} (i.e., to obtain its optimal solution in closed-form). However, in practice, it is unnecessary to obtain the global optimum to {\OptProblemPCB} as long as the performance achieved by the obtained local optimum is sufficient (e.g., at least $\SI{45}{dB}$ RF SIC is achieved). In this work, the local optimal solution to (P2) is obtained using a MATLAB solver.
The detailed implementation and performance of the optimized canceller configuration are described in Section~\ref{ssec:exp-node}.



\section{Model Validation}
\label{sec:pcb-validation}


\begin{table}[!t]
\caption{Four $(\PCBTapCFCapNoidx, \PCBTapQFCapNoidx)$ configurations used in the validations.}
\label{table:set-cap}
\vspace{-0.5\baselineskip}
\scriptsize
\begin{center}
\renewcommand{\arraystretch}{1.5}
\begin{tabular}{|c|c|c|}
\hline
& Highest Q-Factor & Lowest Q-Factor \\
\hline
Highest Center Freq. & Set 1: $(\PCBTapCFCapMin, \PCBTapQFCapMin)$ & Set 3: $(\PCBTapCFCapMin, \PCBTapQFCapMax)$ \\
\hline
Lowest Center Freq. & Set 2: $(\PCBTapCFCapMax, \PCBTapQFCapMin)$ & Set 4: $(\PCBTapCFCapMax, \PCBTapQFCapMax)$ \\
\hline
\end{tabular}
\renewcommand{\arraystretch}{1}
\end{center}
\vspace{-0.5\baselineskip}
\end{table}

\subsubsection*{Validation of the PCB BPF}
We first experimentally validate the PCB BPF model, $\BPFTapTF{i}(f_k)$, given by {\eqref{eq:pcb-bpf-tf}}. The ground truth is obtained by measuring the frequency response (using S-Parameters measurements) of the PCB BPF using a test structure, which contains only the BPF, to avoid the effects of other components on the PCB. The measurements are conducted with varying $(\PCBTapCFCapNoidx, \PCBTapQFCapNoidx)$ configurations and the result of each configuration is averaged over $20$ measurement instances.\footnote{We drop the subscript $i$, since both PCB BPFs behave identically.} The BPF center frequency is measured as the frequency with the highest BPF amplitude, and the BPF quality factor is computed as the ratio between the center frequency and the $\SI{3}{dB}$ bandwidth around the center frequency.

The PCB BPF has a \emph{fixed} quality factor of 2.7, achieved by using only the RLC resonance tank. By setting $\PCBTapQFCapNoidx = \PCBTapQFCapMax$ and $\PCBTapQFCapNoidx = \PCBTapQFCapMin$ (see Section~\ref{ssec:impl-pcb}), the measured lowest and highest achievable BPF quality factors are 9.2 and 17.8, respectively. This shows an improvement in the PCB BPF quality factor tuning range of 3.4$\times$--6.6$\times$, achieved by introducing the impedance transformation networks. Similarly, by setting $\PCBTapCFCapNoidx = \PCBTapCFCapMax$ and $\PCBTapCFCapNoidx = \PCBTapCFCapMin$, the PCB BPF has a center frequency tuning range of $\SI{18}{MHz}$.

\begin{figure}[!t]
\centering
\vspace{-\baselineskip}
\subfloat{
\label{fig:eval-pcb-bpf-amp}
\includegraphics[width=0.47\columnwidth]{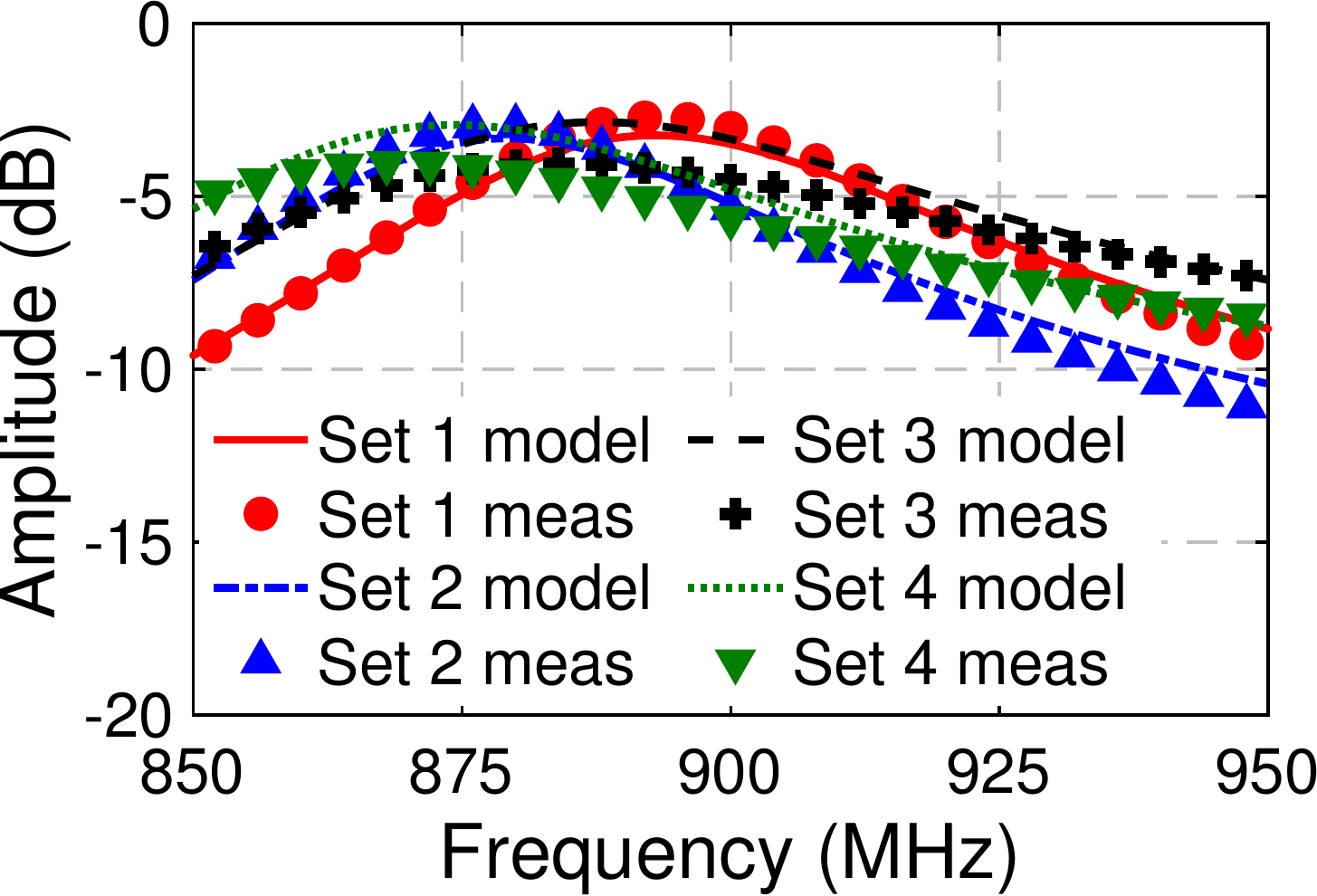}
}
\hfill
\subfloat{
\label{fig:eval-pcb-bpf-phase}
\includegraphics[width=0.47\columnwidth]{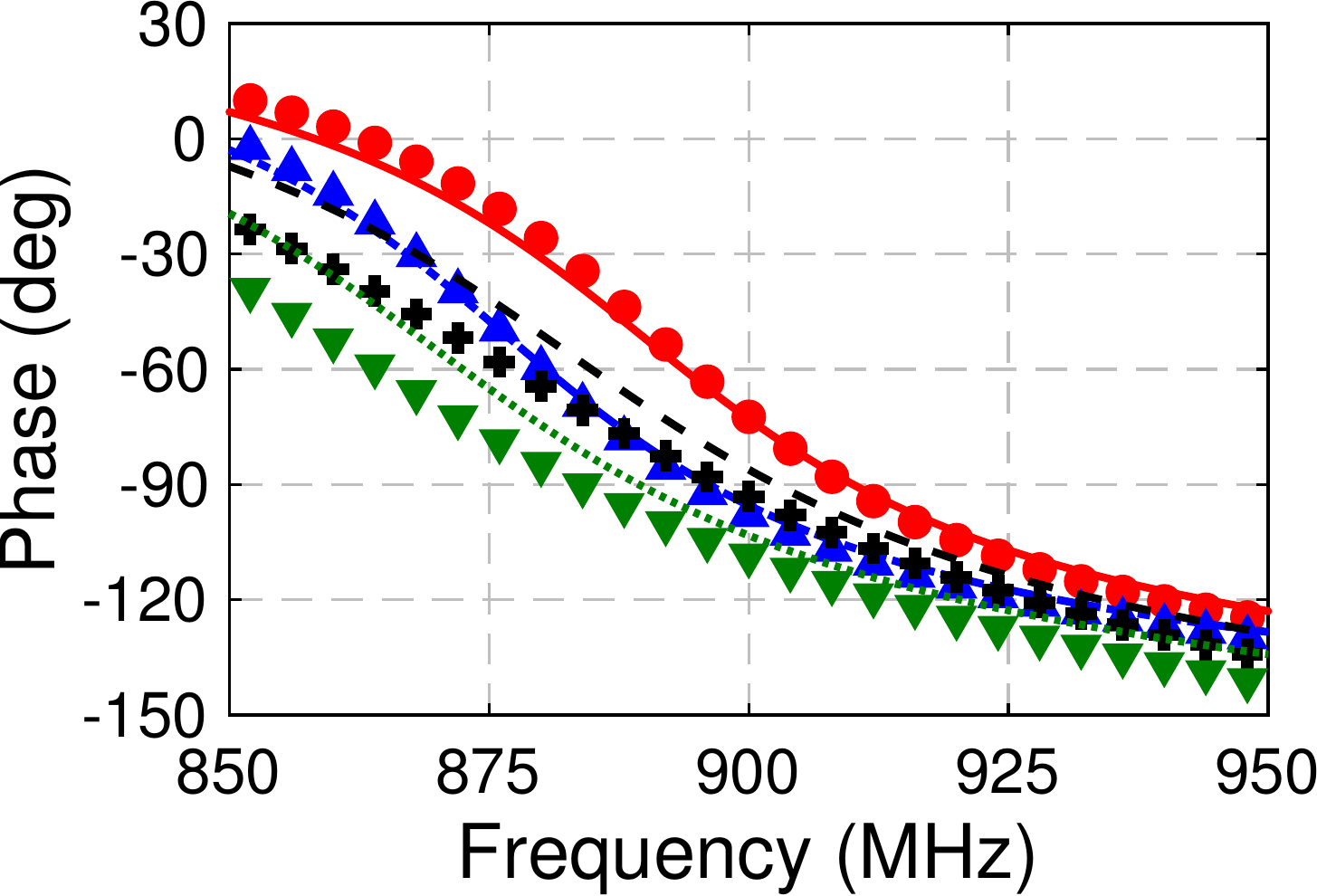}
}
\vspace{-0.5\baselineskip}
\caption{Modeled and measured amplitude and phase responses of the implemented PCB BPF under different $(\PCBTapCFCapNoidx, \PCBTapQFCapNoidx)$ configurations indicated in Table~\ref{table:set-cap}.}
\label{fig:eval-pcb-bpf}
\vspace{-\baselineskip}
\end{figure}

Fig.~\ref{fig:eval-pcb-bpf} presents the modeled and measured amplitude and phase responses of the PCB BPF with 4 $(\PCBTapCFCapNoidx, \PCBTapQFCapNoidx)$ configurations (see Table~\ref{table:set-cap}) which cover the entire tuning range of the BPF center frequency and quality factor. The results show that the PCB BPF model {\eqref{eq:pcb-bpf-tf}} matches very closely with the measurements at the highest BPF quality factor value (Sets 1 and 2). In particular, the maximum differences between the measured and modeled amplitude and phase are $\SI{0.5}{dB}$ and $\SI{7}{\degree}$, respectively. At the lowest BPF quality factor value (Sets 3 and 4), the differences are $\SI{1.2}{dB}$ and $\SI{15}{\degree}$, thereby showing the validity of the PCB BPF model. The same level of accuracy of the PCB BPF model {\eqref{eq:pcb-bpf-tf}} is also observed for other $(\PCBTapCFCapNoidx, \PCBTapQFCapNoidx)$ configurations within their tuning ranges.


\subsubsection*{Validation of the PCB Canceller}
We use the same experiments as in the PCB BPF validation to validate the PCB canceller model with 2 FDE taps, $\PCBTF(f_k)$, given by {\eqref{eq:pcb-tf-calibrated}}. We consider two cases for controlled measurements: (i) only one FDE tap is active, and (ii) both FDE taps are active. Note that the programmable attenuators only have a maximal attenuation of only $\SI{15.5}{dB}$ (see Section~\ref{ssec:impl-pcb}) and at this maximal attenuation, signals can still leak through the FDE tap, resulting in inseparable behaviors between the two FDE taps.

\begin{figure}[!t]
\centering
\vspace{-\baselineskip}
\subfloat{
\label{fig:eval-pcb-amp}
\includegraphics[width=0.47\columnwidth]{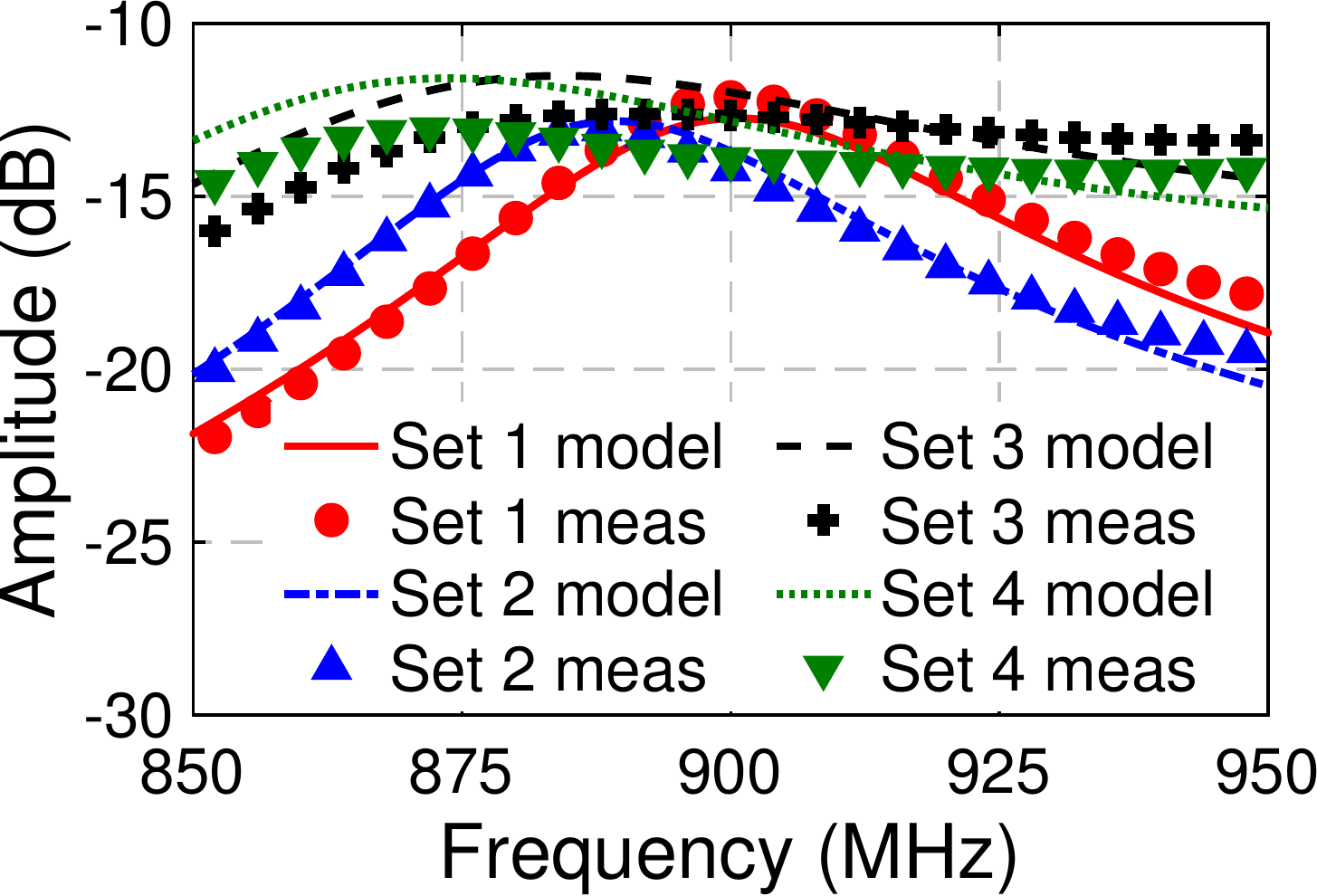}
}
\hfill
\subfloat{
\label{fig:eval-pcb-phase}
\includegraphics[width=0.47\columnwidth]{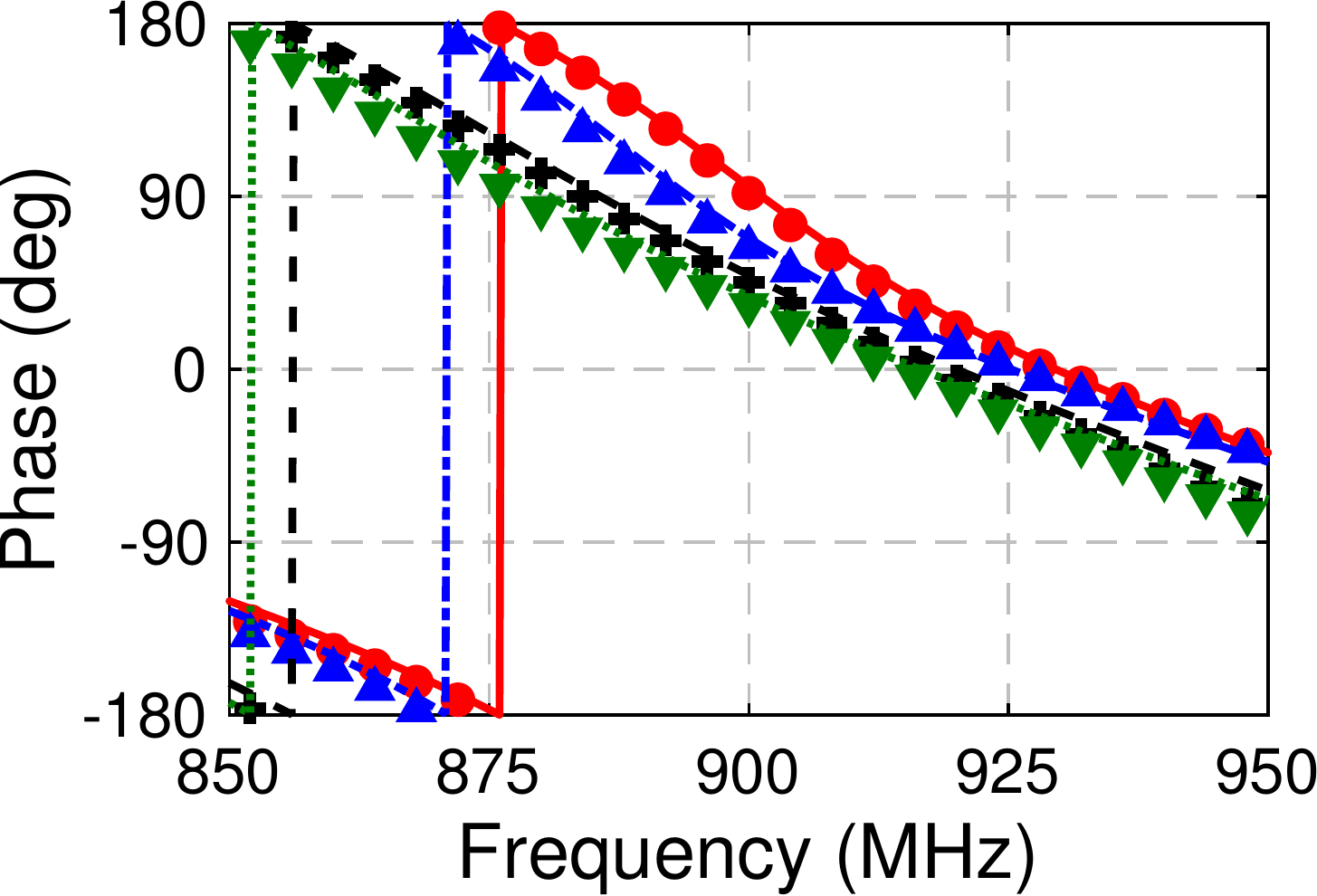}
}
\vspace{-0.5\baselineskip}
\caption{Modeled and measured amplitude and phase responses of the PCB canceller, where only the first FDE tap is active, under different $(\PCBTapCFCapNoidx, \PCBTapQFCapNoidx)$ configurations indicated in Table~\ref{table:set-cap}.}
\label{fig:eval-pcb}
\vspace{-\baselineskip}
\end{figure}

To minimize the effect of the second FDE tap, we set the first FDE tap at its highest amplitude (i.e., lowest attenuation value of $\PCBTapAmp{1}$) with varying values of $(\PCBTapCFCap{1}, \PCBTapQFCap{1})$ while setting the second FDE tap with the lowest amplitude (i.e., highest attenuation value of $\PCBTapAmp{2}$). Fig.~\ref{fig:eval-pcb} shows the modeled and measured amplitude and phase responses of the PCB canceller in this case, i.e., only the first FDE tap is active. At the highest BPF quality factor value (Sets 1 and 2), the maximum differences between the modeled and measured amplitude and phase are $\SI{0.9}{dB}$ and $\SI{8}{\degree}$, respectively. At the lowest BPF quality factor value (Sets 3 and 4), the errors are $\SI{1.5}{dB}$ and $\SI{12}{\degree}$, while still validating the PCB canceller model. We obtain similar results in the case where only the second FDE tap is active by setting highest attenuation value of $\PCBTapAmp{1}$ and lowest attenuation value of $\PCBTapAmp{2}$. The measurements are repeated with different $\{\PCBTapAmp{i}, \PCBTapPhase{i}, \PCBTapCFCap{i}, \PCBTapQFCap{i}\}$ settings for $i=1,2$, and all the results demonstrate the same level of accuracy of the PCB canceller model {\eqref{eq:pcb-tf-calibrated}}.

\section{Experimental Evaluation}
\label{sec:exp}
In this section, we discuss the integration of the PCB canceller described in Section~\ref{sec:impl} with an SDR testbed. Then, we present extensive experimental evaluation of the FDE-based FD radio at the node, link, and network levels.

\subsection{Implementation and Testbed}
\label{ssec:exp-testbed}

\subsubsection*{FDE-based FD Radio and the SDR Testbed}
Figs.~\ref{fig:intro}\subref{fig:intro-fd-radio} and~\ref{fig:intro}\subref{fig:intro-fd-net} depict our FDE-based FD radio design (whose diagram is shown in Fig.~\ref{fig:diagram}) and the SDR testbed. A $698$--$\SI{960}{MHz}$ swivel blade antenna and a coaxial circulator with operating frequency range $860$--$\SI{960}{MHz}$ are used as the antenna interface. We use the NI USRP-2942 SDR with the SBX-120 daughterboard operating at $\SI{900}{MHz}$ carrier frequency, which is the same as the operating frequency of the PCB canceller. As mentioned in Section~\ref{ssec:impl-pcb}, our PCB canceller design can be easily extended to other operating frequencies. and the antenna interface. The USRP has a measured noise floor of $-\SI{85}{dBm}$ at a fixed receiver gain setting.\footnote{This USRP receiver noise floor is limited by the environmental interference at around $\SI{900}{MHz}$. The USRP has a true noise floor of around $-\SI{95}{dBm}$ at the same receiver gain setting, when not connected to an antenna.}

We implemented a full OFDM-based PHY layer using NI LabVIEW on a host PC.\footnote{We consider a general OFDM-based PHY but do not consider the specific frame/packet structure defined by the standards (e.g., LTE or WiFi PHY).} A real-time RF bandwidth of $\BW=\SI{20}{MHz}$ is used through our experiments. The baseband complex (IQ) samples are streamed between the USRP and the host PC through a high-speed PCI-Express interface. The OFDM symbol size is 64 samples (subcarriers) with a cyclic-prefix ratio of 0.25 (16 samples). Throughout the evaluation, $\{f_k\}_{k=1}^{K=52}$ is used to represent the center frequency of the 52 non-zero subcarriers. The OFDM PHY layer supports various modulation and coding schemes (MCSs) with constellations from BPSK to 64QAM and coding rates of 1/2, 2/3, and 3/4, resulting in a highest (HD) data rate of $\SI{54}{Mbps}$.
The digital SIC algorithm with a highest non-linearity order of 7 is also implemented in LabVIEW to further suppress the residual SI signal after RF SIC.\footnote{The digital SIC algorithm is based on Volterra series and a least-square problem, which is similar to that presented in~\cite{bharadia2013full}. We omit the details here due to limited space.}

In total, our testbed consists of 3 FDE-based FD radios, whose performance is experimentally evaluated at the node, link, and network levels. Regular USRPs (without the PCB canceller) are also included in scenarios where additional HD users are needed.

\subsubsection*{Optimized PCB Canceller Configuration}
The optimized PCB canceller configuration scheme is implemented on the host PC and the canceller is configured by a SUB-20 controller through the USB interface. For computational efficiency, the PCB canceller response {\eqref{eq:pcb-tf-calibrated}} (which is validated in Section~\ref{sec:pcb-validation} and is independent of the environment) is pre-computed and stored. The detailed steps of the canceller configuration are as follows.
\begin{enumerate}[leftmargin=*,topsep=3pt]
\item[1.]
Measure the real-time antenna interface response, $\AntTF(f_k)$, using a preamble (2 OFDM symbols) by dividing the received preamble by the known transmitted preamble in the frequency domain;
\item[2.]
Solve for an initial PCB canceller configuration using optimization {\OptProblemPCB} based on the measured $\AntTF(f_k)$ and the canceller model {\eqref{eq:pcb-tf-calibrated}} (see Section~\ref{ssec:impl-opt}). The returned configuration parameters are rounded to their closest possible values based on hardware resolutions (see Section~\ref{ssec:impl-pcb});
\item[3.]
Perform a finer-grained local search and record the optimal canceller configuration (usually \texttt{\char`\~}10 iterations).
\end{enumerate}
In our design, the optimized PCB canceller configuration can be obtained in less than $\SI{10}{ms}$ on a regular PC with quad-core Intel i7 CPU via a non-optimized MATLAB solver.\footnote{Assuming that the canceller needs to be configured once per second, this is only a $1\%$ overhead. We note that a C-based optimization solver and/or an implementation based on FPGA/look-up table can significantly improve the performance of the canceller configuration and is left for future work.}


\subsection{Node-Level: Microbenchmarks}
\label{ssec:exp-node}

\subsubsection*{Optimized PCB Canceller Response and RF SIC}

\begin{figure}[!t]
\centering
\vspace{-\baselineskip}
\subfloat[]{
\label{fig:exp-usrp-algo-iq}
\includegraphics[width=0.47\columnwidth]{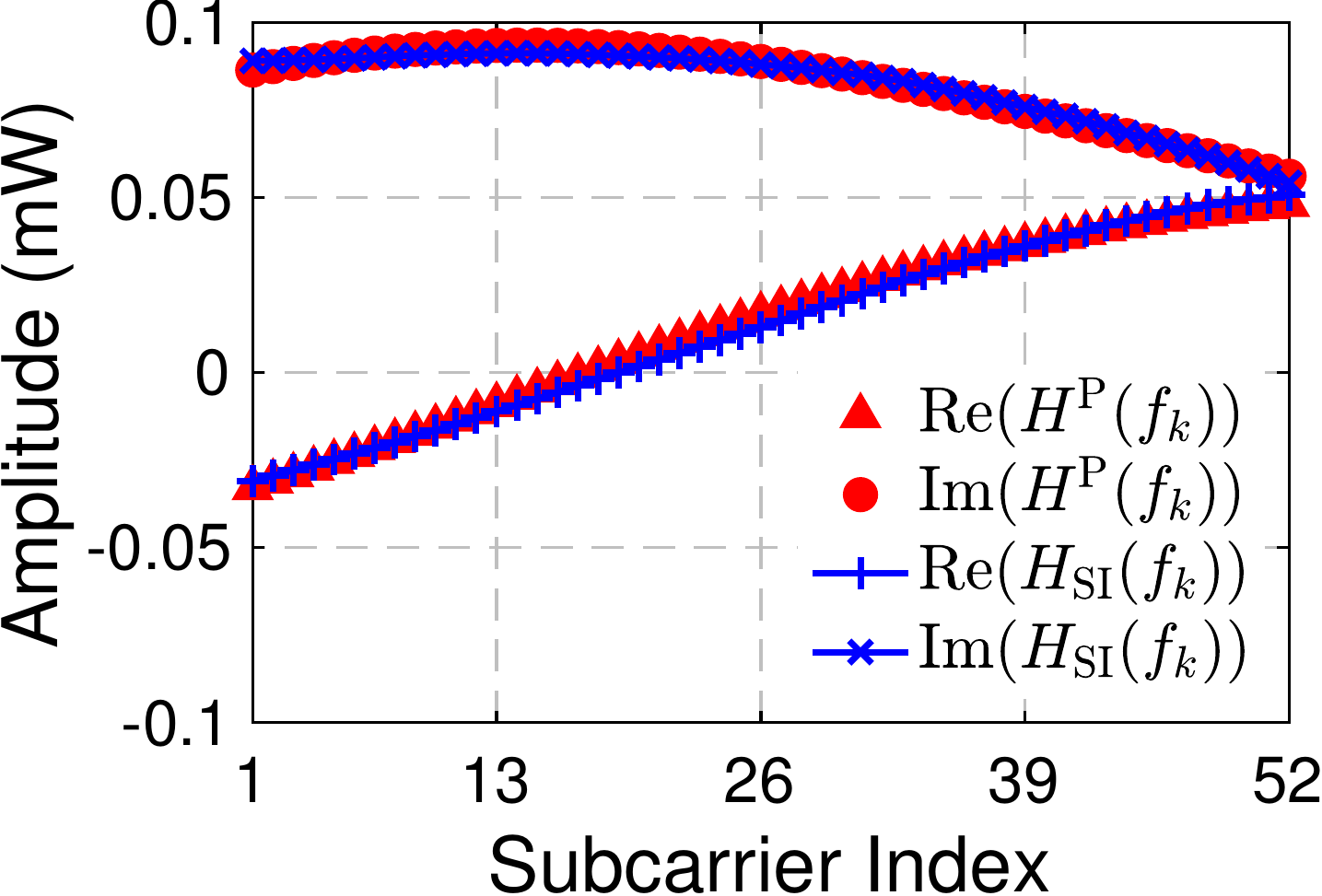}
}
\hfill
\subfloat[]{
\label{fig:exp-usrp-algo-res}
\includegraphics[width=0.47\columnwidth]{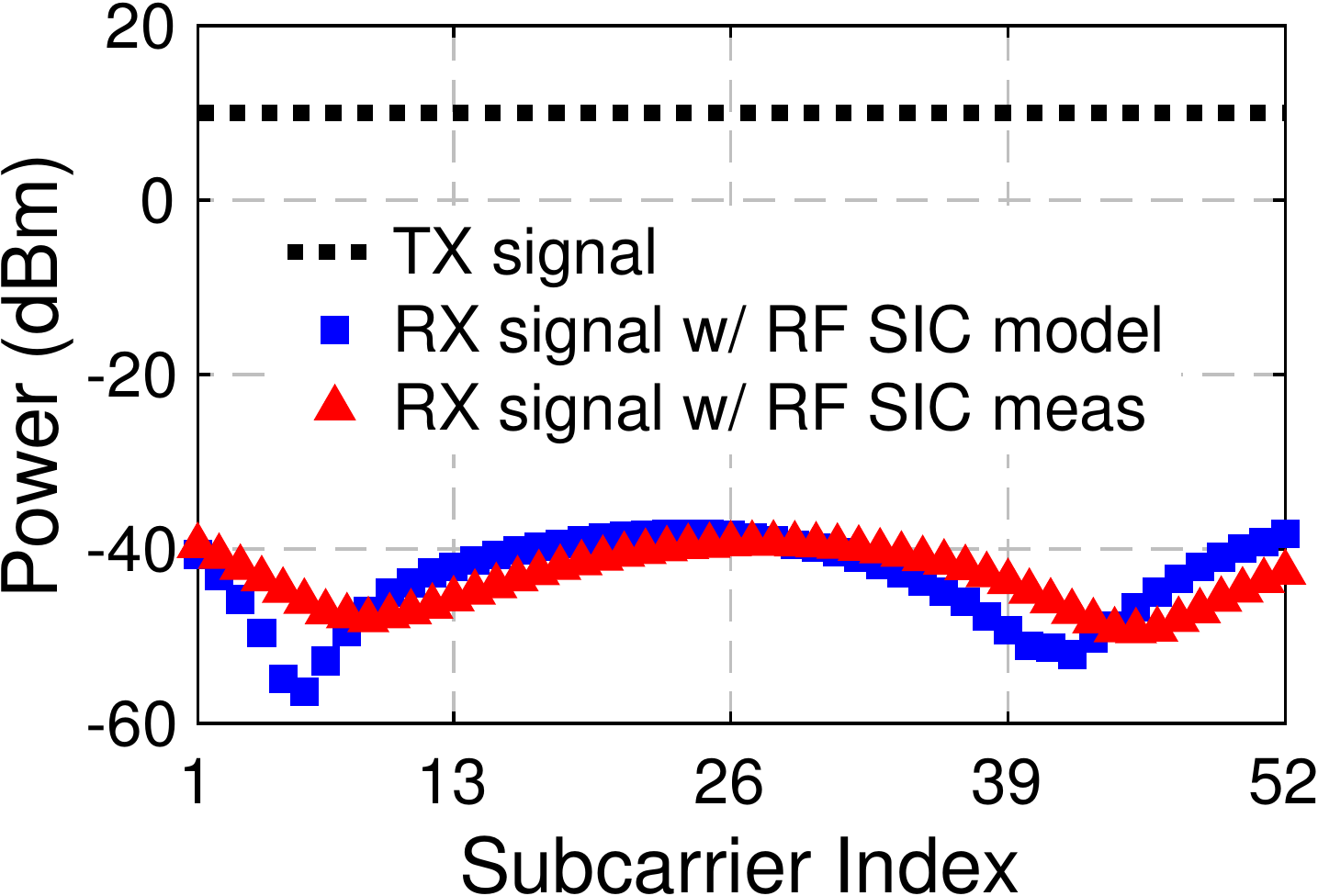}
}
\vspace{-0.5\baselineskip}
\caption{(a) Real and imaginary parts of the optimized PCB canceller response, $\PCBTF(f_k)$, vs. real-time SI channel measurements, $\AntTF(f_k)$, and (b) modeled and measured RX signal power after RF SIC at $\SI{10}{dBm}$ TX power. An average $\SI{52}{dB}$ RF SIC across $\SI{20}{MHz}$ is achieved in the experiments.}
\label{fig:exp-usrp-algo}
\vspace{-\baselineskip}
\end{figure}

We set up an FDE-based FD radio running the optimized PCB canceller configuration scheme and record the canceller configuration, measured $\AntTF(f_k)$, and measured residual SI power after RF SIC. The recorded canceller configuration is then used to compute the PCB canceller response using {\eqref{eq:pcb-tf-calibrated}}.

Fig.~\ref{fig:exp-usrp-algo}\subref{fig:exp-usrp-algo-iq} shows an example of the optimized PCB canceller response, $\PCBTF(f_k)$, and the measured antenna interface response, $\AntTF(f_k)$, in real and imaginary parts (or I and Q). It can be seen that $\PCBTF(f_k)$ closely matches with $\AntTF(f_k)$ with maximal amplitude and phase differences of only $\SI{0.5}{dB}$ and $\SI{2.5}{\degree}$, respectively. This confirms the accuracy of the PCB canceller model and the performance of the optimized canceller configuration. Fig.~\ref{fig:exp-usrp-algo}\subref{fig:exp-usrp-algo-res} shows the modeled (computed by subtracting the modeled canceller response from the measured $\AntTF(f_k)$) and measured RX signal power after RF SIC at $\SI{10}{dBm}$ TX power. The results show that the FDE-based FD radio achieves an average $\SI{52}{dB}$ RF SIC across $\SI{20}{MHz}$ bandwidth, from which $\SI{20}{dB}$ is obtained from the antenna interface isolation. Similar performance is observed in various experiments throughout the experimental evaluation.

\subsubsection*{Overall SIC}

\begin{figure}[!t]
\centering
\includegraphics[width=0.85\columnwidth]{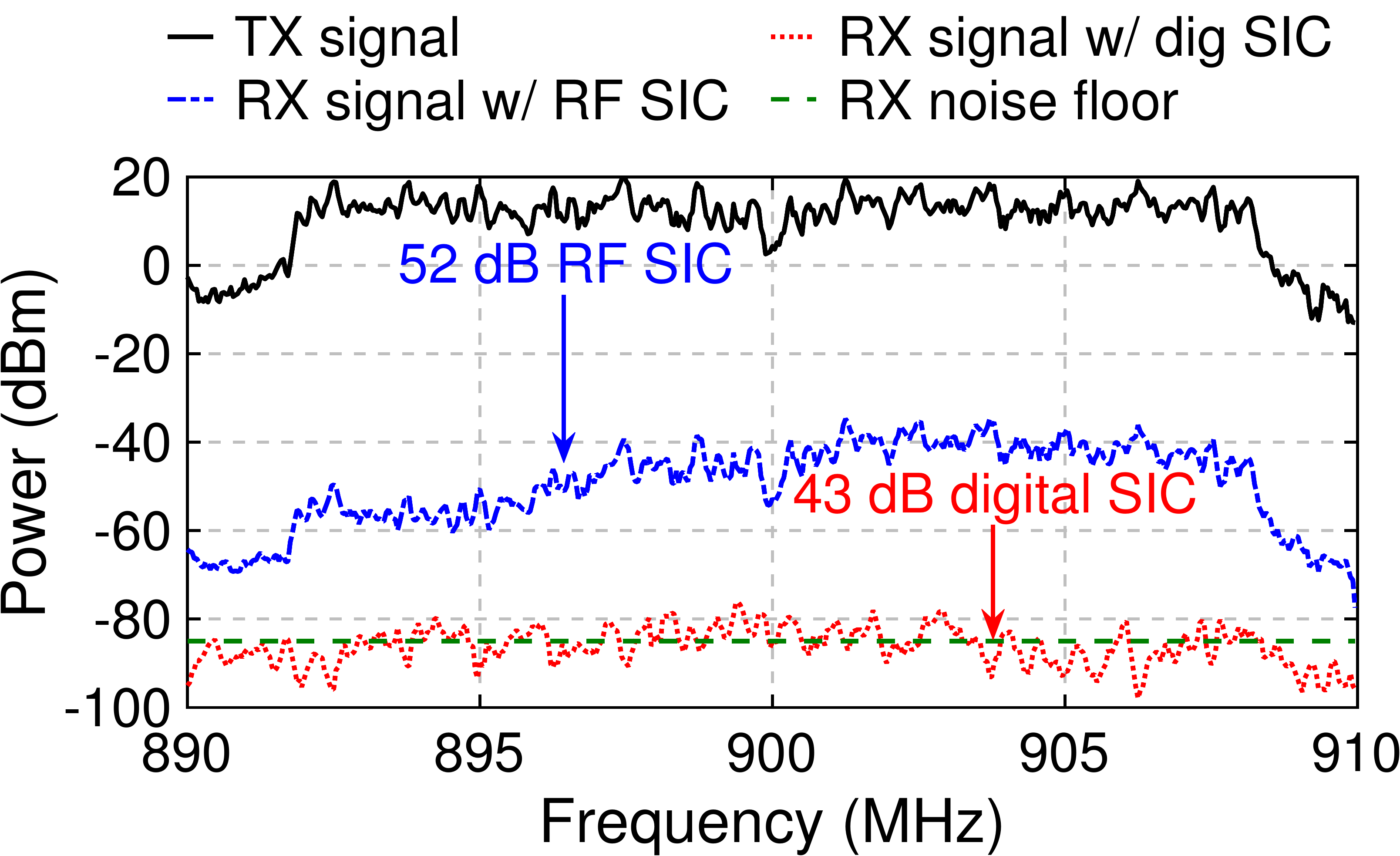}
\vspace{-0.5\baselineskip}
\caption{Power spectrum of the received signal after SIC in the RF and digital domains with $\SI{10}{dBm}$ average TX power, $\SI{20}{MHz}$ bandwidth, and $-\SI{85}{dBm}$ receiver noise floor.}
\label{fig:eval-usrp-spec-20mhz}
\vspace{-\baselineskip}
\end{figure}

We measure the overall SIC achieved by the FDE-based FD radio including the digital SIC in the same setting as described above, and the results are presented in Fig.~\ref{fig:eval-usrp-spec-20mhz}. It can be seen that the FDE-based FD radio achieves an average $\SI{95}{dB}$ overall SIC across $\SI{20}{MHz}$, from which $\SI{52}{dB}$ and $\SI{43}{dB}$ are obtained in the RF and digital domains, respectively. Recall from Section~\ref{ssec:exp-testbed} that the USRP has noise floor of $\SI{-85}{dBm}$, the FDE-based RF radio supports a maximal average TX power of $\SI{10}{dBm}$ (where the peak TX power can go as high as $\SI{20}{dBm}$). We use TX power levels lower than or equal to $\SI{10}{dBm}$ throughout the experiments, where all the SI can be canceled to below the RX noise floor.

\subsection{Link-Level: SNR-PRR Relationship}
\label{ssec:exp-snr-prr-relationship}
We now evaluate the relationship between link SNR and link packet reception ratio (PRR). We setup up a link with two FDE-based FD radios at a fixed distance of 5 meters with equal TX power. In order to evaluate the performance of our FD radios with the existence of the PCB canceller, we set an FD radio to operate in HD mode by turning on only its transmitter or receiver. We conduct the following experiment for each of the $12$ MCSs in both FD and HD modes, with varying TX power levels. In particular, the packets are sent over the link simultaneously in FD mode or in alternating directions in HD mode (i.e., the two radios take turns and transmit to each other). In each experiment, both radios send a sequence of 50 OFDM streams, each OFDM stream contains 20 OFDM packets, and each OFDM packet is 800-Byte long.

We consider two metrics. The \emph{HD (resp.\ FD) link SNR} is measured as the ratio between the average RX signal power in both directions and the RX noise floor when both radios operate in HD (resp. FD) mode. The \emph{HD (resp.\ FD) link PRR} is computed as the fraction of packets successfully sent over the HD (resp. FD) link in each experiment. We observe from the experiments that the HD and FD link SNR and PRR values in both link directions are similar. Similar experiments and results were presented in~\cite{zhou2016basic} for HD links.

\begin{figure}[!t]
\centering
\vspace{-\baselineskip}
\subfloat[Code rate 1/2]{
\label{fig:exp-prr-vs-snr-1-2}
\includegraphics[width=0.47\columnwidth]{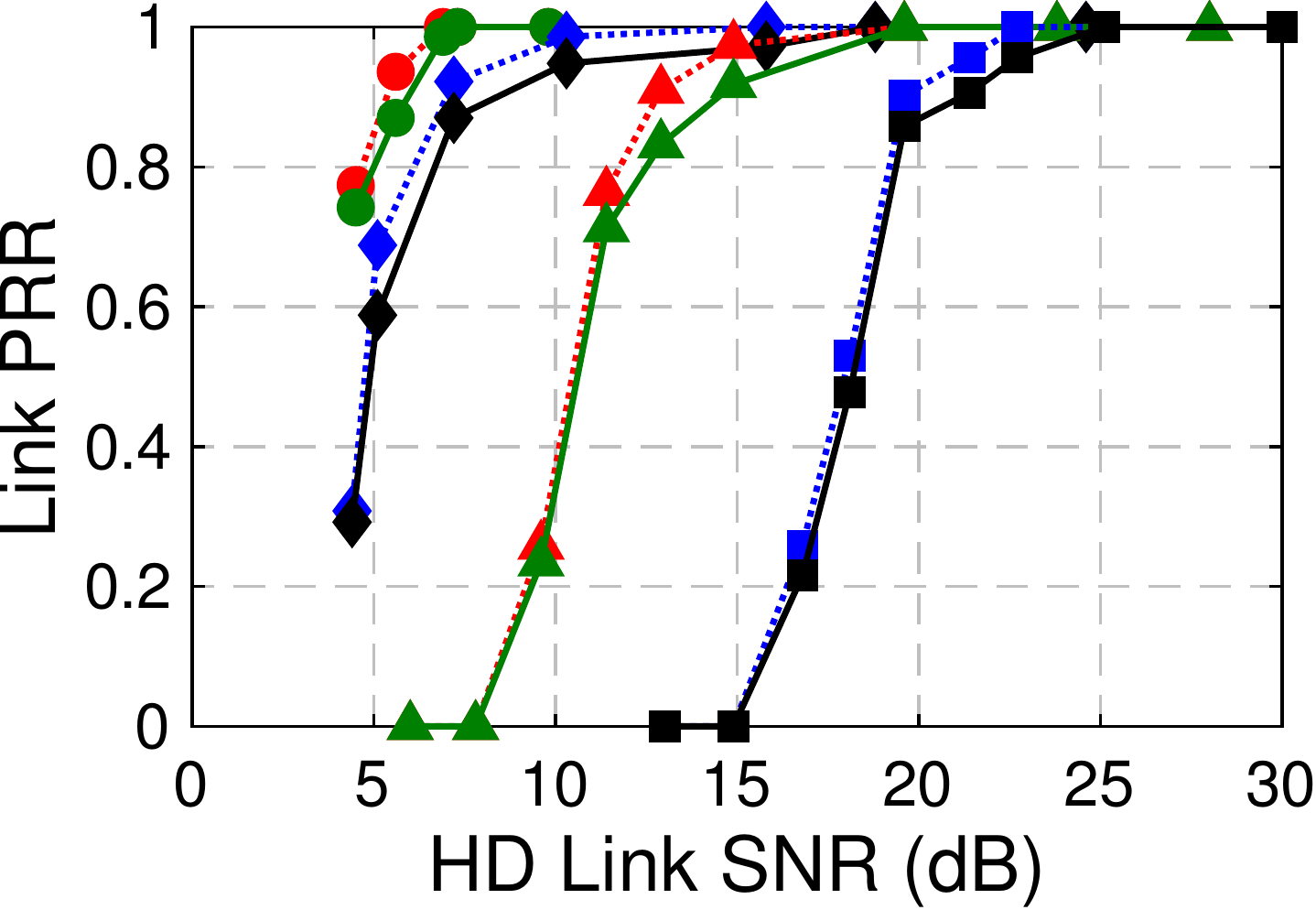}
}
\hfill
\subfloat[Code rate 3/4]{
\label{fig:exp-prr-vs-snr-3-4}
\includegraphics[width=0.47\columnwidth]{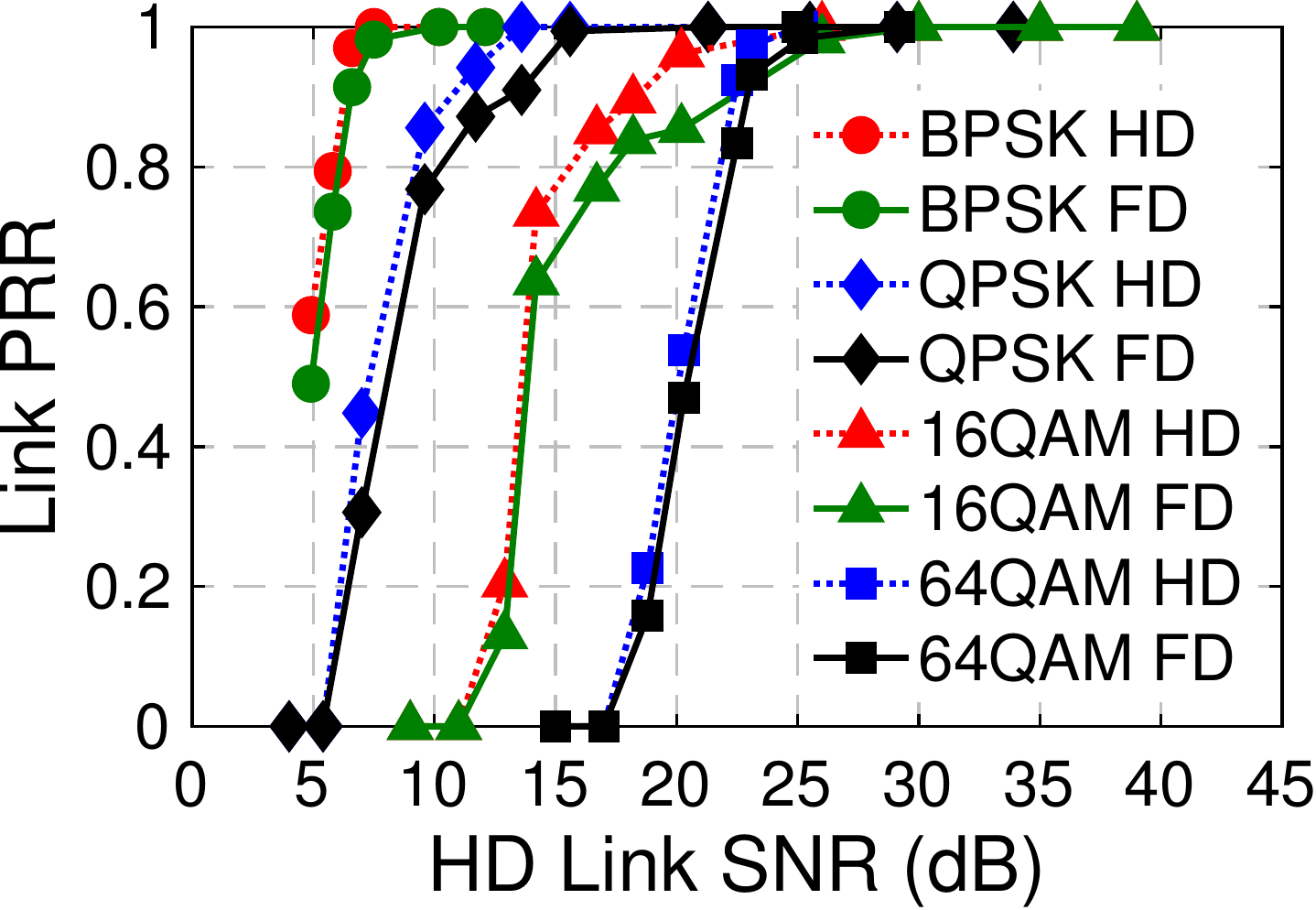}
}
\vspace{-0.5\baselineskip}
\caption{HD and FD link packet reception ratio (PRR) with varying HD link SNR and modulation and coding schemes (MCSs).}
\label{fig:exp-prr-vs-snr}
\vspace{-\baselineskip}
\end{figure}

Fig.~\ref{fig:exp-prr-vs-snr} shows the relationship between link PRR values and HD link SNR values with varying MCSs. The results show that with sufficient link SNR values (e.g., $\SI{8}{dB}$ for BPSK-1/2 and $\SI{28}{dB}$ for 64QAM-3/4), the FDE-based FD radio achieves a link PRR of $100\%$. With insufficient link SNR values, the average FD link PRR is $6.5\%$ lower than the HD link PRR across varying MCSs. This degradation is caused by the link SNR difference when the radios operate in HD or FD mode, which is described later in Section~\ref{ssec:exp-link}. Since packets are sent simultaneously in both directions on an FD link, this average PRR degradation is equivalent to an average FD link throughput gain of $1.87\times$ under the same MCS.

\subsection{Link-Level: SNR Difference and FD Gain}
\label{ssec:exp-link}

\begin{figure}[!t]
\centering
\vspace{-0.5\baselineskip}
\subfloat[LOS deployment and an FD radio in a hallway]{
\label{fig:exp-map-los}
\includegraphics[width=\columnwidth]{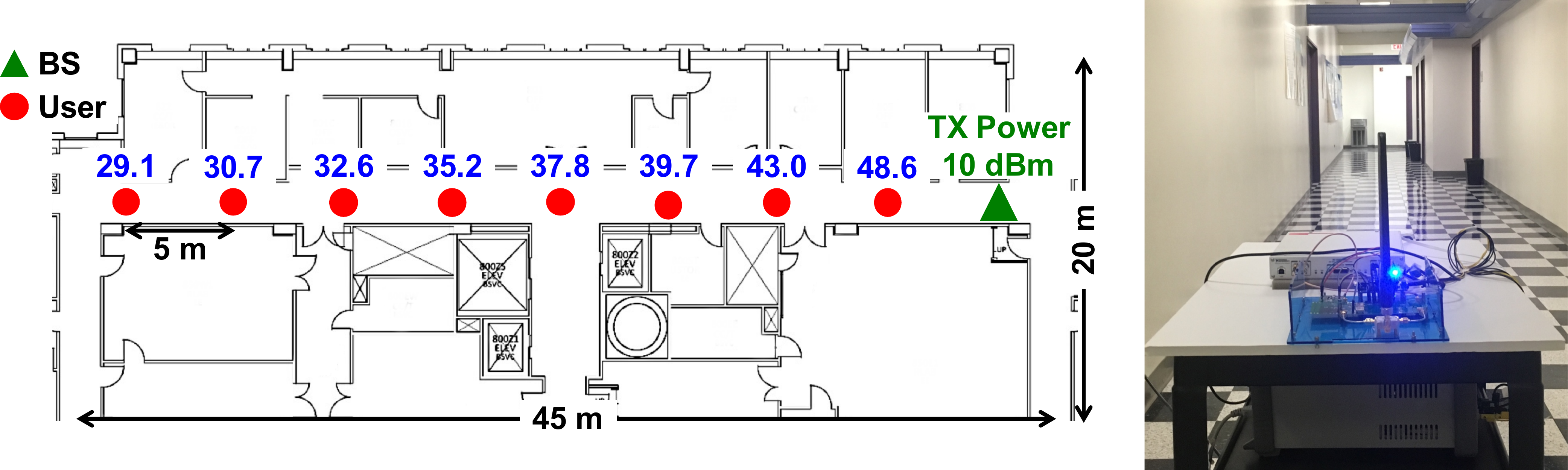}
} \\
\vspace{-0.5\baselineskip}
\subfloat[NLOS deployment and an FD radio in a lab environment]{
\label{fig:exp-map-nlos}
\includegraphics[width=\columnwidth]{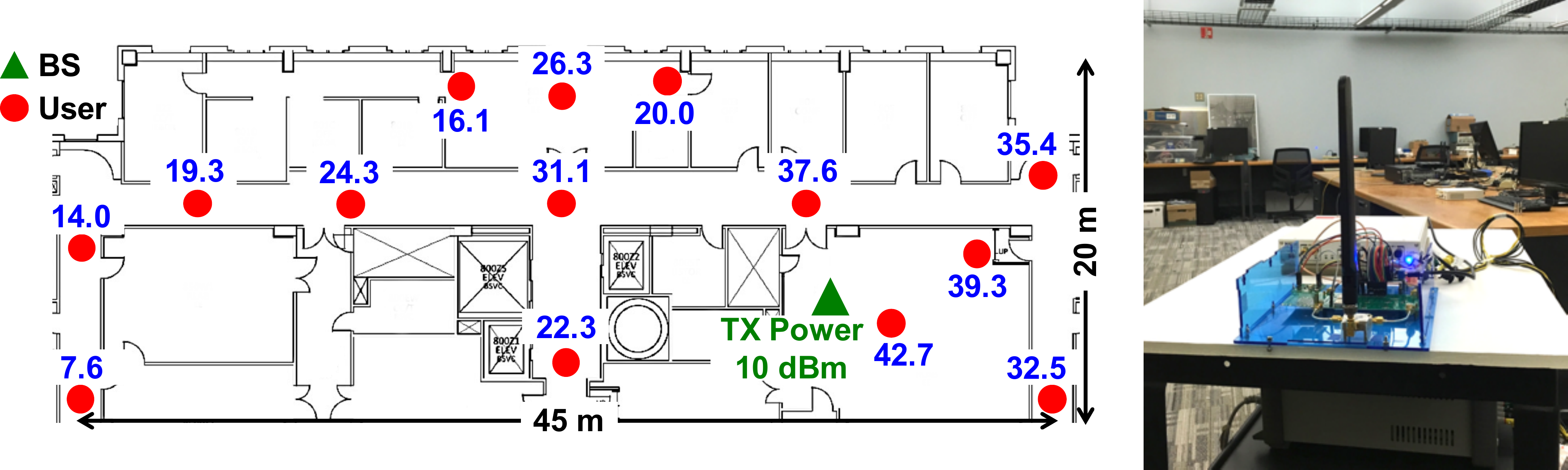}
}
\vspace{-0.5\baselineskip}
\caption{(a) Line-of-sight (LOS), and (b) non-line-of-sight (NLOS) deployments, and the measured HD link SNR values (dB).}
\label{fig:exp-map}
\vspace{-\baselineskip}
\end{figure}

\subsubsection*{Experimental Setup}
To thoroughly evaluate the link level FD throughput gain achieved by our FD radio design, we conduct experiments with two FD radios with $\SI{10}{dBm}$ TX power, one emulating a base station (BS) and one emulating a user. We consider both line-of-sight (LOS) and non-line-of-sight (NLOS) experiments as shown in Fig.~\ref{fig:exp-map}. In the LOS setting, the BS is placed at the end of a hallway and the user is moved away from the BS at stepsizes of 5 meters up to a distance of 40 meters. In the NLOS setting, the BS is placed in a lab environment with regular furniture and the user is placed at various locations (offices, labs, and corridors). We place the BS and the users at about the same height across all the experiments.\footnote{In this work, we emulate the BS and users without focusing on specific deployment scenarios. The impacts of different antenna heights and user densities, as mentioned in~\cite{lopez2015towards}, will be considered in future work.} The measured HD link SNR values are also included in Fig.~\ref{fig:exp-map}. Following the methodology of~\cite{bharadia2013full}, for each user location, we measure the \emph{link SNR difference}, which is defined as the absolute difference between the average HD and FD link SNR values. Throughout the experiments, link SNR values between $0$--$\SI{50}{dB}$ are observed.

\subsubsection*{Difference in HD and FD Link SNR Values}
\begin{figure}[!t]
\centering
\vspace{-\baselineskip}
\subfloat[LOS Experiment]{
\label{fig:exp-link-snr-loss-los}
\includegraphics[width=0.47\columnwidth]{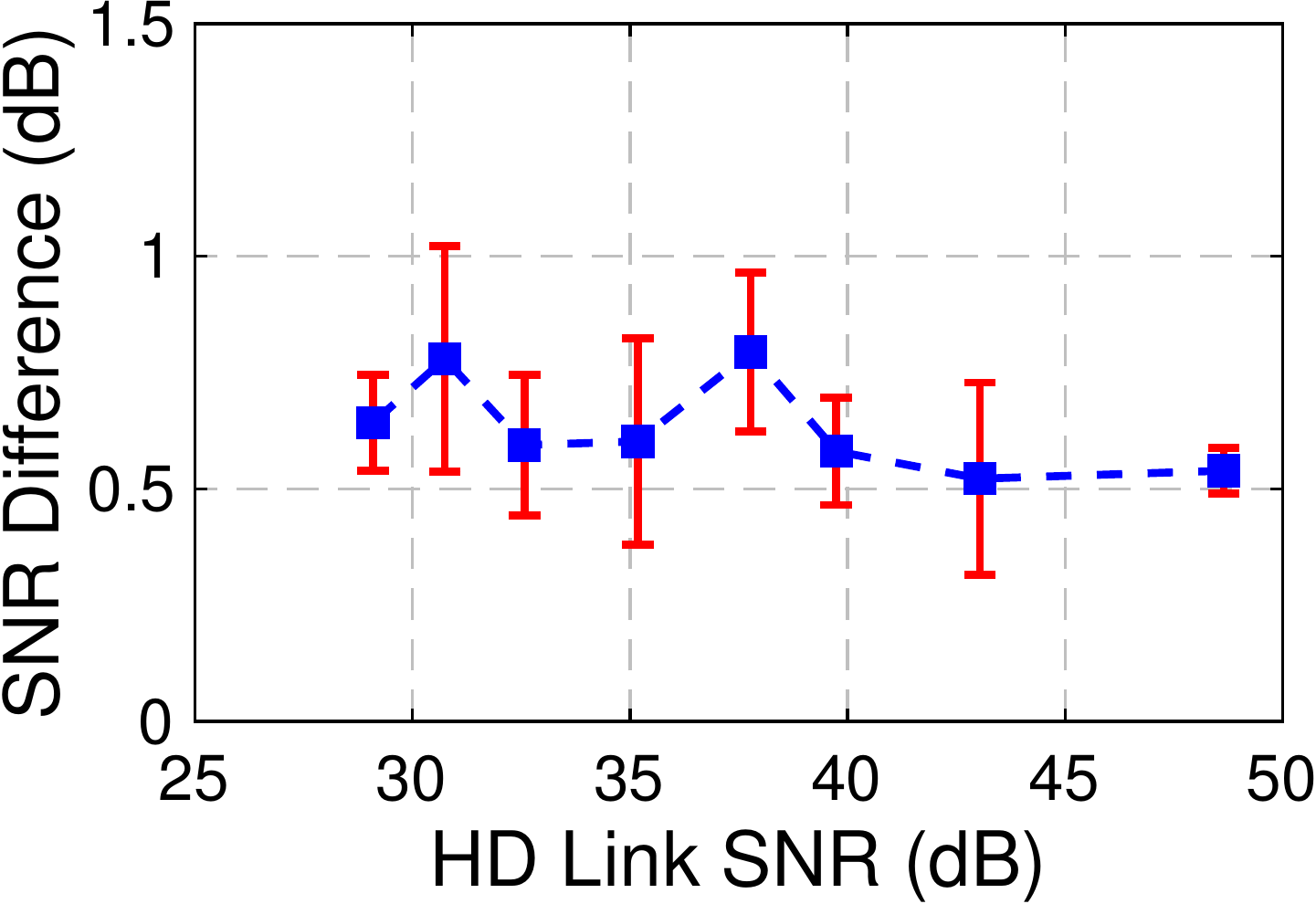}
}
\hfill
\subfloat[NLOS Experiment]{
\label{fig:exp-link-snr-loss-nlos}
\includegraphics[width=0.47\columnwidth]{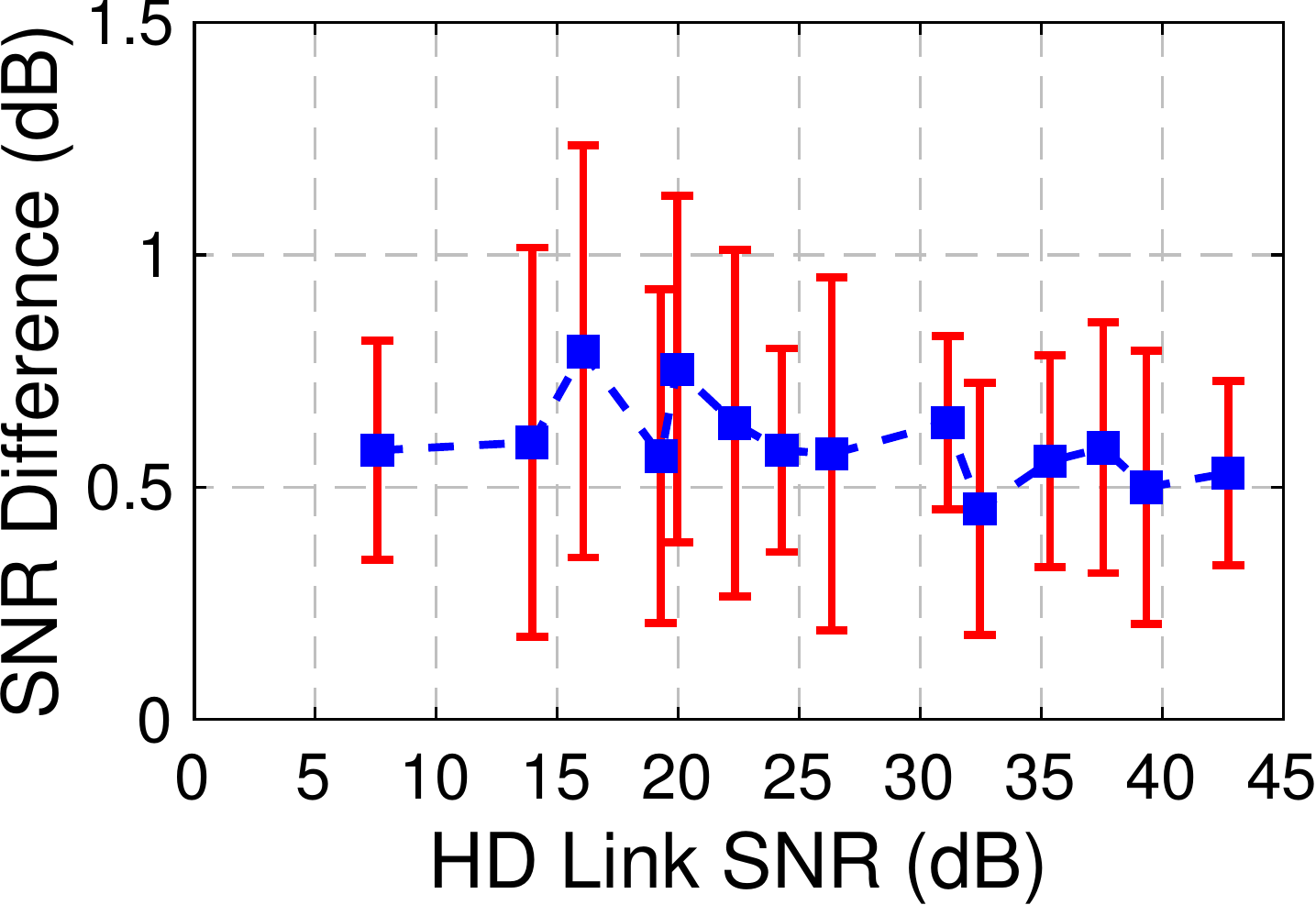}
}
\vspace{-0.5\baselineskip}
\caption{Difference between HD and FD link SNR values in the (a) LOS, and (b) NLOS experiments, with $\SI{10}{dBm}$ TX power and 64QAM-3/4 MCS.}
\label{fig:exp-link-snr-loss}
\vspace{-\baselineskip}
\end{figure}

Fig.~\ref{fig:exp-link-snr-loss} shows the measured link SNR difference as a function of the HD link SNR (i.e., for different user locations) in the LOS and NLOS experiments, respectively, with 64QAM-3/4 MCS. For the LOS experiments, the average link SNR difference is $\SI{0.6}{dB}$ with a standard deviation of $\SI{0.16}{dB}$. For the NLOS experiments, the average link SNR difference is $\SI{0.63}{dB}$ with a standard deviation of $\SI{0.31}{dB}$. The SNR difference has a higher variance in the NLOS experiments, due to the complicated environments (e.g., wooden desks and chairs, metal doors and bookshelves, etc.). In both cases, the link SNR difference is minimal and uncorrelated with user locations, showing the robustness of the FDE-based FD radio.

\subsubsection*{Impact of Constellations}
\begin{figure}[!t]
\centering
\vspace{-\baselineskip}
\subfloat[]{
\label{fig:exp-constellation-snr-loss}
\includegraphics[width=0.47\columnwidth]{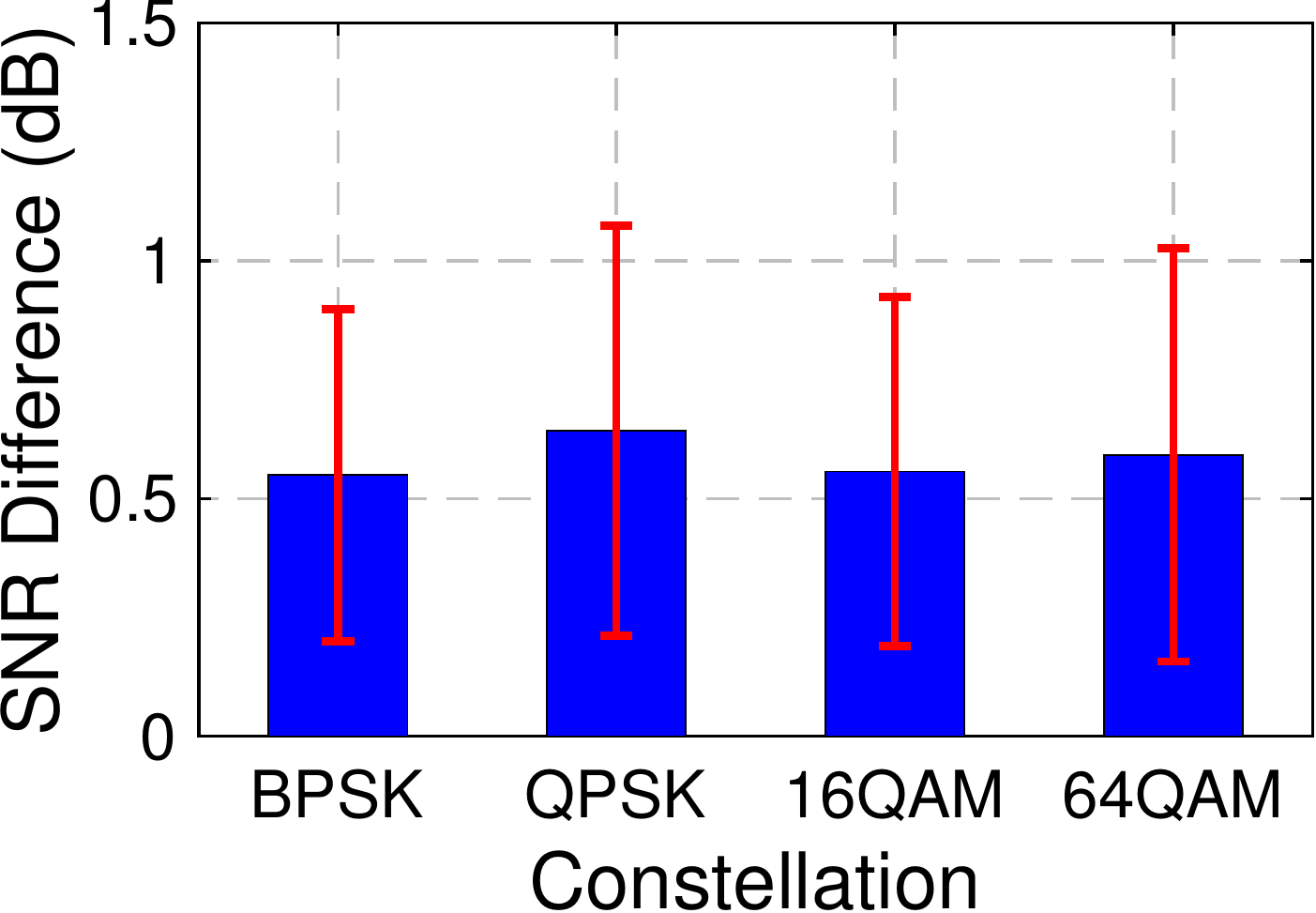}
}
\hfill
\subfloat[]{
\label{fig:exp-constellation-tput-gain}
\includegraphics[width=0.47\columnwidth]{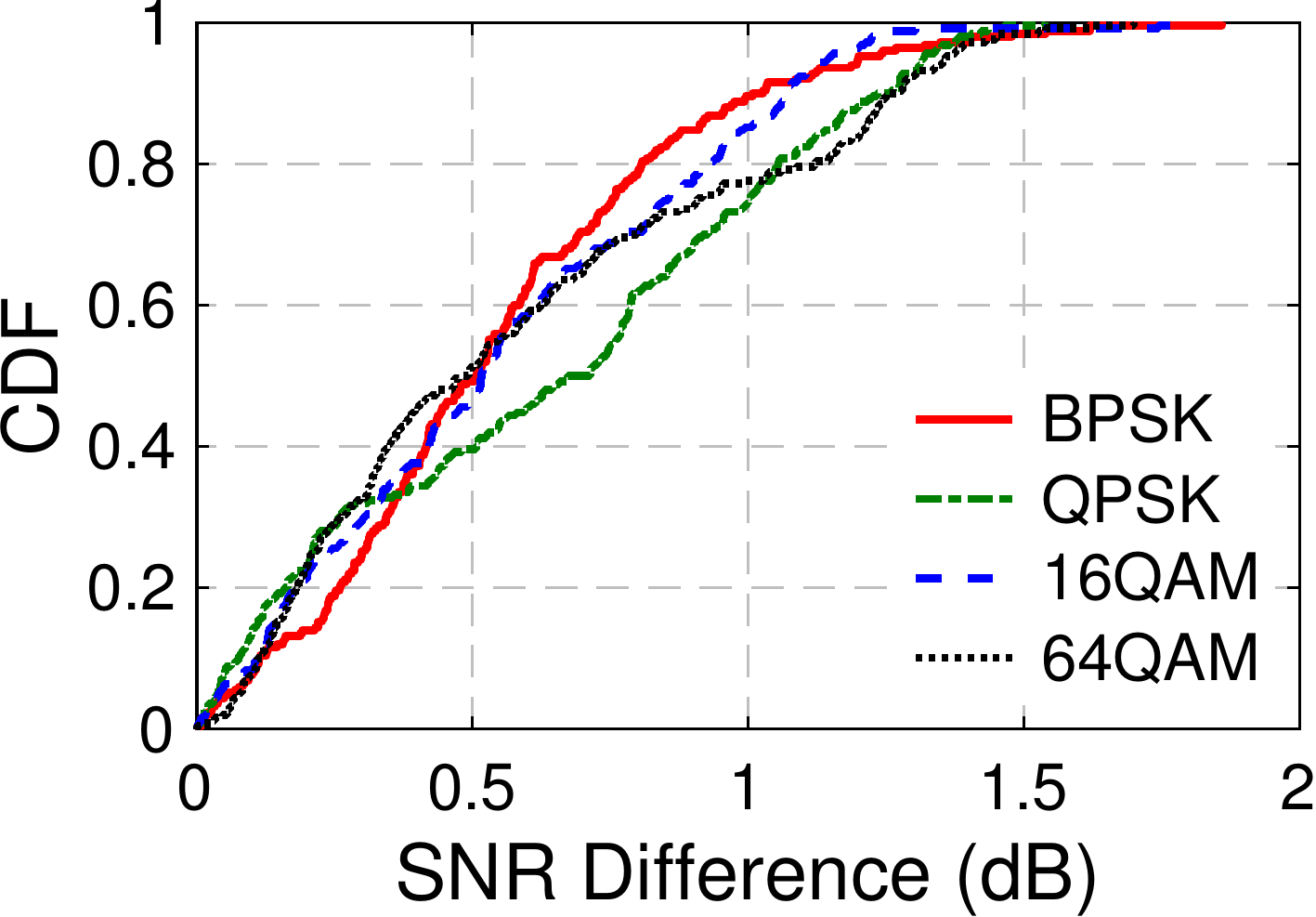}
}
\vspace{-0.5\baselineskip}
\caption{Difference between HD and FD link SNR values with $\SI{10}{dBm}$ TX power under varying constellations: (a) mean and standard deviation, and (b) CDF.}
\label{fig:exp-constellation}
\vspace{-0.5\baselineskip}
\end{figure}

Fig.~\ref{fig:exp-constellation} shows the measured link SNR difference and its CDF with varying constellations and 3/4 coding rate. It can be seen that the link SNR difference has a mean of $\SI{0.58}{dB}$ and a standard deviation of $\SI{0.4}{dB}$, both of which are uncorrelated with the constellations.

\subsubsection*{FD Link Throughput and Gain}
For each user location in the LOS and NLOS experiments, the HD (resp.\ FD) link throughput is measured as the highest average data rate across all MCSs achieved by the link when both nodes operate in HD (resp.\ FD) mode . The FD gain is computed as the ratio between FD and HD throughput values. Recall that the maximal HD data rate is $\SI{54}{Mbps}$, an FD link data rate of $\SI{108}{Mbps}$ can be achieved with an FD link PRR of 1.

\begin{figure}[!t]
\centering
\vspace{-\baselineskip}
\subfloat[LOS Experiment]{
\label{fig:exp-link-tput-los}
\includegraphics[width=0.47\columnwidth]{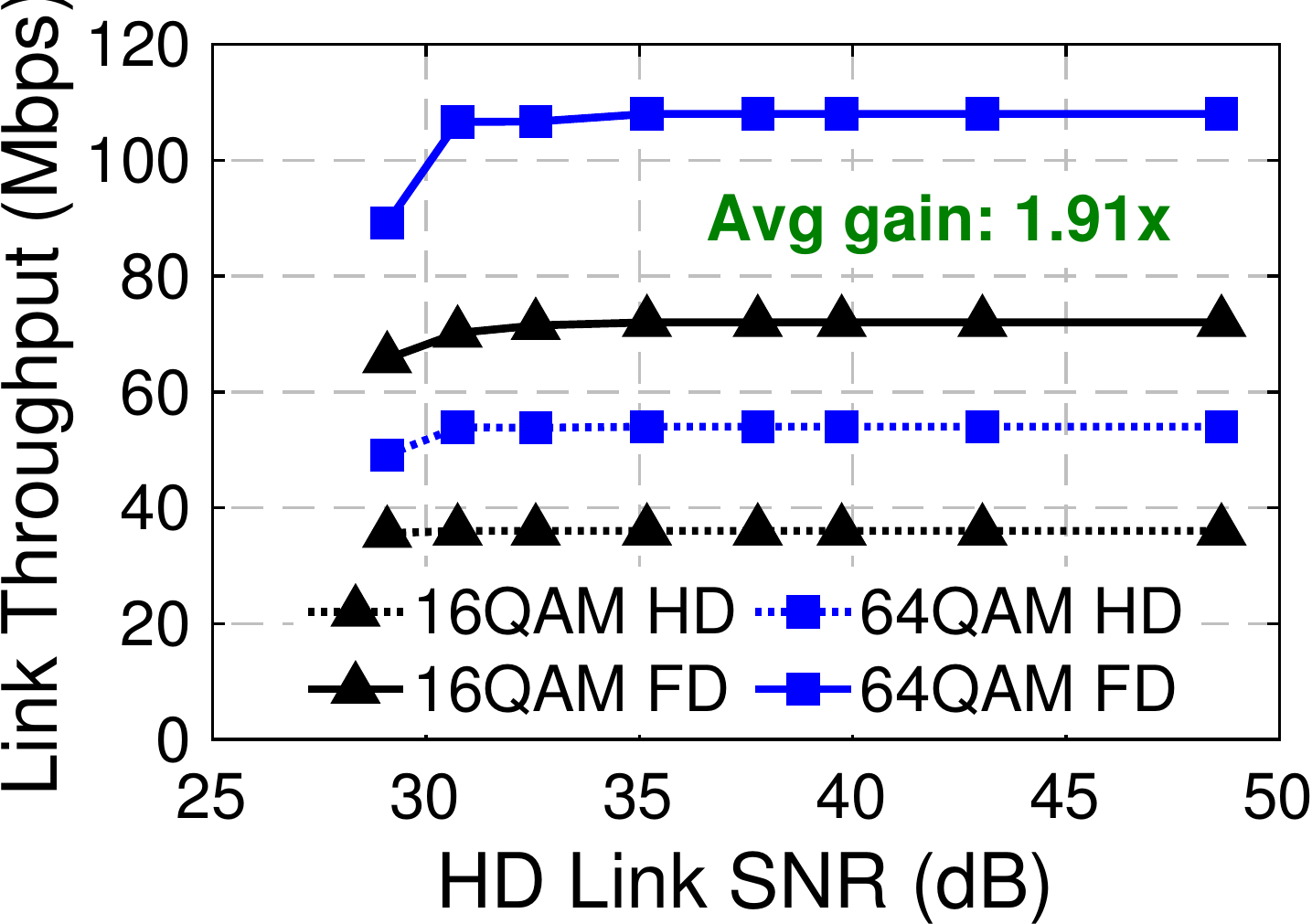}
}
\hfill
\subfloat[NLOS Experiment]{
\label{fig:exp-link-tput-nlos}
\includegraphics[width=0.47\columnwidth]{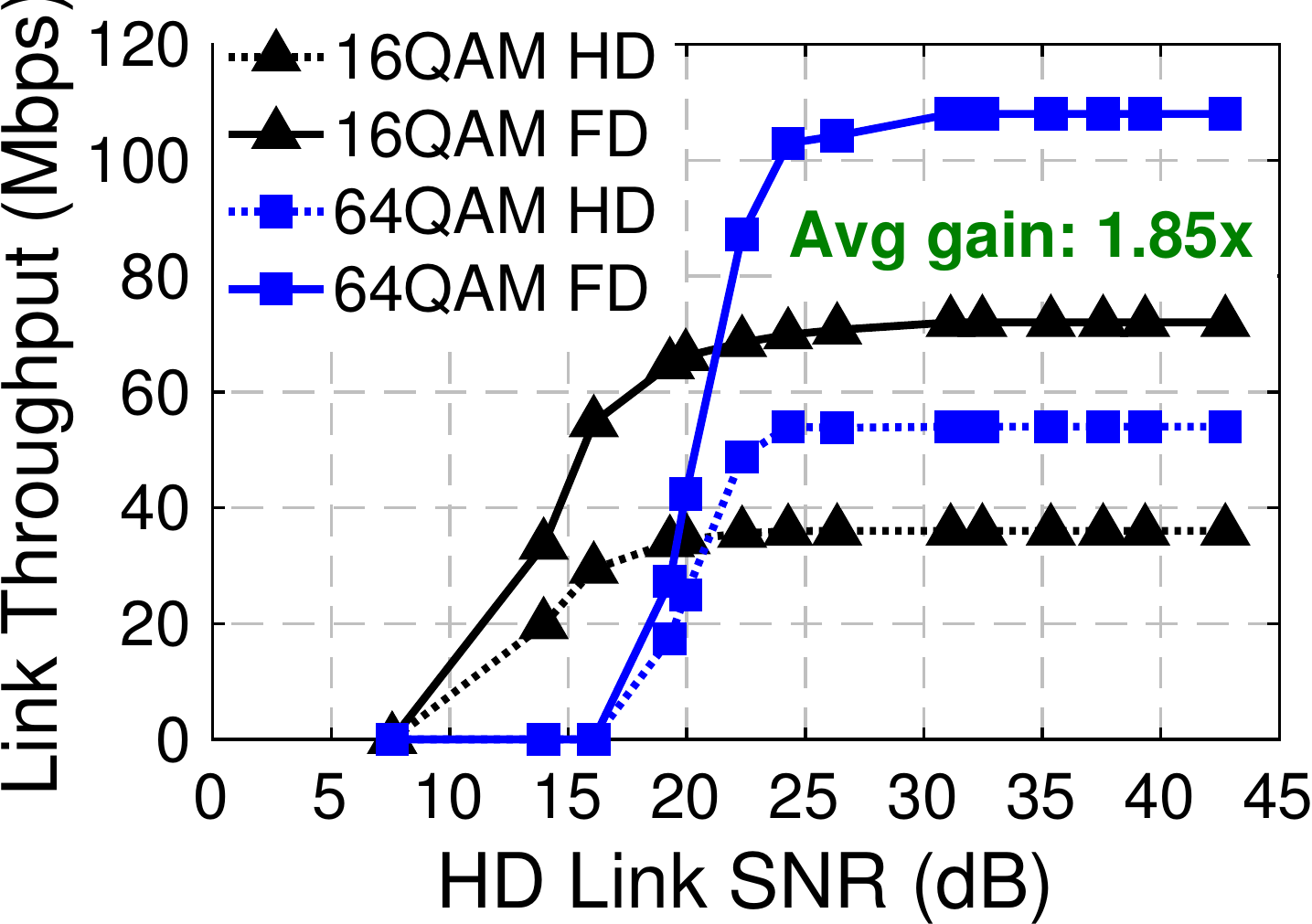}
}
\vspace{-0.5\baselineskip}
\caption{HD and FD link throughput in the (a) LOS, and (b) NLOS experiments, with $\SI{10}{dBm}$ TX power and 16QAM-3/4 and 64QAM-3/4 MCSs.}
\label{fig:exp-link-tput}
\vspace{-0.5\baselineskip}
\end{figure}

Fig.~\ref{fig:exp-link-tput} shows the average HD and FD link throughput with varying 16QAM-3/4 and 16QAM-3/4 MCSs, where each point represents the average throughput across 1,000 packets. The results show that with sufficient link SNR (e.g., $\SI{30}{dB}$ for 64QAM-3/4 MCS), the FDE-based FD radios achieve an \emph{exact} link throughput gain of $2\times$. In these scenarios, the HD/FD link always achieves a link PRR of 1 which results in the maximum achievable HD/FD link data rate. With medium link SNR values, where the link PRR less than 1, the average FD link throughput gains across different MCSs are $1.91\times$ and $1.85\times$ for the LOS and NLOS experiments, respectively. We note that if higher modulation schemes (e.g., 256QAM) are considered and the corresponding link SNR values are high enough for these schemes, the HD/FD throughput can increase (compared to the values in Fig.~\ref{fig:exp-link-tput}). However, considering such schemes is not required in order to evaluate the FDE-based cancellers and the FD gain.

\subsection{Network-Level FD Gain}
\label{ssec:exp-net}
\begin{figure}[!t]
\centering
\vspace{-\baselineskip}
\subfloat[]{
\label{fig:exp-net-setup-ul-dl}
\includegraphics[width=0.32\columnwidth]{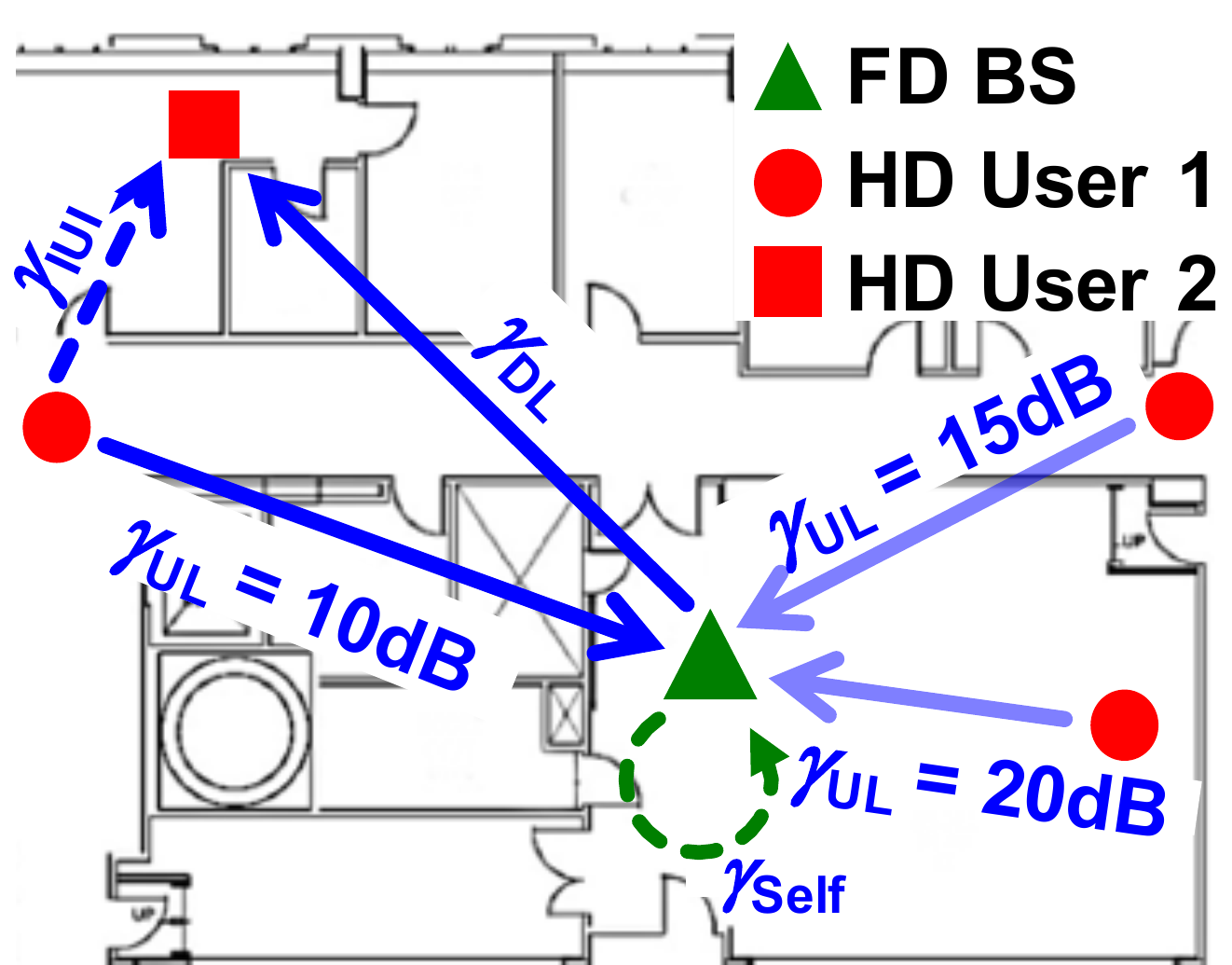}
}
\hspace{-6pt} \hfill
\subfloat[]{
\label{fig:exp-net-setup-two-users}
\includegraphics[width=0.32\columnwidth]{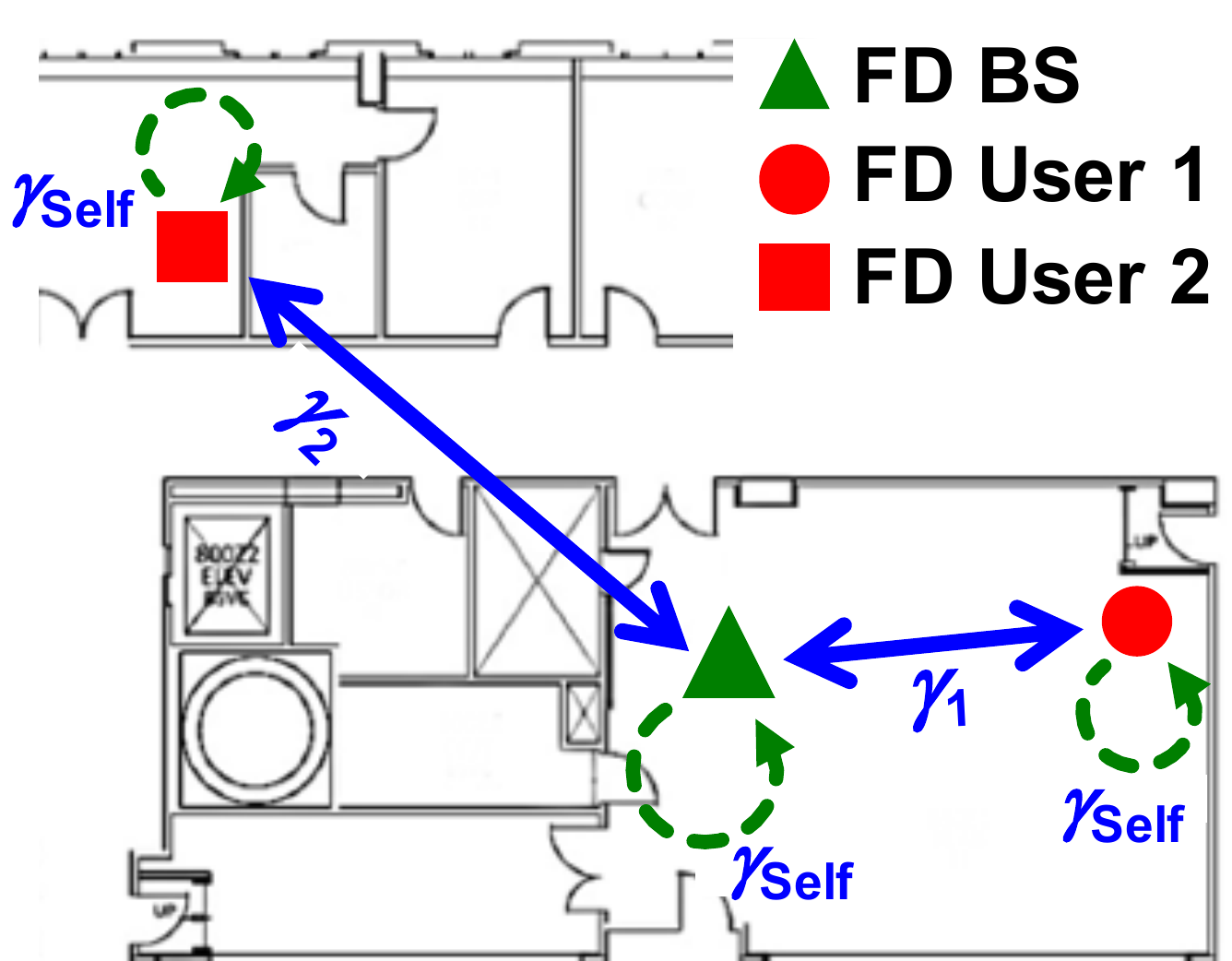}
}
\hspace{-6pt} \hfill
\subfloat[]{
\label{fig:exp-net-setup-three-users}
\includegraphics[width=0.32\columnwidth]{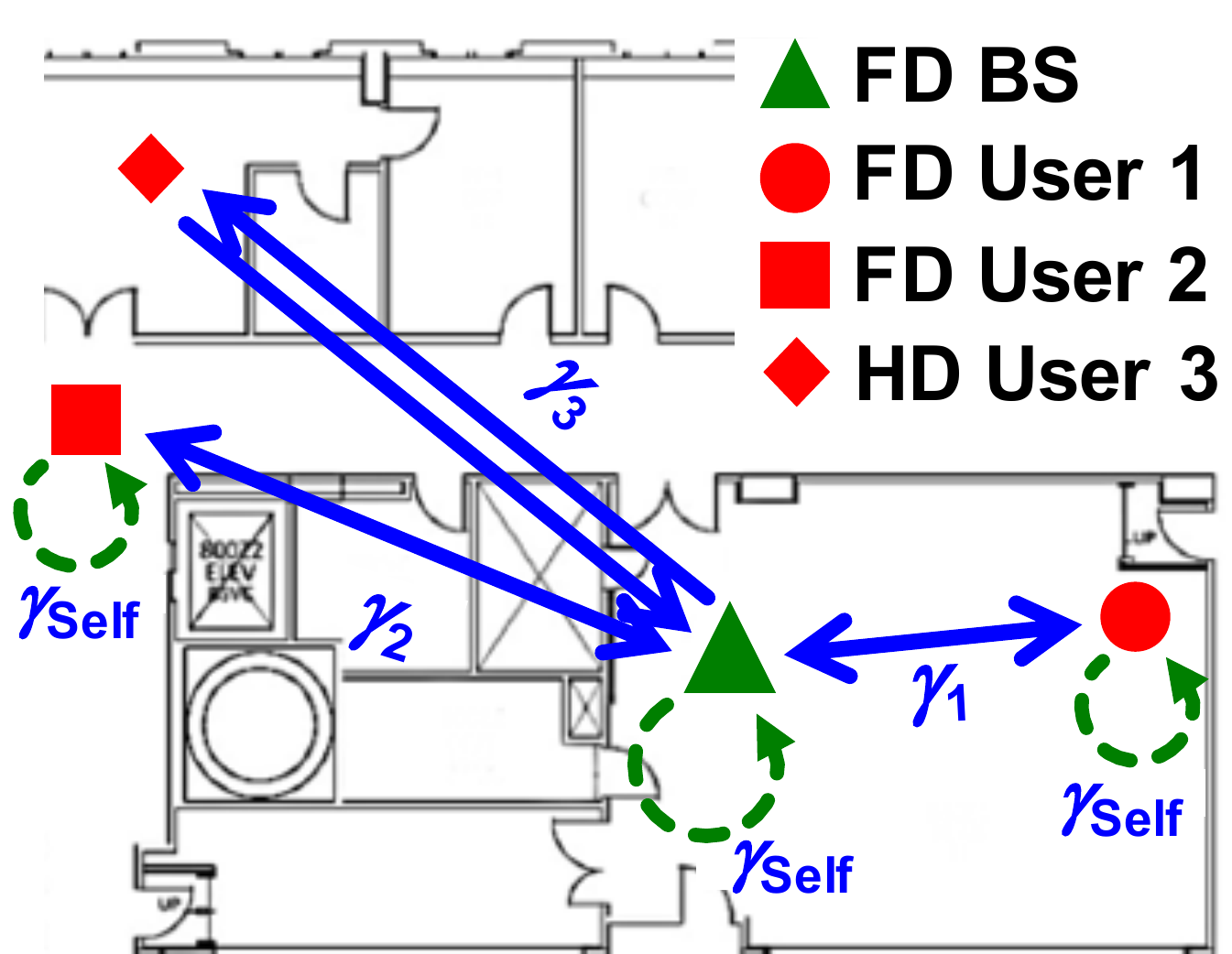}
}
\vspace{-0.5\baselineskip}
\caption{An example experimental setup for: (a) the UL-DL networks with varying $\SNRUL$ and $\SNRDL$, (b) heterogeneous 3-node network with one FD BS and 2 FD users, and (c) heterogeneous 4-node networks with one FD BS, 2 FD users, and one HD user.}
\label{fig:exp-net-setup}
\vspace{-0.5\baselineskip}
\end{figure}

We now experimentally evaluate the network-level throughput gain introduced by FD-capable BS and users. The users can significanlty benefit from the FDE-based FD radio suitable for hand-held devices. We compare experimental results to the analysis (e.g.,~\cite{marasevic2017resource}) and demonstrate practical FD gain in different network settings. Specifically, we consider two types of networks as depicted in Fig.~\ref{fig:exp-net-setup}: (i) \emph{UL-DL networks} with one FD BS and two HD users with inter-user interference (IUI), and (ii) \emph{heterogeneous HD-FD networks} with HD and FD users. Due to software challenge with implementing a real-time MAC layer using a USRP, we apply a TDMA setting where each (HD or FD) user takes turn to be activated for the same period of time.

\subsubsection{UL-DL Networks with IUI}
\label{sssec:exp-net-ul-dl}
We first consider UL-DL networks consisting of one FD BS and two HD users (indexed 1 and 2). Without loss of generality, in this setting, user 1 transmits on the UL to the BS, and the BS transmits to user 2 on the DL (see Fig.~\ref{fig:exp-net-setup}\subref{fig:exp-net-setup-ul-dl}).

\noindent\textbf{Analytical FD gain}.
We use Shannon's capacity formula $\DataRate{}(\SNR{}) = \BW \cdot \log_{2}(1+\SNR{})$ to compute the \emph{analytical throughput} of a link under bandwidth $\BW$ and link SNR $\SNR{}$. If the BS is only HD-capable, the network throughput in a UL-DL network when the UL and DL share the channel in a TDMA manner with equal fraction of time is given by
\begin{align}
\label{eq:net-ul-dl-tput-hd}
\DataRate{\textrm{UL-DL}}^{\textrm{HD}} = \frac{\BW}{2}\log_{2}\left(1+\SNRUL\right) + \frac{\BW}{2}\log_{2}\left(1+\SNRDL\right),
\end{align}
where $\SNRUL$ and $\SNRDL$ are the UL and DL SNRs, respectively. If the BS is FD-capable, the UL and DL can be simultaneously activated with an analytical network throughput of
\begin{align}
\label{eq:net-ul-dl-tput-fd}
\DataRate{\textrm{UL-DL}}^{\textrm{FD}} = \BW\log_{2}\left(1+\frac{\SNRUL}{1+\XINR}\right) + \BW\log_{2}\left(1+\frac{\SNRDL}{1+\IUI}\right),
\end{align}
where: (i) $\left(\frac{\SNRDL}{1+\IUI}\right)$ is the signal-to-interference-plus-noise ratio (SINR) at the DL HD user, and (ii) $\XINR$ is the residual self-interference-to-noise ratio (XINR) at the FD BS. We set $\XINR=1$ when computing the analytical throughput. Namely, the residual SI power is no higher than the RX noise floor (which can be achieved by the FDE-based FD radio, see Section~\ref{ssec:exp-node}). The \emph{analytical FD gain} is then defined as the ratio $\left(\DataRate{\textrm{UL-DL}}^{\textrm{FD}}/\DataRate{\textrm{UL-DL}}^{\textrm{HD}}\right)$. Note that the FD gain depends on the coupling between $\SNRUL$, $\SNRDL$, and $\IUI$, which depend on the BS/user locations, their TX power, etc. 

\begin{figure}[!t]
\centering
\vspace{-\baselineskip}
\subfloat[$\SNRUL=\SI{10}{dB}$]{
\label{fig::exp-net-ul-dl-gain-10db}
\includegraphics[height=1.35in]{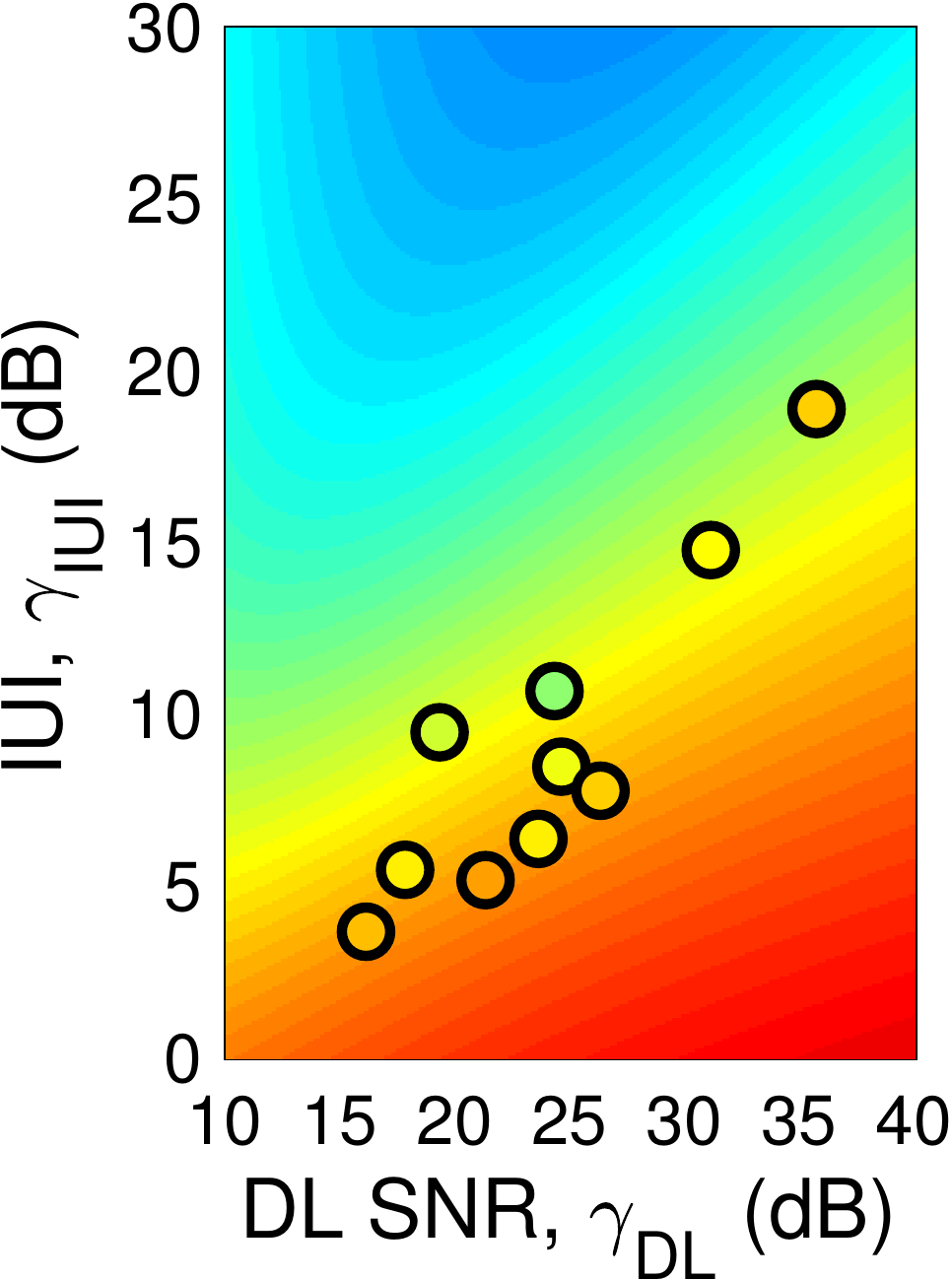}
}
\subfloat[$\SNRUL=\SI{15}{dB}$]{
\label{fig::exp-net-ul-dl-gain-15db}
\includegraphics[height=1.35in]{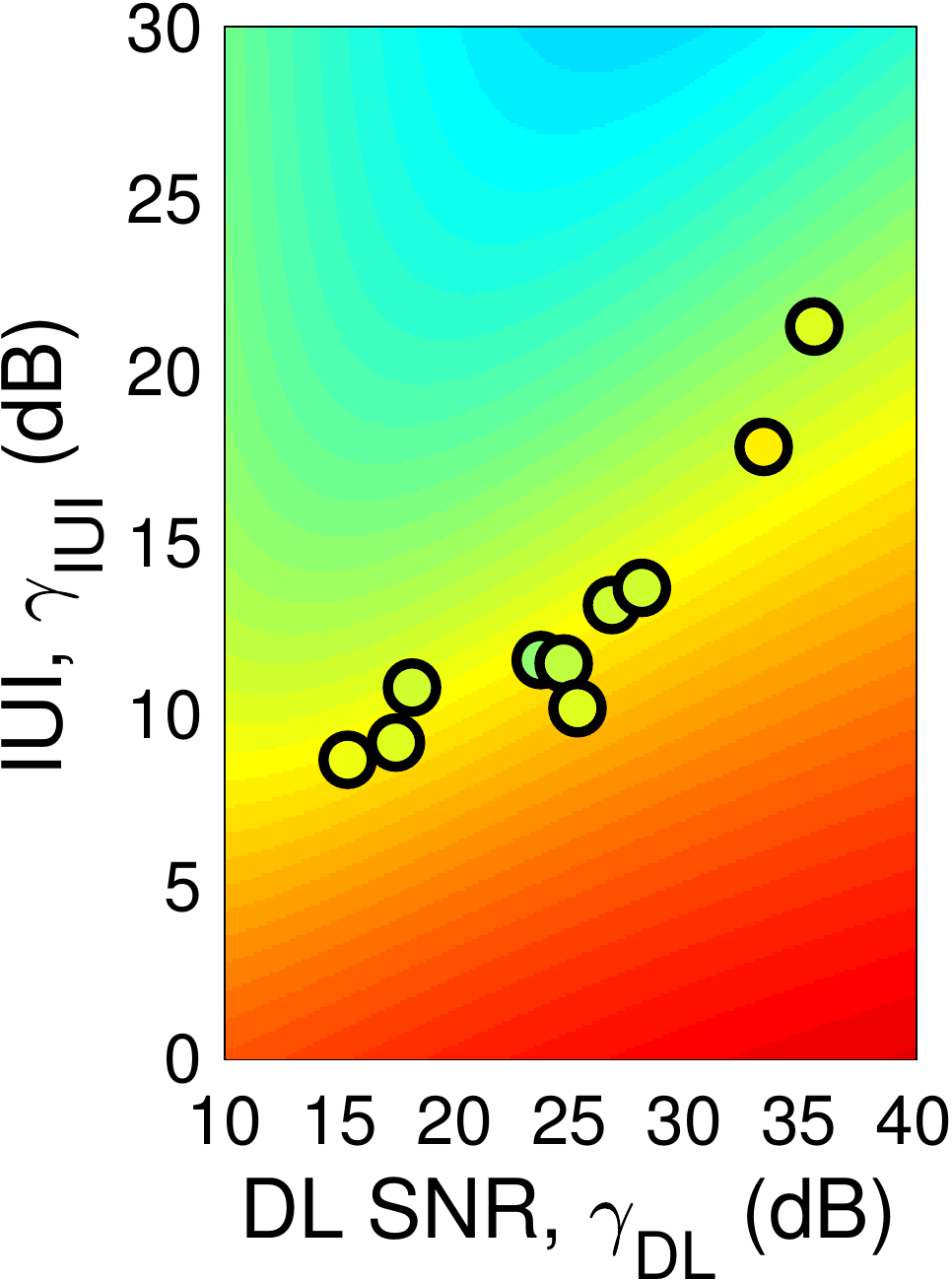}
}
\subfloat[$\SNRUL=\SI{20}{dB}$]{
\label{fig::exp-net-ul-dl-gain-20db}
\includegraphics[height=1.35in]{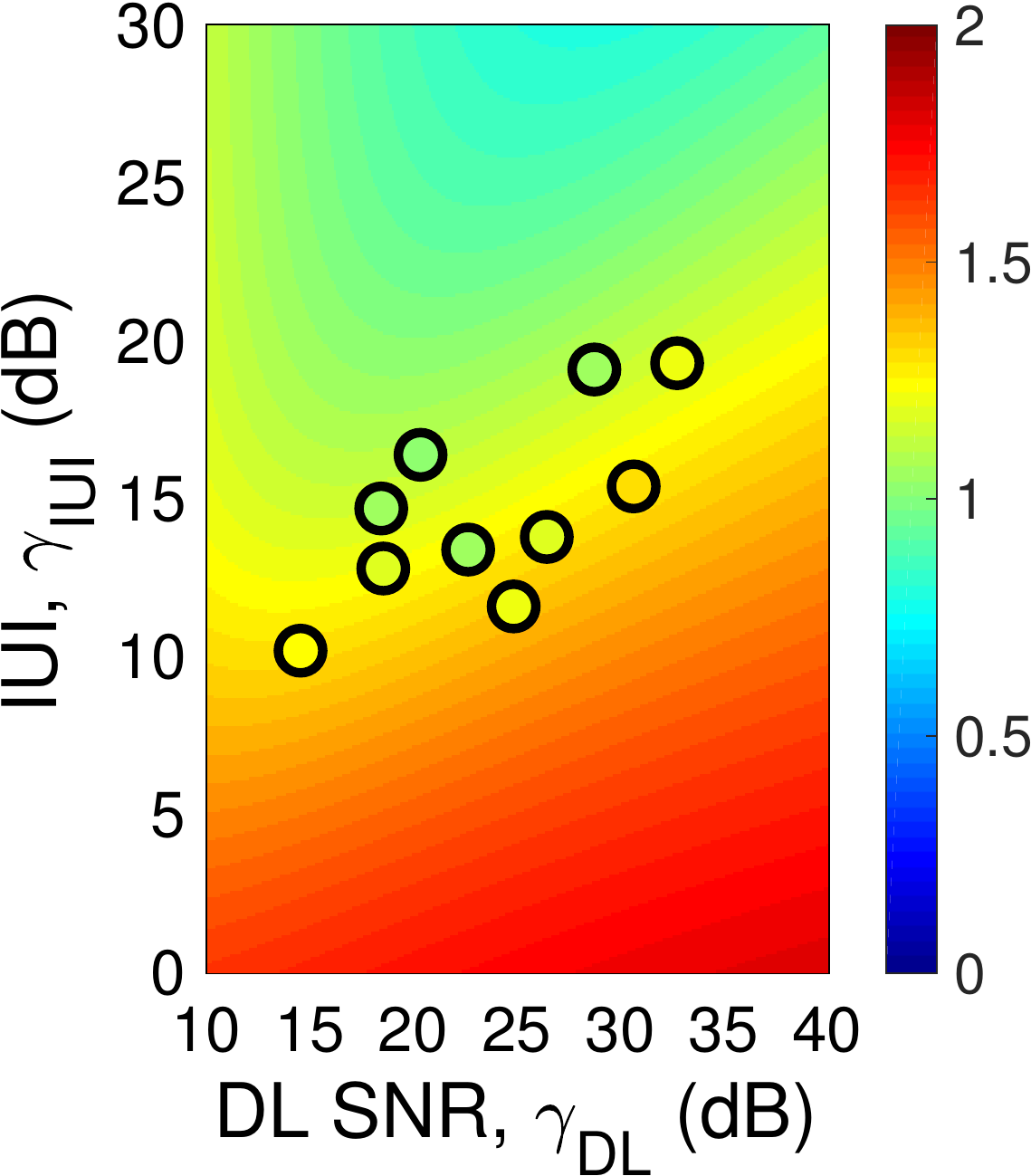}
}
\vspace{-0.5\baselineskip}
\caption{Analytical (colored surface) and experimental (filled circles) network throughput gain for UL-DL networks consisting of one FD BS and two HD users with varying UL and DL SNR values, and inter-user interference (IUI) levels: (a) $\SNRUL=\SI{10}{dB}$, (b) $\SNRUL=\SI{15}{dB}$, and (c) $\SNRUL=\SI{20}{dB}$. The baseline is the network throughput when the BS is HD.}
\label{fig:exp-net-ul-dl-gain}
\vspace{-\baselineskip}
\end{figure}

\noindent\textbf{Experimental FD gain}.
The experimental setup is depicted in Fig.~\ref{fig:exp-net-setup}\subref{fig:exp-net-setup-ul-dl}, where the TX power levels of the BS and user 1 are set to be $\SI{10}{dBm}$ and $\SI{-10}{dBm}$, respectively. We fix the location of the BS and consider different UL SNR values of $\SNRUL = 10/15/20\thinspace\SI{}{dB}$ by placing user 1 at three different locations. For each value of $\SNRUL$, user 2 is placed at 10 different locations, resulting in varying $\SNRDL$ and $\IUI$ values.

\begin{table}[!t]
\caption{Average FD Gain in UL-DL Networks with IUI.}
\label{table:exp-net-ul-dl-gain}
\vspace{-0.5\baselineskip}
\scriptsize
\begin{center}
\begin{tabular}{|c|c|c|}
\hline
UL SNR, $\SNRUL$ & Analytical FD Gain & Experimental FD Gain \\
\hline
$\SI{10}{dB}$ & $1.30\times$ & $1.25\times$ \\
\hline
$\SI{15}{dB}$ & $1.23\times$ & $1.16\times$ \\
\hline
$\SI{20}{dB}$ & $1.22\times$ & $1.14\times$ \\
\hline
\end{tabular}
\end{center}
\vspace{-0.5\baselineskip}
\end{table}

Fig.~\ref{fig:exp-net-ul-dl-gain} shows the analytical (colored surface) and experimental (filled circles) FD gain, where the analytical gain is extracted using {\eqref{eq:net-ul-dl-tput-hd}} and {\eqref{eq:net-ul-dl-tput-fd}}, and the experimental gain is computed using the measured UL and DL throughput. It can be seen that smaller values of $\SNRUL$ and lower ratios between $\SNRDL$ and $\IUI$ lead to higher throughput gains in both analysis and experiments. The average analytical and experimental FD gains are summarized in Table~\ref{table:exp-net-ul-dl-gain}. Due to practical reasons such as the link SNR difference and its impact on link PRR (see Section~\ref{ssec:exp-snr-prr-relationship}), the experimental FD gain is $7\%$ lower than the analytical FD gain. The results confirm the analysis in~\cite{marasevic2017resource} and demonstrate the practical FD gain achieved in wideband UL-DL networks without any changes in the current network stack (i.e., only bringing FD capability to the BS). Moreover, performance improvements are expected through advanced power control and scheduling schemes.

\subsubsection{Heterogeneous 3-Node Networks}
\label{sssec:exp-net-two-users}

We consider heterogeneous HD-FD networks with 3 nodes: one FD BS and two users that can operate in either HD or FD mode (see an example experimental setup in Figs.\ref{fig:intro}\subref{fig:intro-fd-net} and~\ref{fig:exp-net-setup}\subref{fig:exp-net-setup-two-users}). All 3 nodes have the same $\SI{0}{dBm}$ TX power so that each user has symmetric UL and DL SNR values of $\SNR{i}$ ($i=1,2$).
We place user 1 at 5 different locations and place user 2 at 10 different locations for each location of user 1, resulting in a total number of 50 pairs of $(\SNR{1},\SNR{2})$.

\noindent\textbf{Analytical FD gain}.
We set the users to share the channel in a TDMA manner. The analytical network throughput in a 3-node network when zero, one, and two users are FD-capable is respectively given by
\begin{align}
& \DataRate{}^{\textrm{HD}} = \frac{\BW}{2}\log_{2}\left(1+\SNR{1}\right) + \frac{\BW}{2}\log_{2}\left(1+\SNR{2}\right), \label{eq:net-two-users-tput-hd} \\
& \DataRate{\textrm{User}~i~\textrm{FD}}^{\textrm{HD-FD}} = \BW\log_{2}\left(1+\frac{\SNR{i}}{1+\XINR}\right) + \frac{\BW}{2}\log_{2}\left(1+\SNR{\overline{i}}\right),\ \label{eq:net-two-users-tput-hd-fd} \\
& \DataRate{}^{\textrm{FD}} = \BW\log_{2}\left(1+\frac{\SNR{1}}{1+\XINR}\right) + \BW\log_{2}\left(1+\frac{\SNR{2}}{1+\XINR}\right), \label{eq:net-two-users-tput-fd}
\end{align}
where $\XINR=1$ is set (similar to Section~\ref{sssec:exp-net-ul-dl}). We consider both FD gains of $\left(\DataRate{\textrm{User}~i~\textrm{FD}}^{\textrm{HD-FD}}/\DataRate{}^{\textrm{HD}}\right)$ (i.e., user $i$ is FD and user $\overline{i} \ne i$ is HD), and $\left(\DataRate{}^{\textrm{FD}}/\DataRate{}^{\textrm{HD}}\right)$ (i.e., both users are FD).

\begin{figure}[!t]
\centering
\vspace{-\baselineskip}
\subfloat[Only user 1 FD]{
\label{fig:exp-net-two-users-user1-fd}
\includegraphics[height=1.35in]{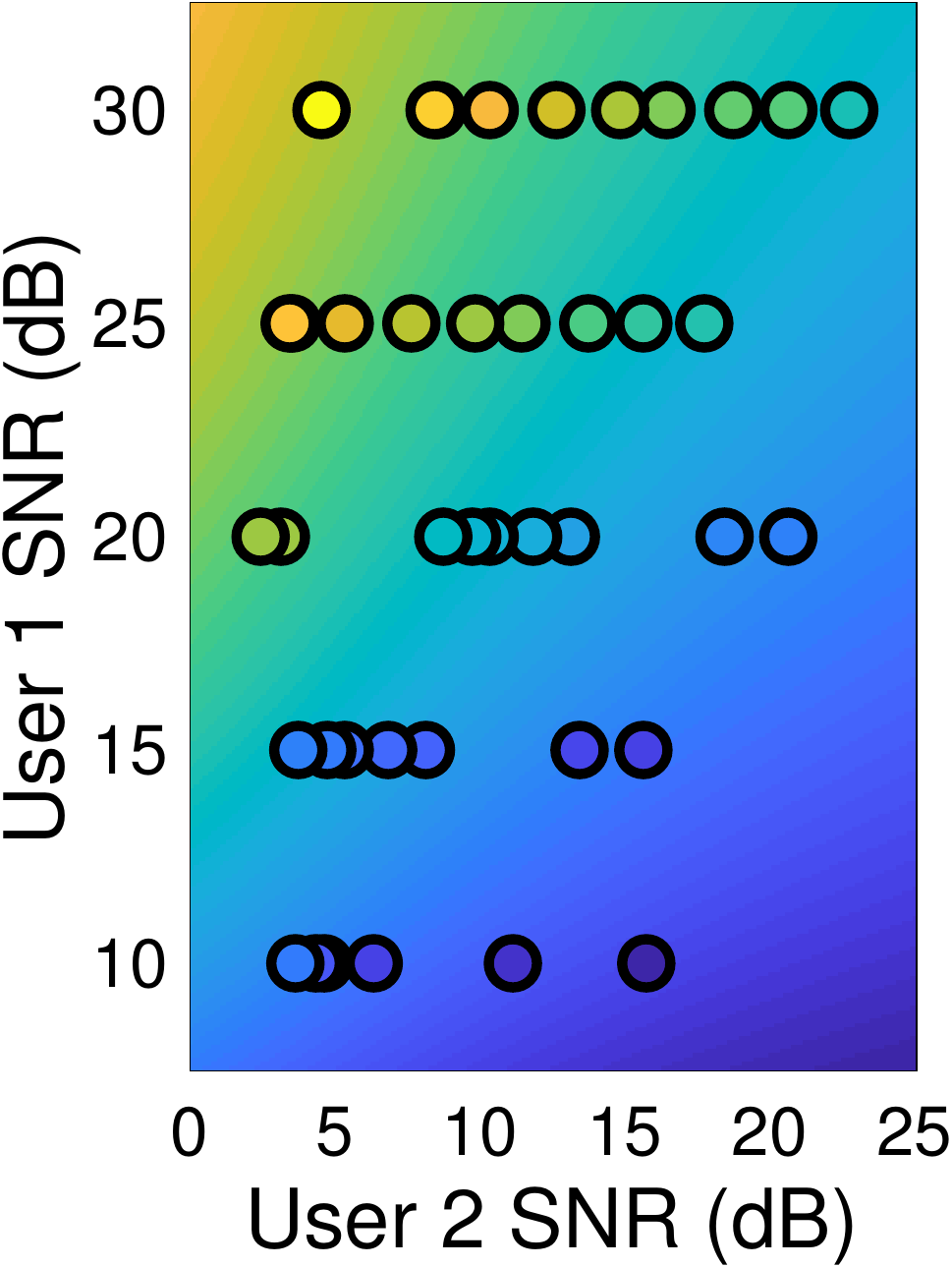}
}
\subfloat[Only user 2 FD]{
\label{fig:exp-net-two-users-user2-fd}
\includegraphics[height=1.35in]{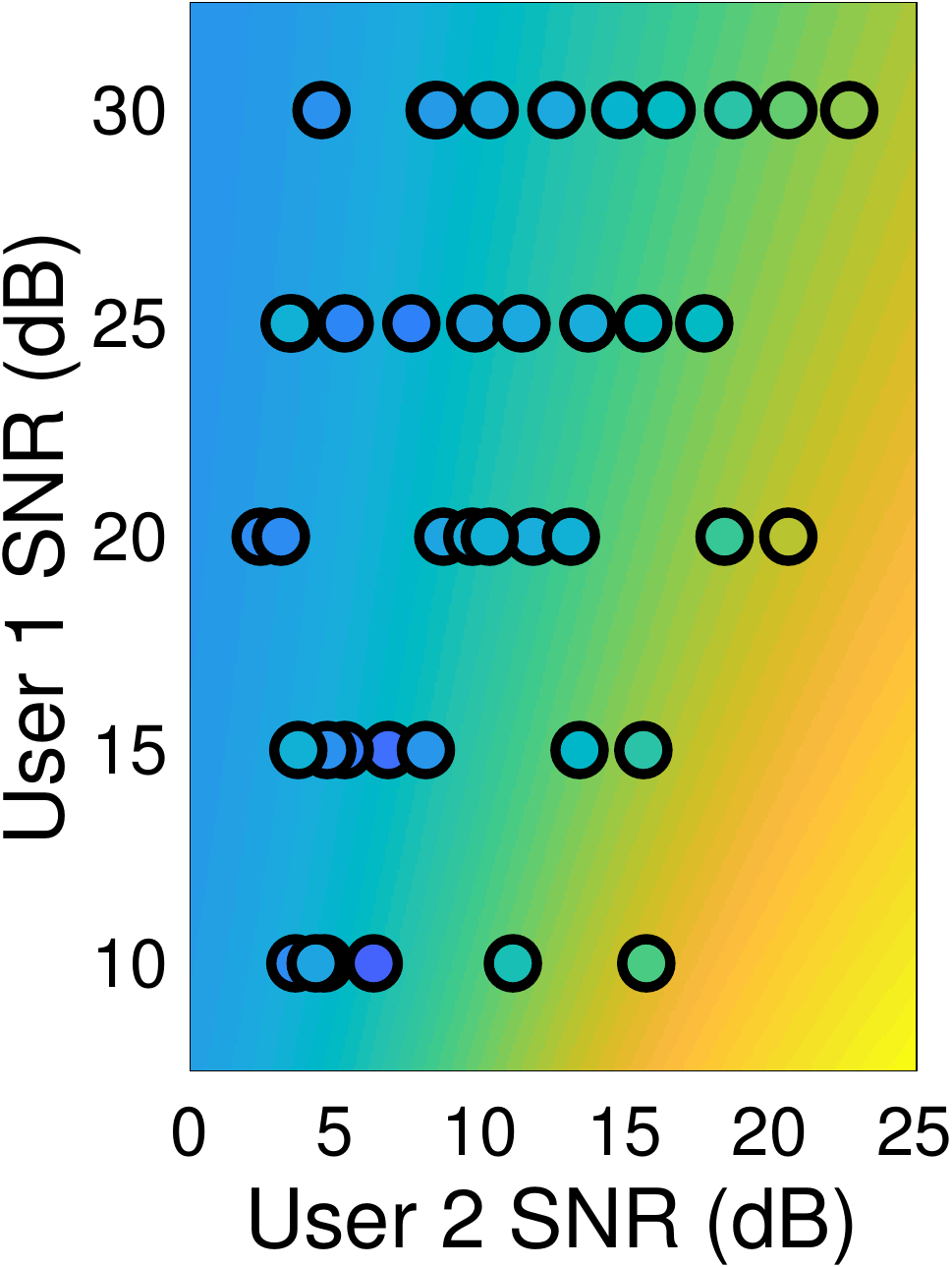}
}
\subfloat[Both users FD]{
\label{fig:exp-net-two-users-both-fd}
\includegraphics[height=1.35in]{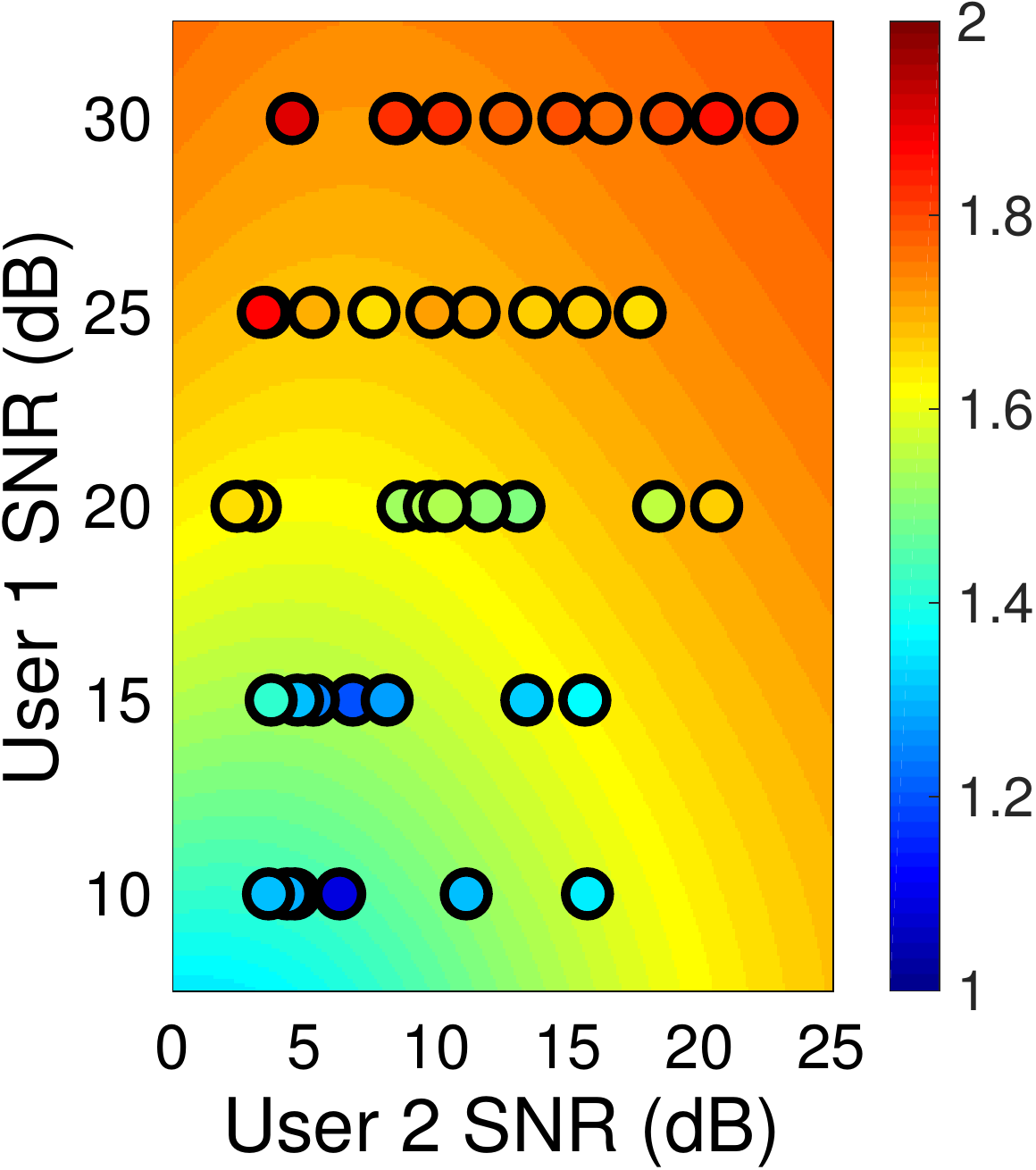}
}
\vspace{-0.5\baselineskip}
\caption{Analytical (colored surface) and experimental (filled circles) network throughput gain for 3-node networks consisting of one FD BS and two users with varying link SNR values: (a) only user 1 is FD, (b) only user 2 is FD, and (c) both users are FD. The baseline is the network throughput when both users are HD.}
\label{fig:exp-net-two-users}
\vspace{-\baselineskip}
\end{figure}

\noindent\textbf{Experimental FD gain}.
For each pair of $(\SNR{1},\SNR{2})$, experimental FD gain is measured in three cases: (i) only user 1 is FD, (ii) only user 2 is FD, and (iii) both users are FD. Fig.~\ref{fig:exp-net-two-users} shows the analytical (colored surface) and experimental (filled circles) FD gain for each case. We exclude the results with $\SNR{i}<\SI{3}{dB}$ since the packets cannot be decoded, resulting in a throughput of zero (see Fig.~\ref{fig:exp-prr-vs-snr}).

\begin{figure}[!t]
\centering
\includegraphics[width=0.9\columnwidth]{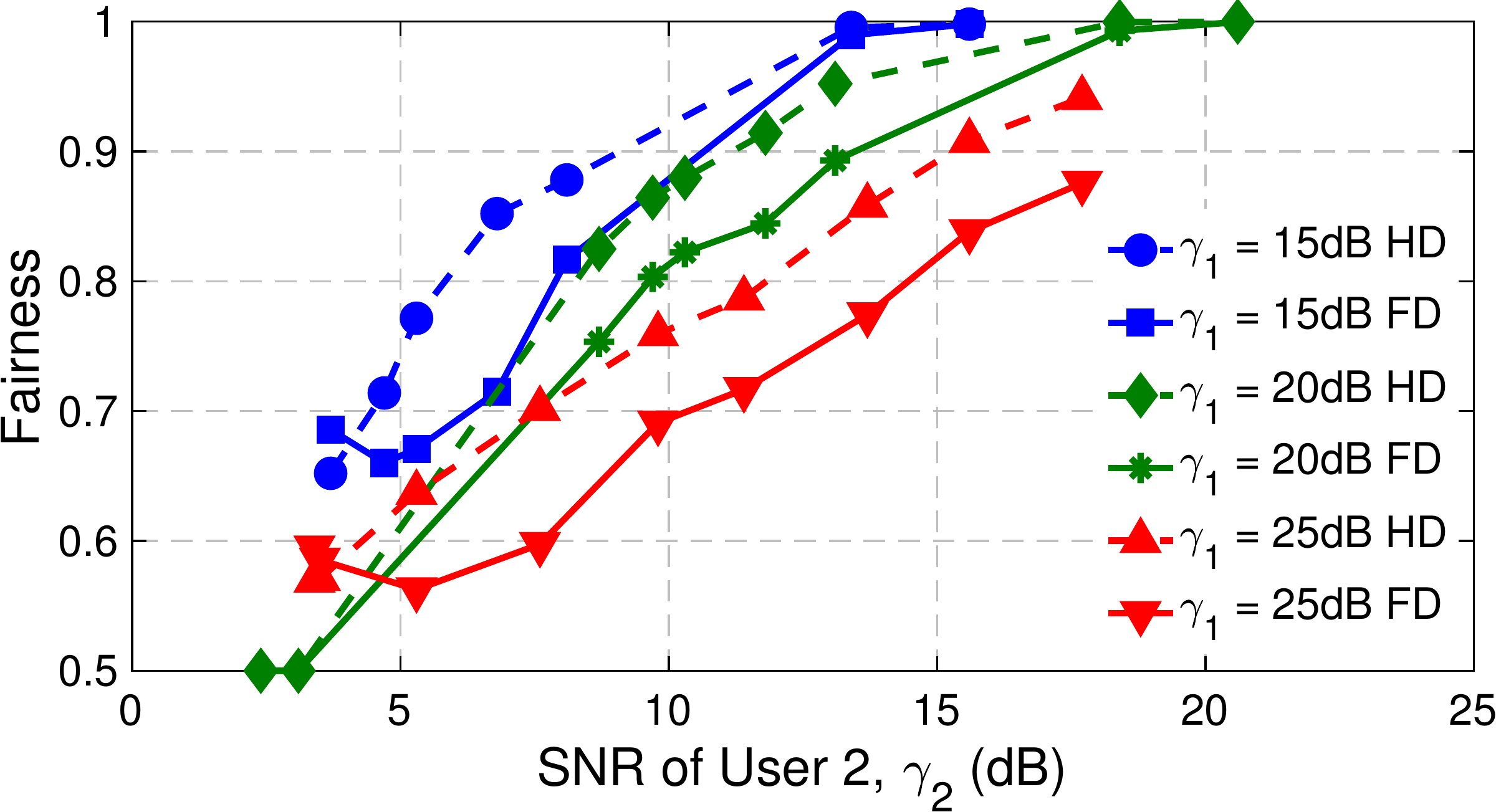}
\vspace{-0.5\baselineskip}
\caption{Measured Jain's fairness index (JFI) in 3-node networks when both users are HD and FD with varying $(\SNR{1},\SNR{2})$.}
\label{fig:exp-net-two-users-fairness}
\vspace{-\baselineskip}
\end{figure}

The results show that with small link SNR values, the experimental FD gain is lower than the analytical value due to the inability to decode the packets. On the other hand, with sufficient link SNR values, the experimental FD gain exceeds the analytical FD gain. This is because setting $\XINR=1$ in {\eqref{eq:net-two-users-tput-hd-fd}} and {\eqref{eq:net-two-users-tput-fd}} results in a $\SI{3}{dB}$ SNR loss in the analytical FD link SNR, and thereby in a lower throughput. However, in practice, the packets can be decoded with a link PRR of 1 with sufficient link SNRs, resulting in exact twice number of packets being successfully sent over an FD link. Moreover, the FD gain is more significant when enabling FD capability for users with higher link SNR values.

Another important metric we consider is the fairness between users, which is measured by the Jain's fairness index (JFI). In the considered 3-node networks, the JFI ranges between 1/2 (worst case) and 1 (best case). Fig.~\ref{fig:exp-net-two-users-fairness} shows the measured JFI when both users operate in HD or FD mode. The results show that introducing FD capability results in an average degradation in the network JFI of only $5.6\%/4.4\%/7.4\%$ for $\SNR{1} = 15/20/25\thinspace\SI{}{dB}$, while the average network FD gains are $1.32\times$/$1.58\times$/$1.73\times$, respectively. In addition, the JFI increases with higher and more balanced user SNR values, which is as expected.

\subsubsection{Heterogeneous 4-Node Networks}
\label{sssec:exp-net-three-users}

We experimentally study 4-node networks consisting of an FD BS and three users with $\SI{10}{dBm}$ TX power (see an example experimental setup in Fig.~\ref{fig:exp-net-setup}\subref{fig:exp-net-setup-three-users}).
The experimental setup is similar to that described in Section~\ref{sssec:exp-net-two-users}. 100 experiments are conducts where the 3 users are placed at different locations with different user SNR values. For each experiment, the network throughput is measured in three cases where: (i) zero, (ii) one, and (iii) two users are FD-capable.

\begin{figure}[!t]
\centering
\includegraphics[width=0.9\columnwidth]{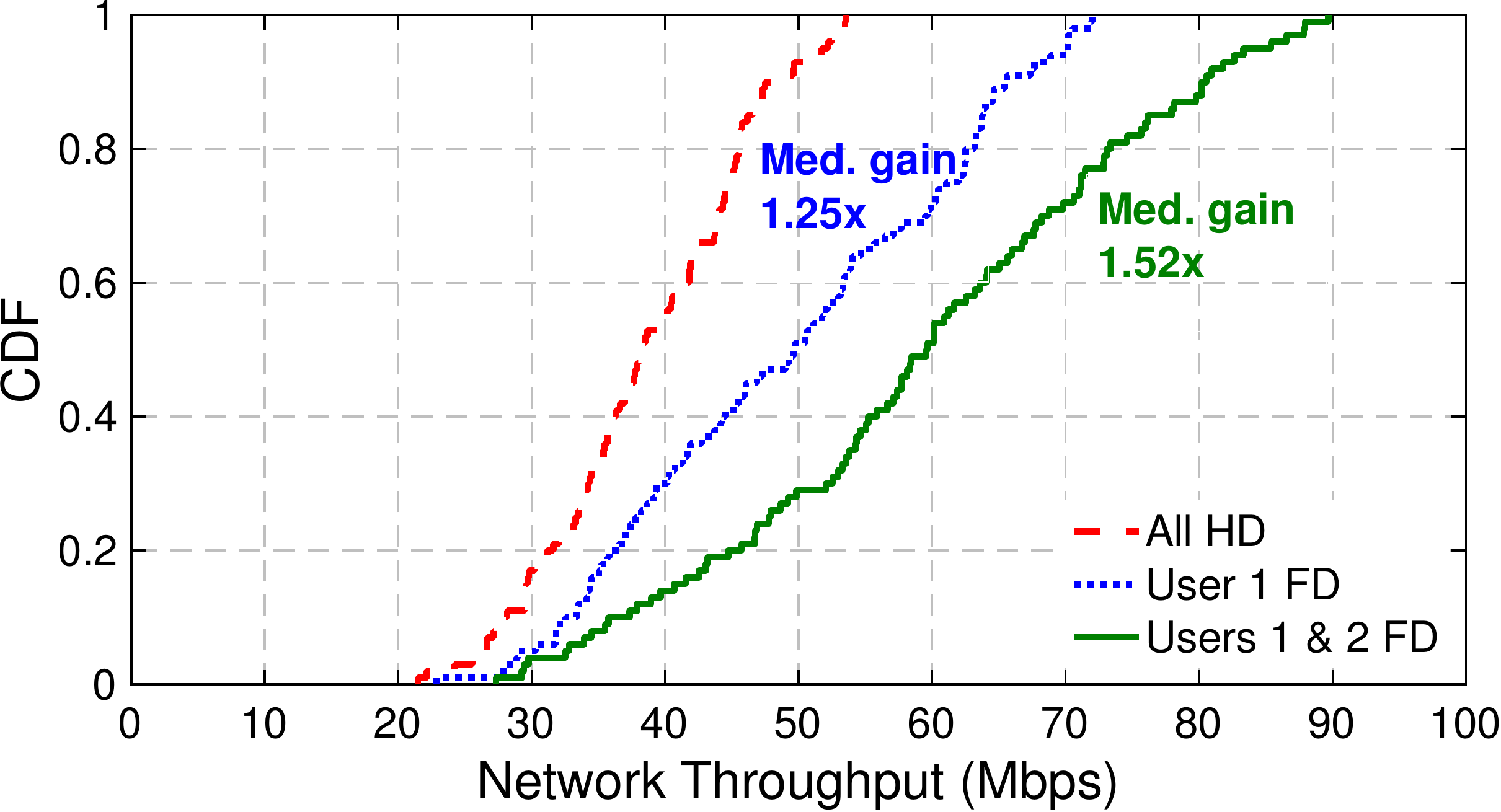}
\vspace{-0.5\baselineskip}
\caption{Experimental network throughput gain for 4-node networks when zero, one, or two users are FD-capable, with $\SI{10}{dBm}$ TX power and varying user locations.}
\label{fig:exp-net-three-users}
\vspace{-\baselineskip}
\end{figure}

Fig.~\ref{fig:exp-net-three-users} shows the CDF of the network throughput of the three cases, where the measured link SNR varies between $5$--$\SI{45}{dB}$. Overall, median network FD gains of $1.25\times$ and $1.52\times$ are achieved in cases with one and two FD users, respectively. The trend shows that in a real-world environment, the total network throughput increases as more users become FD-capable, and the improvement is more significant with higher user SNR values. Note that we only apply a TDMA scheme and a more advanced MAC layer (e.g.,~\cite{kim2013janus}) has the potential to improve the FD gain in these networks.

\section{Numerical Evaluation}
\label{sec:sensitivity}
In this section, we numerically evaluate and compare the performance of the FDE-based RFIC~\cite{Zhou_WBSIC_JSSC15} and PCB cancellers \emph{based on measurements and validated models}. We confirm that the PCB canceller emulates its RFIC counterpart and show that the optimized canceller configuration scheme can significantly improve the performance of the RFIC canceller. We also evaluate the performance of FDE-based cancellers with respect to the number of FDE taps, $\NumTap$, and desired RF SIC bandwidth, $\BW$, and discuss various design tradeoffs.

\subsection{Setup}
We use a real, practical antenna interface response, $\AntTF(f_k)$, measured in the same setting as described in Section~\ref{ssec:exp-testbed}, and consider $\NumTap \in \{1,2,3,4\}$ and $\BW \in \{20,40,80\}\thinspace\SI{}{MHz}$. We only report the RF SIC performance with up to 4 FDE taps since, as we will show, this case can achieve sufficient RF SIC up to $\SI{80}{MHz}$ bandwidth.\footnote{We select typical values of $20/40/80\thinspace\SI{}{MHz}$ as the desired RF SIC bandwidth, since the circulator has a frequency range of $\SI{100}{MHz}$.}

We use {\eqref{eq:rfic-tf}} to both model and evaluate the RFIC canceller with configuration parameters $\{\ICTapAmp{i}, \ICTapPhase{i}, \ICTapCF{i}, \ICTapQF{i}\}$, since it is shown that a $2^{\textrm{nd}}$-order BPF can accurately model the FDE $N$-path filter~\cite{ghaffari2011tunable,Zhou_WBSIC_JSSC15}. Similar to {\OptProblemPCB} (see Section~\ref{ssec:impl-opt}), the optimized RFIC canceller configuration can be obtained by solving {\OptProblemIC} with $\ICTF(f_k)$ given by {\eqref{eq:rfic-tf}}.
\begin{align*}
\OptProblemIC\ & \min: \littlesum_{k=1}^{\NumChnl} \NormTwo{\ICResTF(f_k)} = \littlesum_{k=1}^{\NumChnl} \NormTwo{ \AntTF(f_k) - \ICTF(f_k) }^2 \\
\textrm{s.t.:}\ & \ICTapAmp{i} \in [\ICTapAmpMin, \ICTapAmpMax],\ \ICTapPhase{i} \in [-\pi, \pi], \\
& \ICTapCF{i} \in [\ICTapCFMin, \ICTapCFMax],\ \ICTapQF{i} \in [\ICTapQFMin, \ICTapQFMax],\ \forall i.
\end{align*}

\noindent Note that in~\cite{Zhou_WBSIC_JSSC15}, there is no optimization of the RFIC canceller configuration, and the canceller is configured based on a heuristic approach. As we will show, the optimized canceller scheme outperforms the heuristic approach by an order of magnitude in terms of the amount of RF SIC achieved.

The implemented PCB canceller includes only $\NumTap=2$ FDE taps due to its design (see Section~\ref{sec:impl}). However, it is practically feasible to include more parallel FDE taps. For numerical evaluation purposes, we model the PCB canceller with $\NumTap>2$ FDE taps by extending the validated model {\eqref{eq:pcb-tf-calibrated}} with symmetric FDE taps (i.e., all BPFs in the FDE taps behave identically). Although the canceller configuration scheme has a computational complexity of $4^{\NumTap}$ (i.e., four DoF per FDE tap), we will show that $\NumTap = 4$ taps can achieve sufficient amount of RF SIC in realistic scenarios.

In practice, the canceller configuration parameters cannot be arbitrarily selected from a continuous range as described in {\OptProblemPCB} and {\OptProblemIC}. Instead, they are often restricted to discrete values given the resolution of the corresponding hardware components. To address this problem, we evaluate the canceller models in both the \emph{ideal} case and the case \emph{with practical quantization constraints}. The canceller configuration with quantization constraints are obtained by rounding the configuration parameters returned by solving {\OptProblemPCB} or {\OptProblemIC} to their closest quantized values.

In particular, the RFIC canceller has the following constraints: $\forall i$, $\ICTapAmp{i} \in [-40, -10]\thinspace\SI{}{dB}$, $\ICTapPhase{i} \in [-\pi, \pi]$, $\ICTapCF{i} \in [875, 925]\thinspace\SI{}{MHz}$, and $\ICTapQF{i} \in [1, 50]$. When adding practical quantization constraints, we assume that the amplitude $\ICTapAmp{i}$ has a $\SI{0.25}{dB}$ resolution within its range. For $\ICTapPhase{i}$, $\ICTapCF{i}$, and $\ICTapQF{i}$, an $8$-bit resolution constraint is introduced, which is equivalent to $2^8=256$ discrete values spaced equally in the given range. These constraints are practically selected and can be easily realized in an IC implementation.  The PCB canceller model has following constraints: $\forall i$, $\PCBTapAmp{i} \in [-15.5, 0]\thinspace\SI{}{dB}$, $\PCBTapPhase{i} \in [-\pi, \pi]$, $\PCBTapCFCap{i} \in [0.6,2.4]\thinspace\SI{}{pF}$, and $\PCBTapQFCap{i} \in [2,14]\thinspace\SI{}{pF}$. When adding the quantization constraints, we consider $\SI{0.5}{dB}$, $\SI{0.12}{pF}$, and $\SI{0.39}{pF}$ resolution to $\PCBTapAmp{i}$, $\PCBTapCFCap{i}$, and $\PCBTapQFCap{i}$, respectively. For $\PCBTapPhase{i}$, an 8-bit resolution is introduced. These numbers are consistent with our implementation and experiments (see Sections~\ref{ssec:impl-pcb} and~\ref{sec:exp}).

\subsection{Performance Evaluation and Comparison between the RFIC and PCB Cancellers}

\begin{figure}[!t]
\centering
\vspace{-\baselineskip}
\subfloat{
\includegraphics[width=0.48\columnwidth]{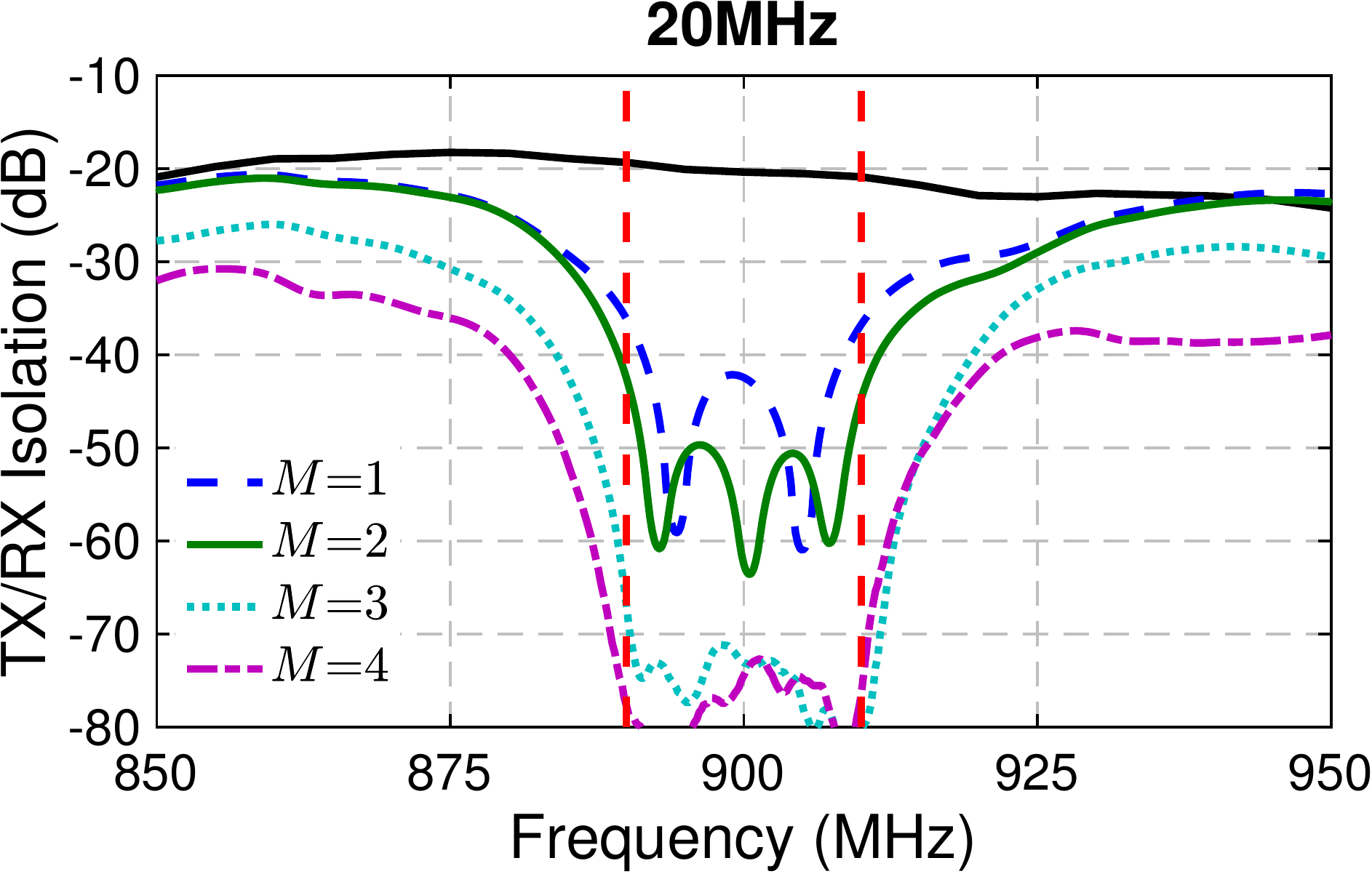}
}
\subfloat{
\includegraphics[width=0.48\columnwidth]{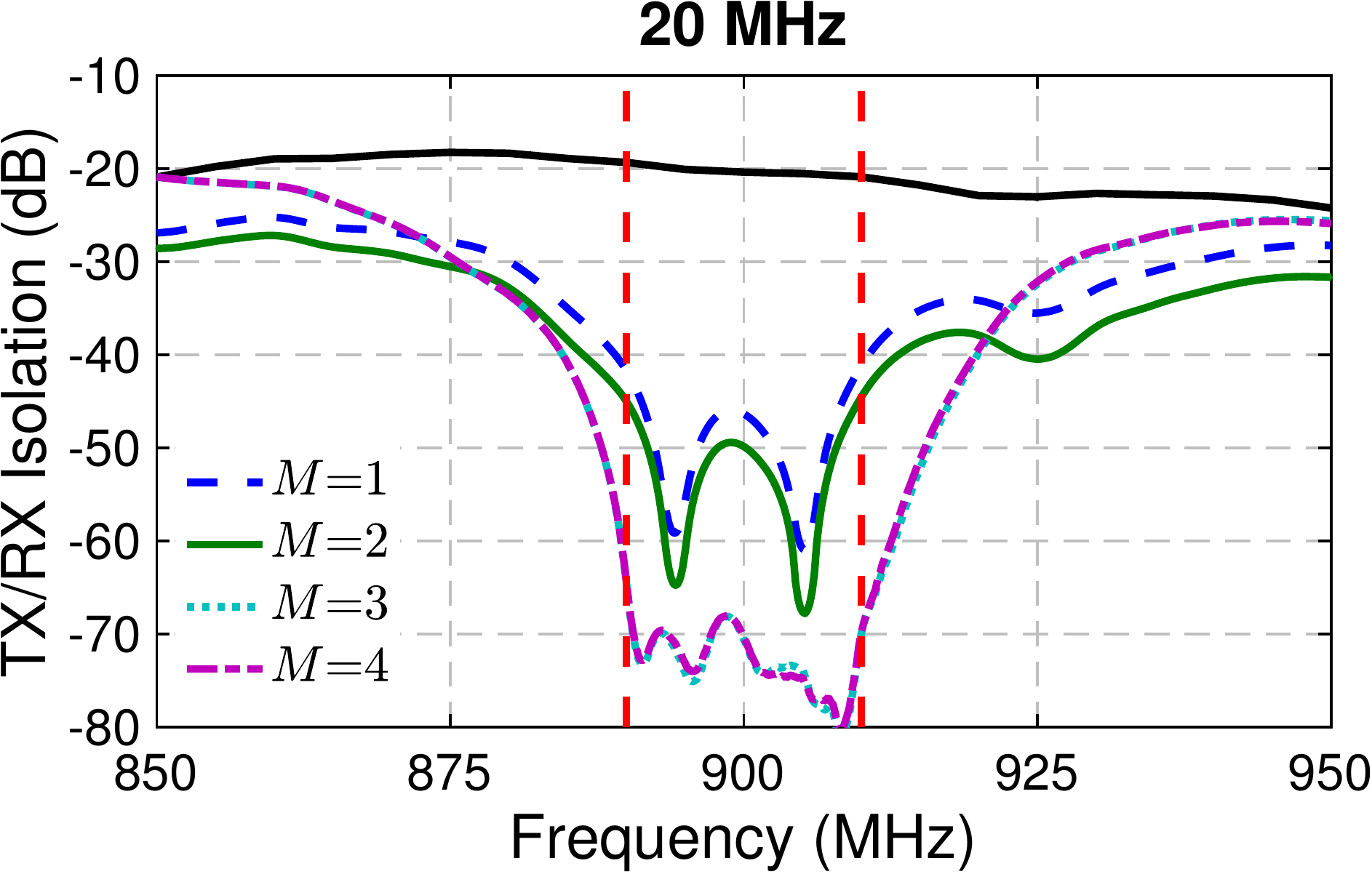}
}
\vspace{-0.5\baselineskip}
\\
\subfloat{
\includegraphics[width=0.48\columnwidth]{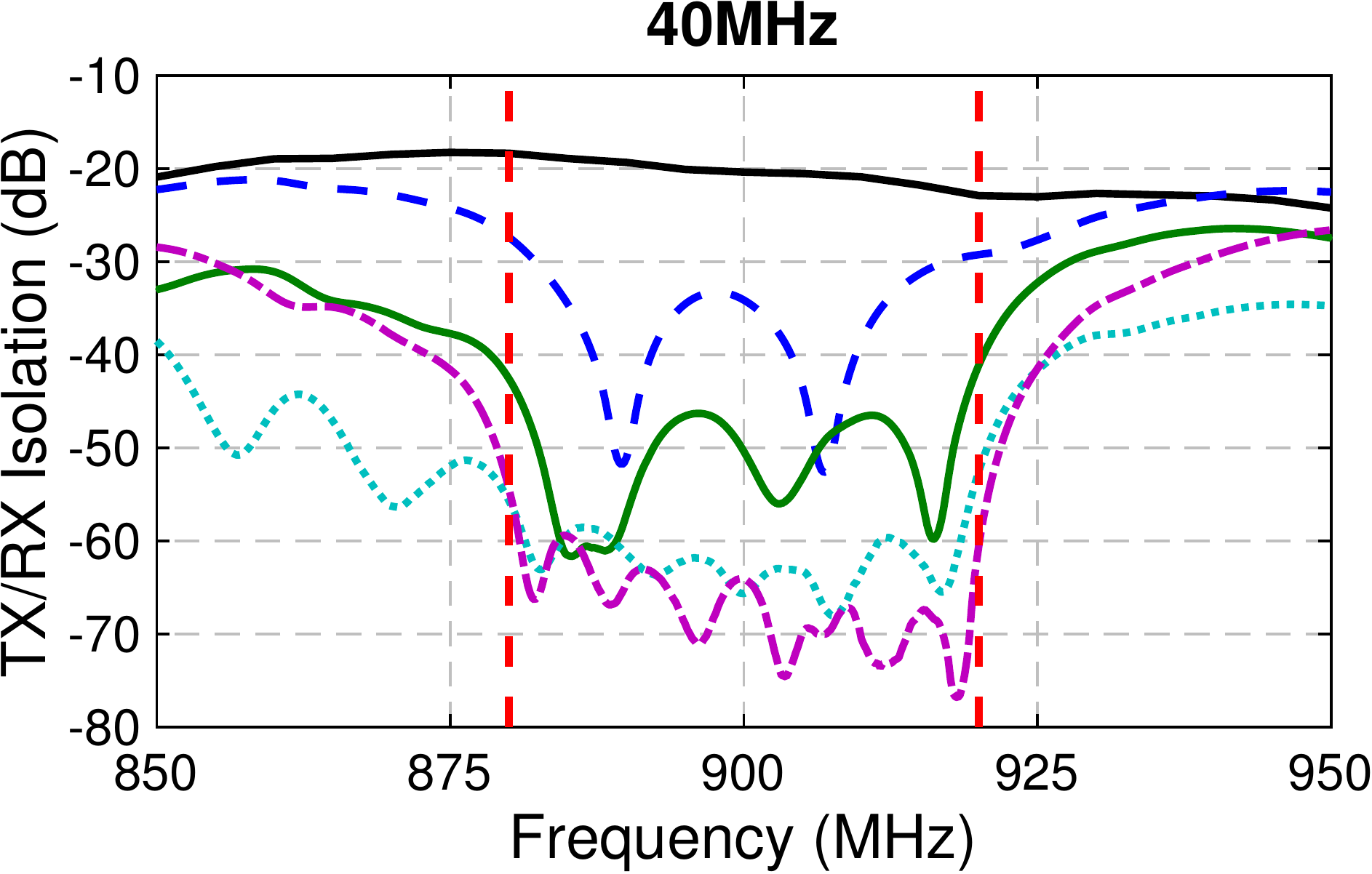}
}
\subfloat{
\includegraphics[width=0.48\columnwidth]{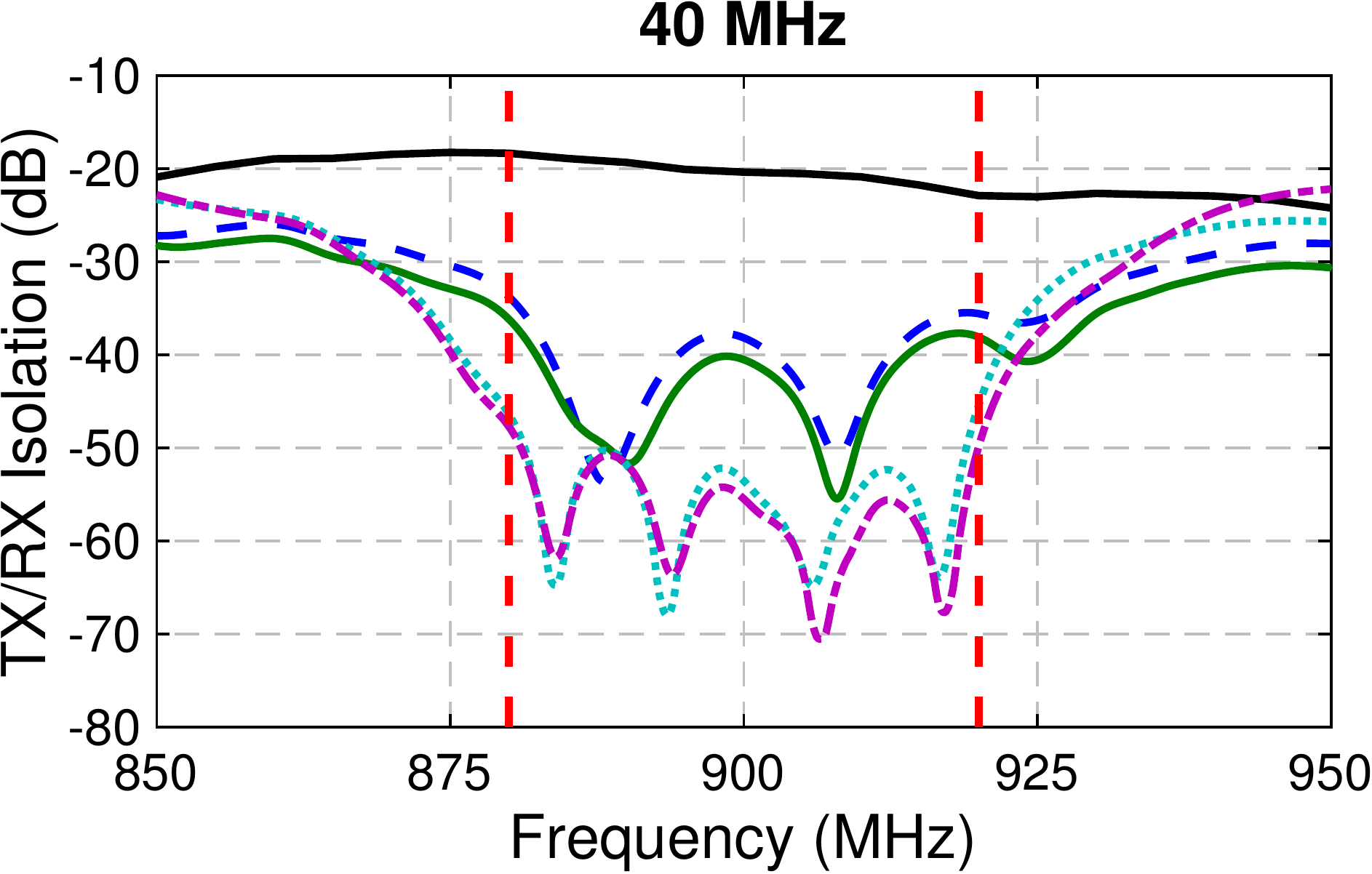}
}
\vspace{-0.5\baselineskip}
\\
\setcounter{subfigure}{0}
\subfloat[RFIC Canceller]{
\label{fig:sim-ideal-rfic}
\includegraphics[width=0.48\columnwidth]{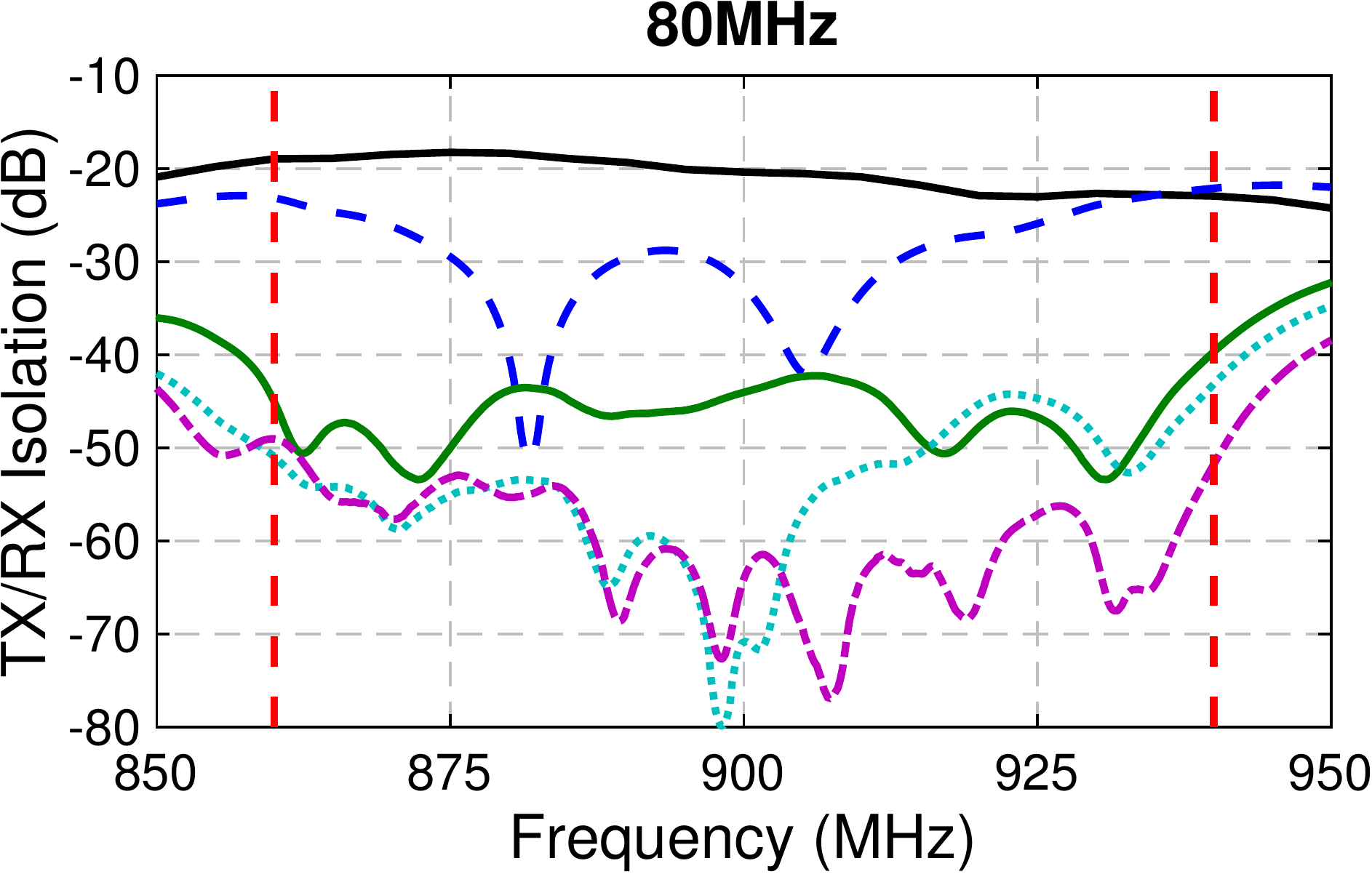}
}
\subfloat[PCB Canceller]{
\label{fig:sim-ideal-pcb}
\includegraphics[width=0.48\columnwidth]{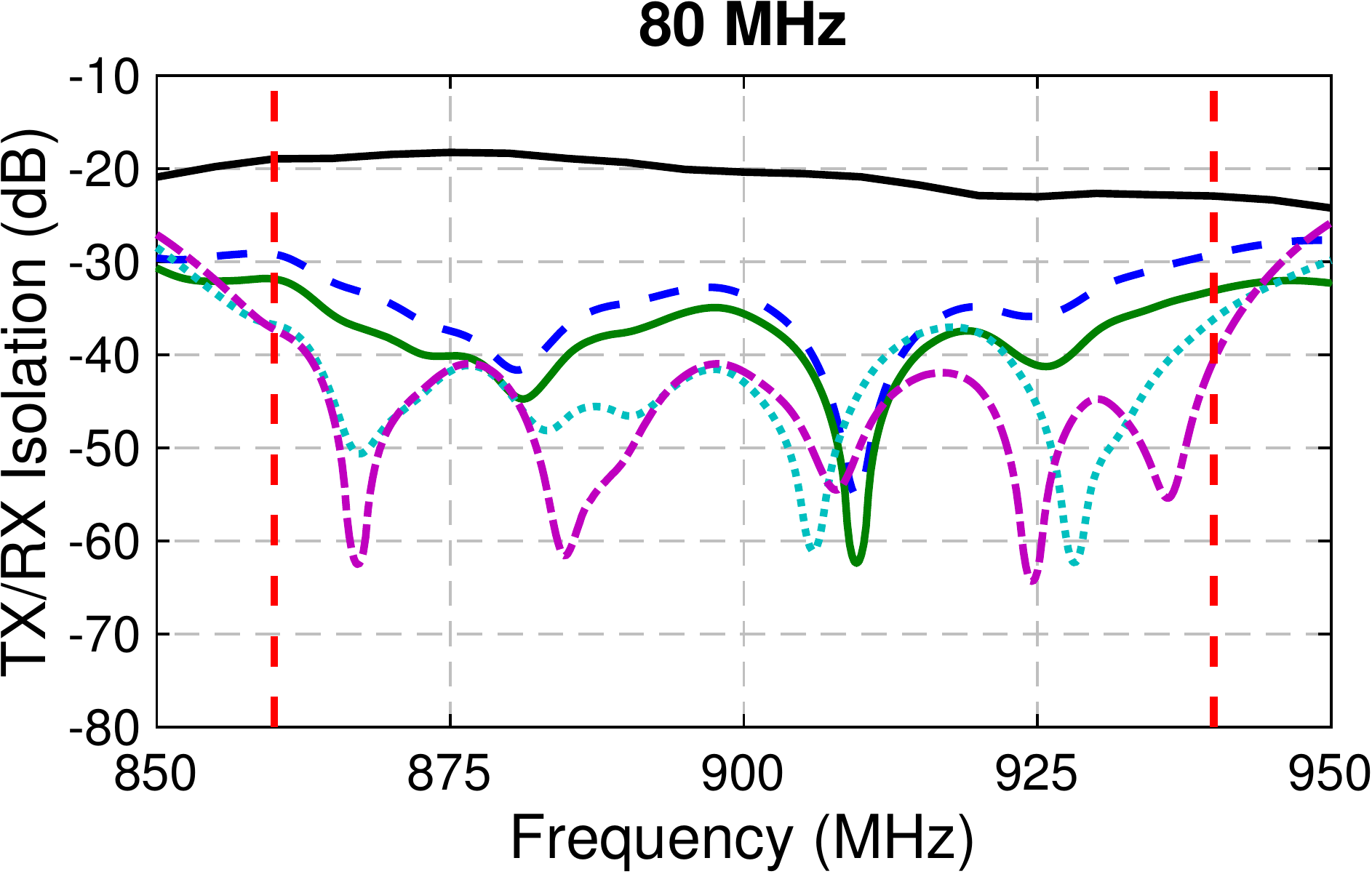}
}
\vspace{-0.5\baselineskip}
\caption{TX/RX isolation of the antenna interface (black curve) and with the RFIC and PCB cancellers with varying number of FDE taps, $\NumTap \in \{1,2,3,4\}$, and desired RF SIC bandwidth, $\BW \in \{20,40,80\}\thinspace\SI{}{MHz}$, in the ideal case.}
\label{fig:sim-ideal}
\vspace{-\baselineskip}
\end{figure}

Fig.~\ref{fig:sim-ideal} shows the TX/RX isolation achieved by the RFIC and PCB cancellers with optimized canceller configuration, with varying $\NumTap$ and $\BW$ in the ideal case (i.e., without quantization constraints). It can be seen that: (i) under a given value of $\BW$, a larger number of FDE taps results in higher average RF SIC, and (ii) for a larger value of $\BW$, more FDE taps are required to achieve sufficient RF SIC. For example, the ideal RFIC and PCB cancellers with 2 FDE taps can achieve an average $50/46/42\thinspace\SI{}{dB}$ and $50/42/35\thinspace\SI{}{dB}$ RF SIC across $20/40/80\thinspace\SI{}{MHz}$ bandwidth, respectively.

\begin{figure}[!t]
\centering
\vspace{-\baselineskip}
\subfloat{
\includegraphics[width=0.48\columnwidth]{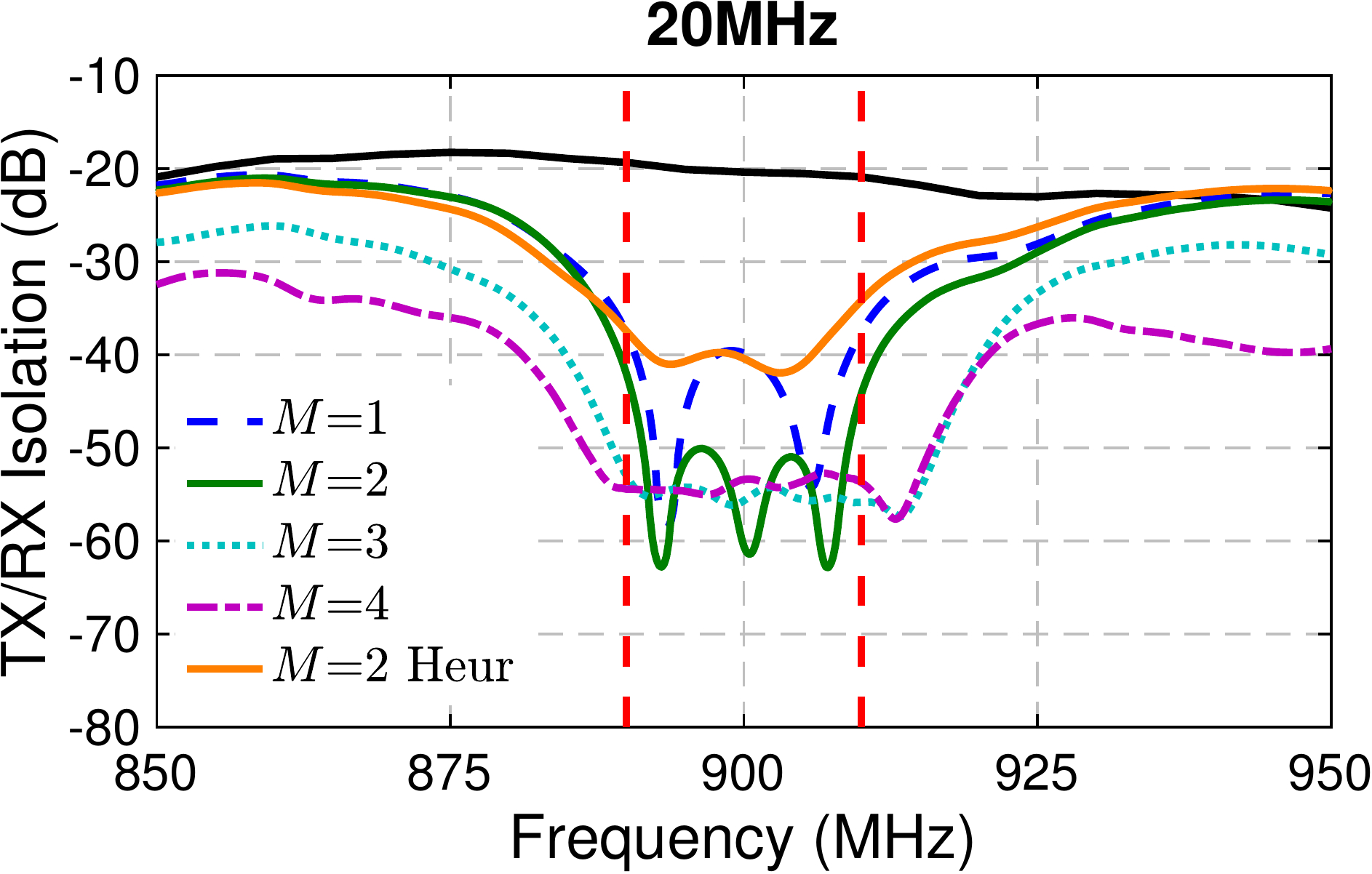}
}
\subfloat{
\includegraphics[width=0.48\columnwidth]{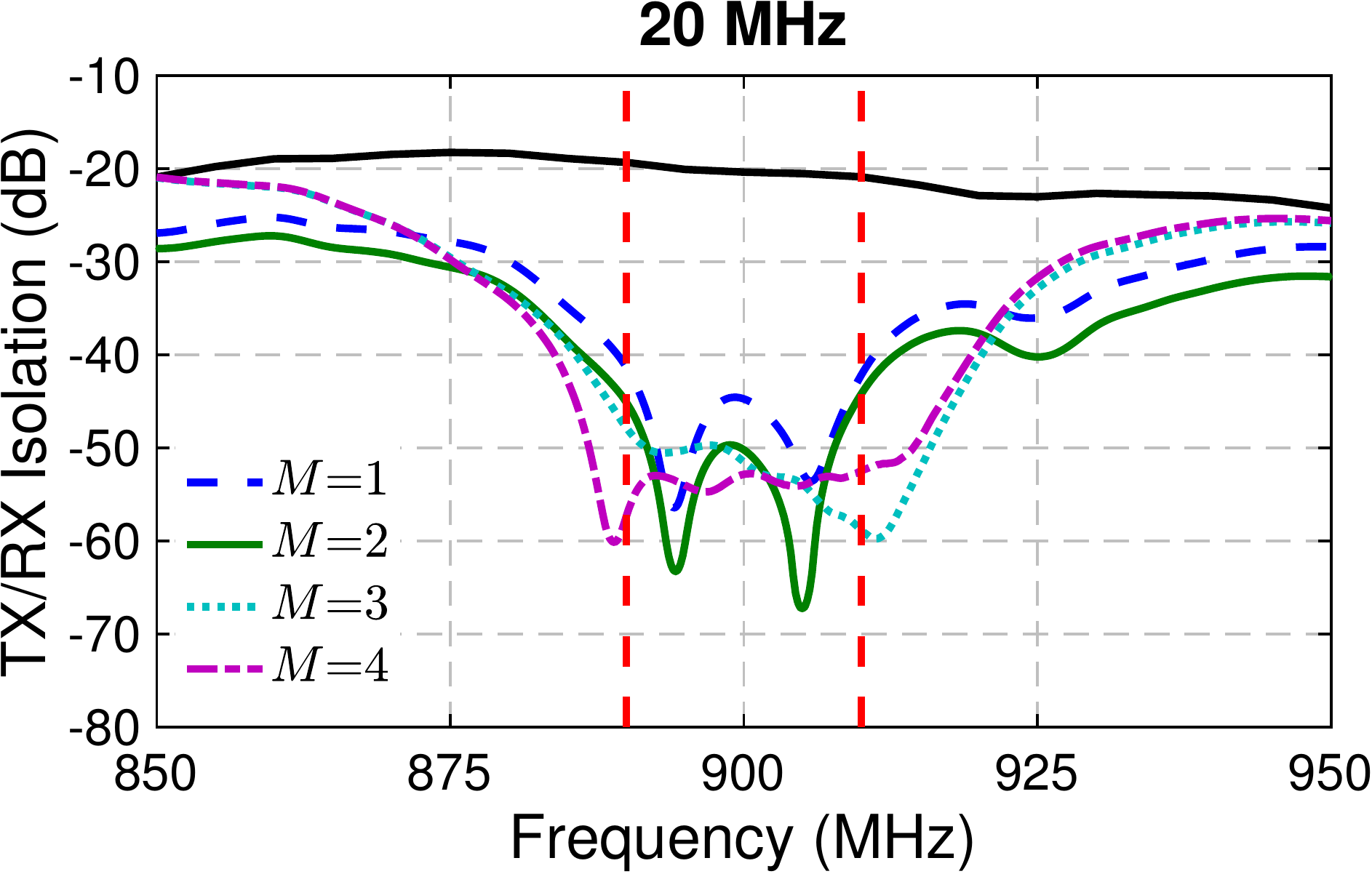}
}
\vspace{-0.5\baselineskip}
\\ 
\subfloat{
\includegraphics[width=0.48\columnwidth]{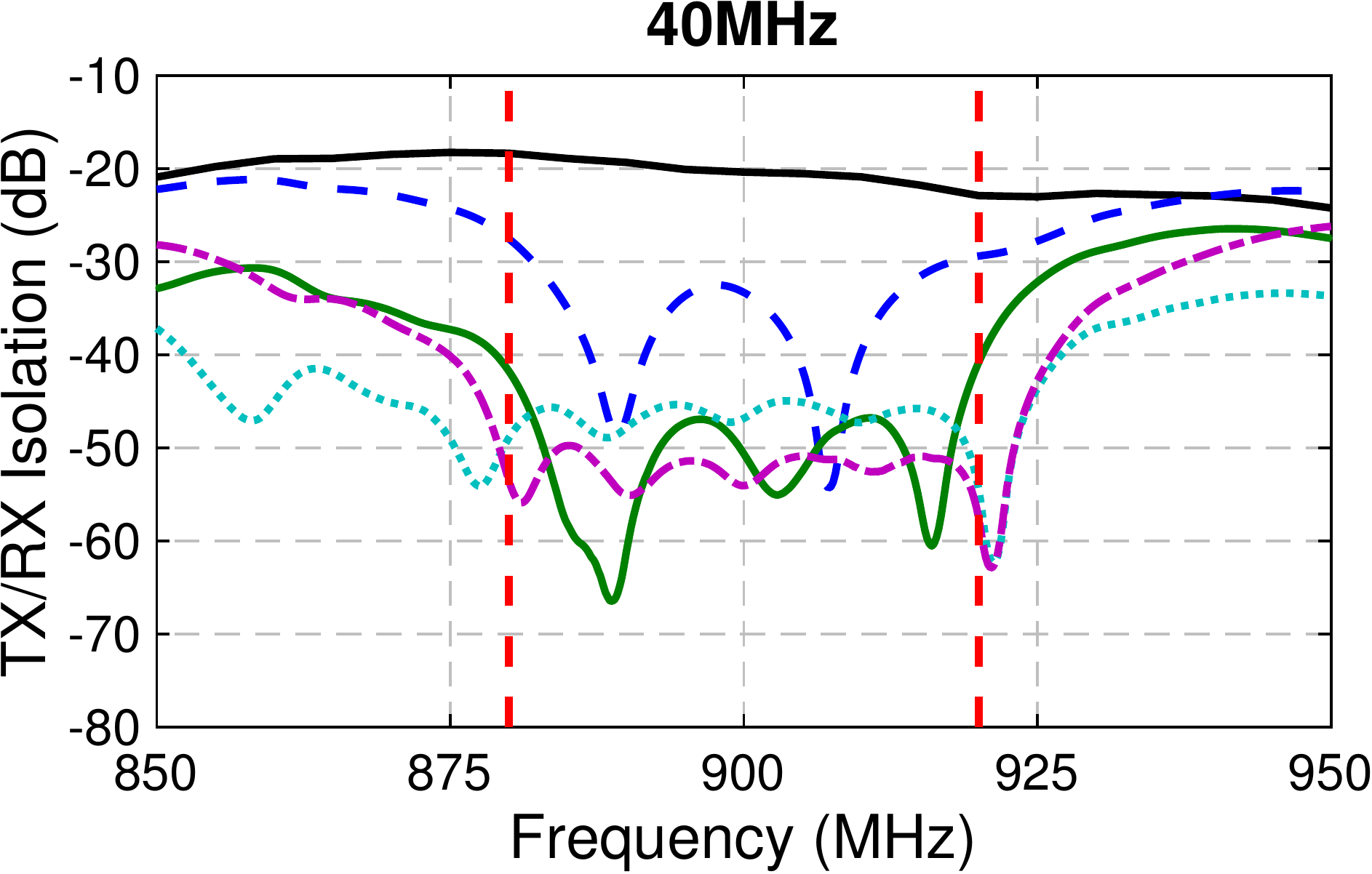}
}
\subfloat{
\includegraphics[width=0.48\columnwidth]{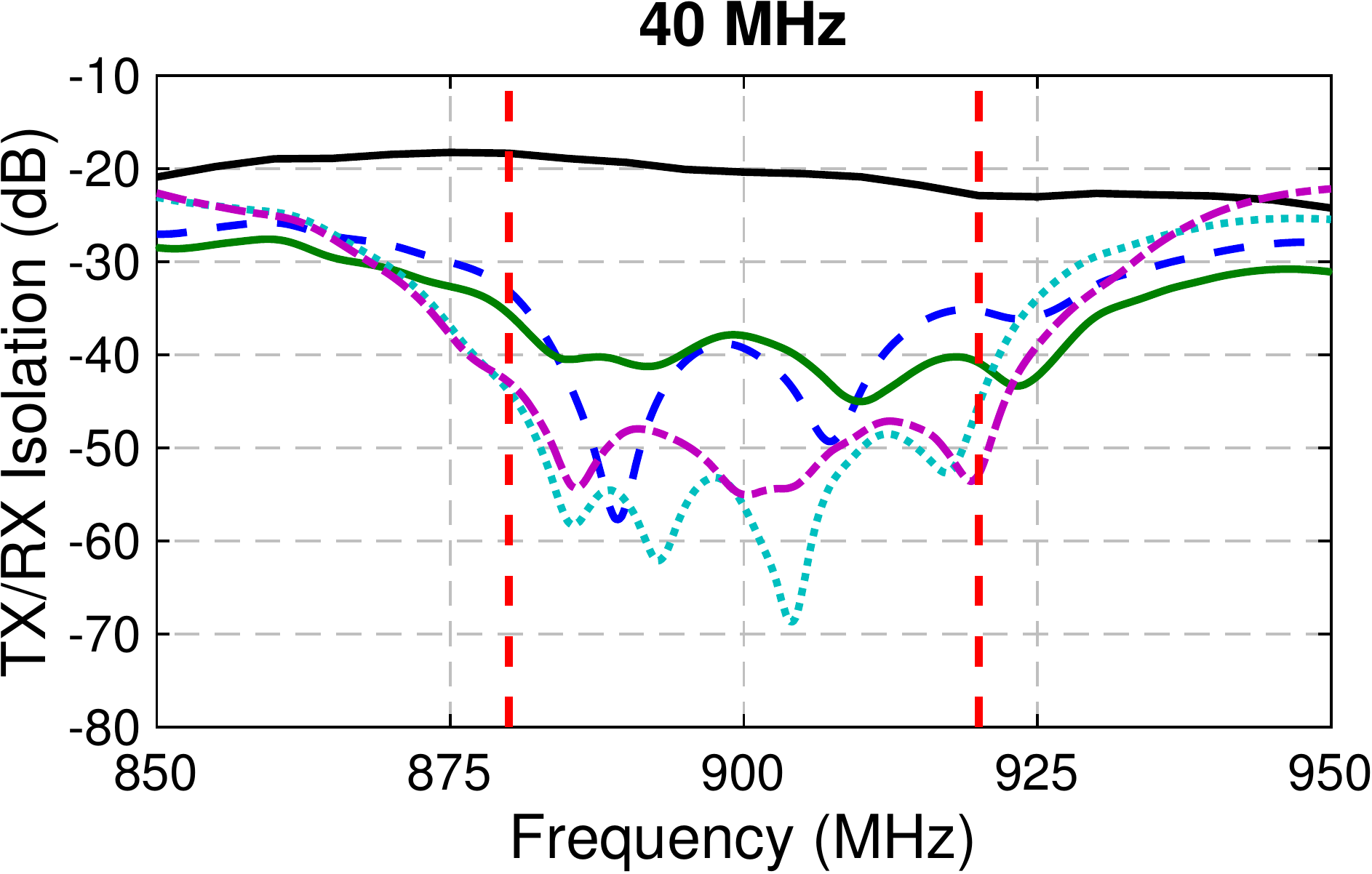}
}
\vspace{-0.5\baselineskip}
\\ 
\setcounter{subfigure}{0}
\subfloat[RFIC Canceller]{
\label{fig:sim-quan-rfic}
\includegraphics[width=0.48\columnwidth]{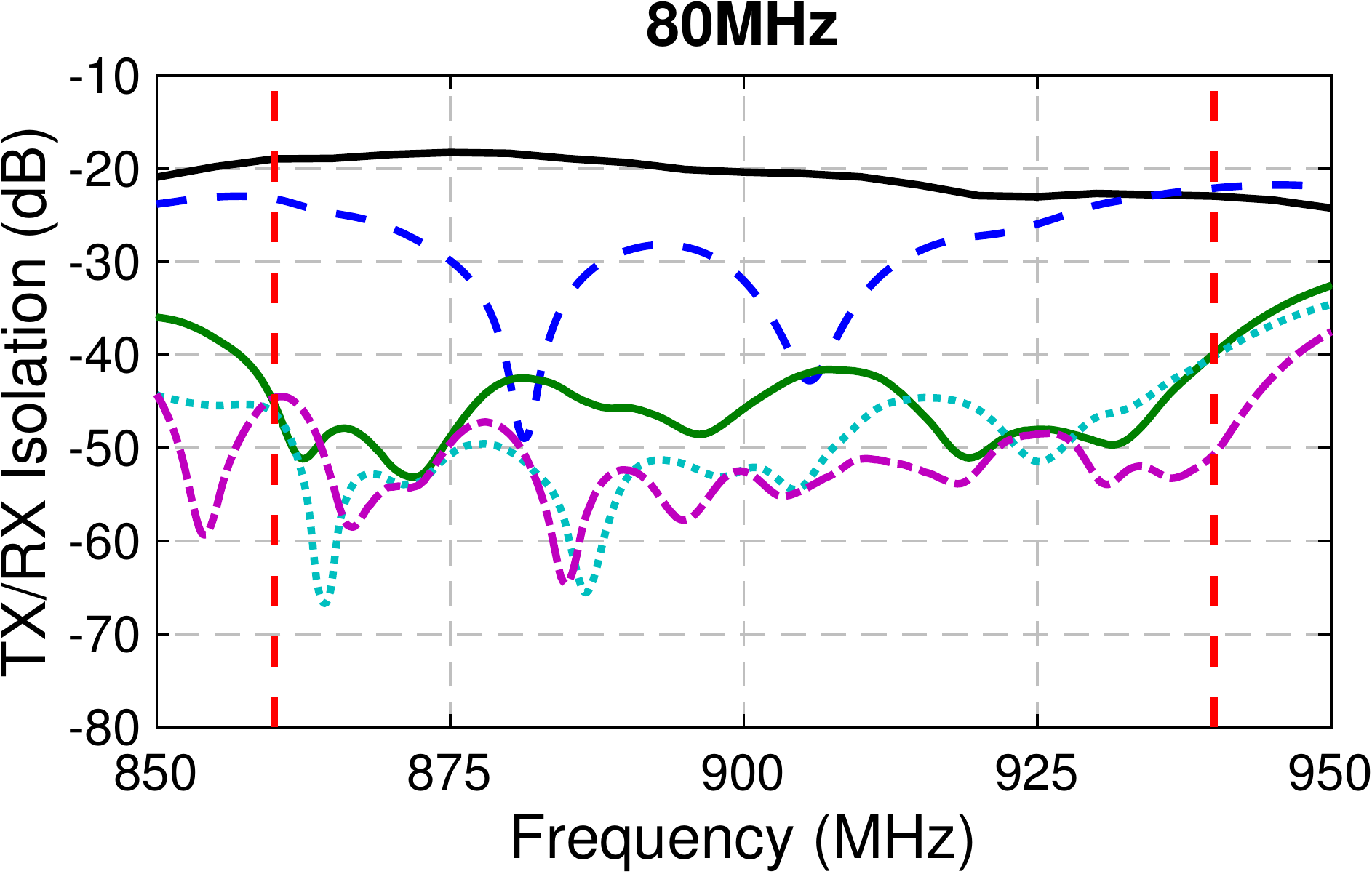}
}
\subfloat[PCB Canceller]{
\label{fig:sim-quan-pcb}
\includegraphics[width=0.48\columnwidth]{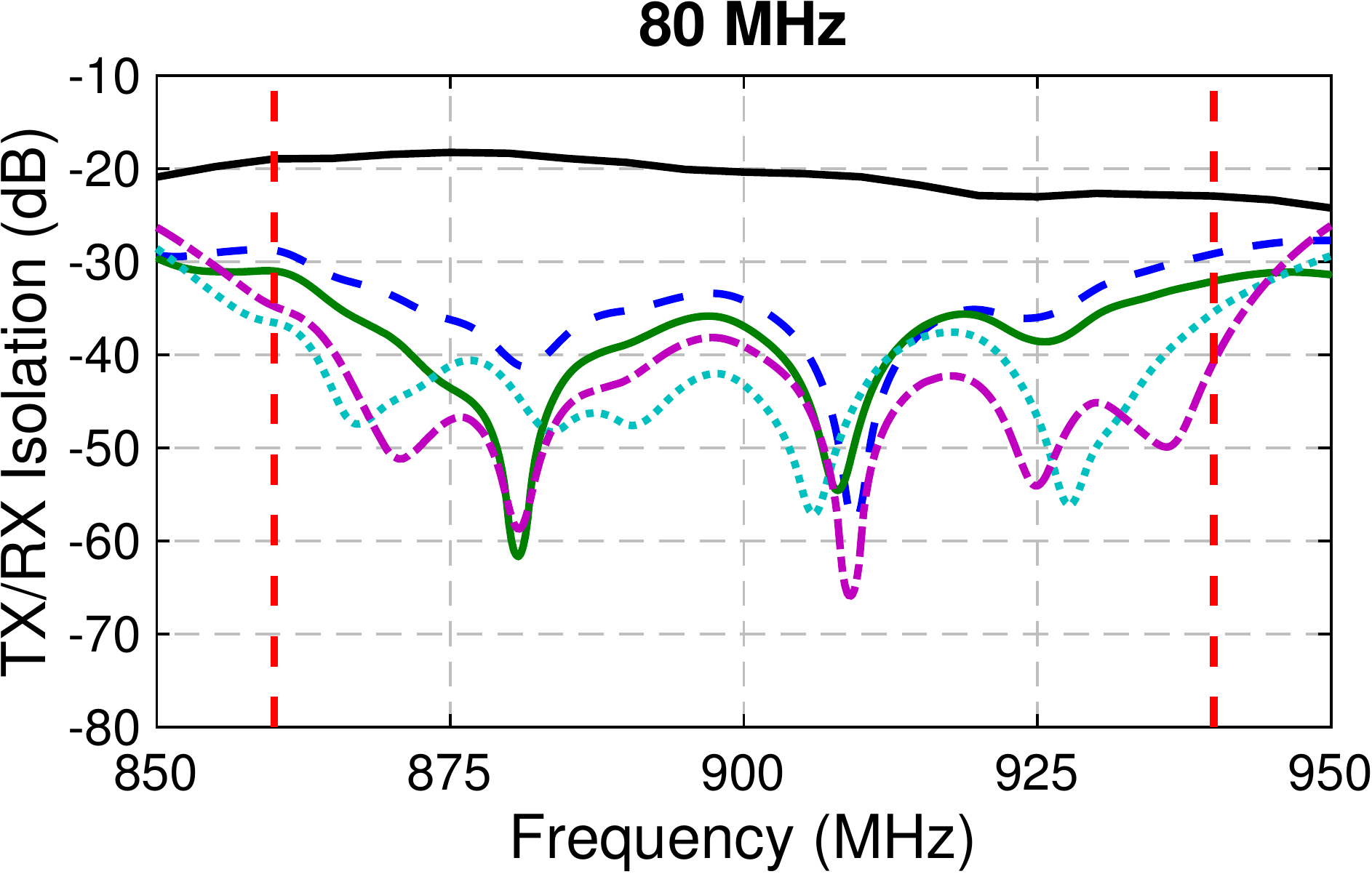}
}
\vspace{-0.5\baselineskip}
\caption{TX/RX isolation of the antenna interface (black curve) and with the RFIC and PCB cancellers with varying number of FDE taps, $\NumTap \in \{1,2,3,4\}$, and desired RF SIC bandwidth, $\BW \in \{20,40,80\}\thinspace\SI{}{MHz}$, under practical quantization constraints.}
\label{fig:sim-quan}
\vspace{-\baselineskip}
\end{figure}

Fig.~\ref{fig:sim-quan} shows the TX/RX isolation achieved by the RFIC and PCB cancellers with optimized canceller configuration under practical quantization constraints. Comparing to Fig.~\ref{fig:sim-ideal}, the results show a performance degradation due to limited hardware resolutions, which is more significant as $\NumTap$ increases. This is because a larger value of $\NumTap$ introduces a higher number of DoF with more canceller parameters that can be flexibly controlled. As a result, the RF SIC performance is more sensitive to the coupling between individual FDE tap responses after quantization. The results show that under practical constraints, the RFIC and PCB cancellers with 4 FDE taps can still achieve an average $54/50/45\thinspace\SI{}{dB}$ and $52/45/39\thinspace\SI{}{dB}$ RF SIC across $20/40/80\thinspace\SI{}{MHz}$ bandwidth, respectively. Fig.~\ref{fig:sim-quan} also shows that the RFIC canceller under the optimized configuration scheme achieves a $\SI{10}{dB}$ higher RF SIC compared with that achieved by the heuristic approach described in~\cite{Zhou_WBSIC_JSSC15} (labeled ``Heur'').

It is interesting to observe that the RF SIC profile of the PCB canceller with 2 FDE taps is very similar to our experimental results (see Fig.~\ref{fig:exp-usrp-algo} in Section~\ref{ssec:exp-node}). It is also worth noting that, in practice, adding more FDE taps cannot improve the amount of RF SIC in some scenarios (e.g., with $\SI{20}{MHz}$ bandwidth), which is limited by the quantization constraints. However, performance improvement is expected by relaxing these constraints (e.g., through using components with higher resolutions and/or wider tuning ranges).

\begin{table}[!t]
\caption{Comparison between the PCB and RFIC cancellers.}
\label{table:comparison-pcb-rfic}
\vspace{-0.5\baselineskip}
\scriptsize
\begin{center}
\begin{tabular}{|c|c|c|}
\hline
& PCB (this work) & RFIC~\cite{Zhou_WBSIC_JSSC15} \\
\hline
Center Frequency & $\SI{900}{MHz}$ & $\SI{1.37}{GHz}$ \\
\hline
\# of FDE Taps & 2 & 2 \\
\hline
Antenna Interface & \makecell{A single antenna \\ and a circulator} & \makecell{A TX/RX \\ antenna pair} \\
\hline
Antenna Isolation & $\SI{20}{dB}$ & $\SI{35}{dB}$ \\
\hline
Canceller SIC ($\SI{20}{MHz}$) & $\SI{32}{dB}$ & $\SI{20}{dB}$ \\
\hline
Canceller Configuration & Optimization {\OptProblemPCB} & Heuristic \\
\hline
Digital SIC & $\SI{43}{dB}$ & N/A \\
\hline
Evaluation & Node/Link/Network & Node \\
\hline
\end{tabular}
\end{center}
\vspace{-\baselineskip}
\end{table}

Table~\ref{table:comparison-pcb-rfic} shows the comparison between our implemented PCB canceller and the RFIC canceller presented in~\cite{Zhou_WBSIC_JSSC15}.
To summarize, we numerically show that the performance of the RFIC and PCB cancellers is similar. The results based on measurements and validated canceller models confirm that the PCB canceller accurately emulates its RFIC counterpart, and that the FDE-based approach is valid and suitable for achieving wideband RF SIC in small-form-factor devices.


\section{Conclusion}
\label{sec:conclusion}
We designed and implemented a PCB canceller using the FDE technique, which was shown to achieve wideband RF SIC in compact nodes. We presented a PCB canceller model and a scheme for optimizing the canceller configuration. We experimentally evaluated the performance of the FDE-based FD radio at the node, link, and network levels using an SDR testbed. We also compared the RFIC and PCB implementations and discussed various design tradeoffs of the FDE-based canceller.
Future directions include: (i) better design and implementation of FDE-based canceller to support higher TX power handling and RF SIC bandwidth, (ii) extension of the FDE technique to multi-antenna systems, (iii) integration in open-access testbeds, and (iv) development and experimental evaluation of resource allocation and scheduling algorithms tailored for FDE-based FD radios.


\section{Acknowledgements}
\label{sec:acks}
This work was supported in part by NSF grants ECCS-1547406, CNS-1650685, and CNS-1827923. We thank Jackson Welles for his contributions to the assembly and testing of the PCB cancellers.

\section{Appendix A: The PCB BPF Model}
\label{append:pcb-model}
We use transmission (ABCD) matrix to derive $\BPFTapTF{i}(f)$, given by {\eqref{eq:pcb-bpf-tf}}. From Fig.~\ref{fig:diagram-pcb} and $Y_{\textrm{F}}(f_k)$ and $Y_{\textrm{Q}}(f_k)$ in {\eqref{eq:pcb-admittance}},
\begin{align}
\begin{bmatrix}
V_{\textrm{in}} \\ I_{\textrm{in}}
\end{bmatrix}
& =
\begin{bmatrix}
1 & 0 \\ Y_{\textrm{Q}}(f_k) & 1
\end{bmatrix}
\mathbf{M}^{\textrm{TL}}
\begin{bmatrix}
1 & 0 \\ Y_{\textrm{F}}(f_k) & 1
\end{bmatrix}
\mathbf{M}^{\textrm{TL}}
\begin{bmatrix}
1 & 0 \\ Y_{\textrm{Q}}(f_k) & 1
\end{bmatrix}
\begin{bmatrix}
V_{\textrm{out}} \\ I_{\textrm{out}}
\end{bmatrix}
\nonumber \\
& :=
\begin{bmatrix}
M_{A}^{\textrm{BPF}}(f_k) & M_{B}^{\textrm{BPF}}(f_k) \\
M_{C}^{\textrm{BPF}}(f_k) & M_{D}^{\textrm{BPF}}(f_k)
\end{bmatrix}
\begin{bmatrix}
V_{\textrm{out}} \\ I_{\textrm{out}}
\end{bmatrix},
\label{eq:pcb-bpf-matrix}
\end{align}
where $\mathbf{M}^{\textrm{TL}}$ is the ABCD matrix of a T-Line with wavenumber $\beta$, characteristic impedance $Z_0$, and length $l$, i.e., 
\begin{align}
\label{eq:pcb-tl-matrix}
\mathbf{M}^{\textrm{TL}} & = 
\begin{bmatrix}
\cos{(\beta l)} & jZ_{0}\sin{(\beta l)} \\
j\sin{(\beta l)}/Z_{0} & \cos{(\beta l)} \\
\end{bmatrix}.
\end{align}
With the parameters described in Section~\ref{ssec:impl-pcb}, the frequency response of the implemented PCB BPF, $\BPFTapTF{i}(f_k)$, is given by
\begin{align}
& \BPFTapTF{i}(f_k) = \frac{V_{\textrm{out}}(f_k)}{V_{\textrm{in}}(f_k)} = \frac{1}{R_{\textrm{S}}} \cdot \frac{V_{\textrm{out}}(f_k)}{I_{\textrm{in}}(f_k)} = \frac{1}{R_{\textrm{S}}} \cdot \frac{1}{M_C^{\textrm{BPF}}(f_k)}. \nonumber 
\end{align}
Plugging {\eqref{eq:pcb-admittance}} and {\eqref{eq:pcb-tl-matrix}} into {\eqref{eq:pcb-bpf-matrix}} yields the model $\BPFTapTF{i}(f)$.

\bibliographystyle{ACM-Reference-Format}
\bibliography{paper_fde} 

\end{document}